\documentclass[pra, reprint,superscriptaddress,floatfix]{revtex4-2}
\usepackage{amsmath}
\usepackage{amsfonts}
\usepackage{graphicx}
 \usepackage{physics}
 \usepackage{bbm}
 \usepackage{comment}
 \usepackage{amssymb}

\def \markColorOne {Blue}
\def \markColorTwo {Green}
\def \markColorThree{Plum}

\def \markColorOne {Black}
\def \markColorTwo {Black}
\def \markColorThree{Black}

 \usepackage[dvipsnames]{xcolor}

 \def\be{\begin{equation}} \def\ee{\end{equation}}
\def\bea{\begin{eqnarray}} \def\eea{\end{eqnarray}}

\begin{document}

\title{Unified light-matter Floquet theory  and its application to quantum communication}

\author{Georg Engelhardt}
\email{engelhardt@sustech.edu.cn}
\affiliation{Shenzhen Institute for Quantum Science and Engineering, Southern University of Science and Technology, Shenzhen 518055, China}
\affiliation{International Quantum Academy, Shenzhen 518048, China}
\affiliation{Guangdong Provincial Key Laboratory of Quantum Science and Engineering, Southern University of Science and Technology, Shenzhen 518055, China}

\author{Sayan Choudhury}
\email{sayan.choudhury@pitt.edu}
\affiliation{Department of Physics and Astronomy, University of Pittsburgh, Pittsburgh, PA 15260, USA}

\affiliation{Harish Chandra Research Institute, A CI of Homi Bhabha National Institute, Chhatnag Road, Jhunsi, Prayagraj, UttarPradesh 211019, India}

\author{W. Vincent Liu}
\email{wvliu@pitt.edu}
\affiliation{Department of Physics and Astronomy, University of Pittsburgh, Pittsburgh, Pennsylvania 15260, USA}
\affiliation{International Quantum Academy, Shenzhen 518048, China}
\affiliation{Shanghai Research Center for Quantum Sciences, Shanghai 201315, China}

\date{\today}

\begin{abstract}
Periodically-driven quantum systems can exhibit a plethora of intriguing non-equilibrium phenomena that can be analyzed using Floquet theory. Naturally, Floquet theory is employed to describe the dynamics of atoms interacting with intense laser fields. However, this semiclassical analysis can not account for quantum-optical phenomena that rely  on the quantized nature of light. In this paper, we take a significant step to go beyond the semiclassical description of atom-photon coupled systems by unifying Floquet theory with quantum optics using the framework of full-counting statistics. This is achieved by introducing counting fields that keep track of the photonic dynamics. This formalism, which we dub  ``photon-resolved Floquet theory" (PRFT), is based on two-point tomographic measurements, instead of the two-point projective measurements used in standard full-counting statistics.  Strikingly, the PRFT predicts the generation of macroscopic light-matter entanglement when atoms interact with multimode electromagnetic fields, thereby leading to complete decoherence of the atomic subsystem in the basis of the Floquet states. This decoherence occurs rapidly in the optical frequency regime, but  is negligible in the radio frequency regime.  Our results thus pave the way for the design of efficient quantum memories and quantum operations.  Finally, employing the PRFT, we propose a quantum communication protocol that can significantly outperform the state-of-art few-photon protocols by two orders of magnitude or better.  The PRFT  potentially leads to insights in various Floquet settings including spectroscopy, thermodynamics, quantum metrology, and quantum simulations.

\end{abstract}
	
\maketitle

\section{Introduction}
\label{sec:introduction}

In recent years, periodic driving has emerged as a powerful tool for the coherent control of many-body systems. This has led to the realization of novel quantum phases of matter like dynamical topological states~\cite{rudner2013anomalous,rechtsman2013photonic,lindner2011floquet,roy2017floquet,lababidi2014counter,rudner2020band,wintersperger2020realization,maczewsky2017observation,Engelhardt2016,Engelhardt2017,Hu2020,huang2020floquet,pyrialakos2022bimorphic,potter2016classification,Engelhardt2013,Ng2019} and discrete time crystals~\cite{sacha2018review,sacha2020book,nayak2019review,khemani2019review,sachapra2015,sondhi2016prb,sondhi2016prl,yao2017prl,nayak2016prl,monroe2017nature,liu2018prl,choudhury2021route,lukin2017nature,Russomanno2017,yang2021dynamical,Pena2022} as well as breakthroughs in applications like spectroscopy~\cite{Engelhardt2019,Gu2018,chen2021floquet,Yan2019}, metrology~\cite{lyu2020eternal,Zhang2023,choi2017quantum}, and quantum simulation~\cite{schweizer2019floquet,gorg2019realization,arnal2020chaos,weitenberg2021tailoring,choudhury2014stability,choudhury2015stability,choudhury2015transverse,geier2021floquet,strater2016floquet,meinert2016floquet,clark2018observation,gorg2018enhancement}. These non-equilibrium quantum systems are generally analyzed using Floquet theory --- a method first developed by Jon Shirley in 1965~\cite{Shirley1965}. Interestingly, Floquet theory is also employed to investigate the dynamics of quantum systems interacting with a single-frequency quantum field. A particularly striking example of this is the case of atoms interacting with an intense laser field~\cite{Guerin1997,sentef2020quantum}. Despite decades of extensive progress in quantum optics, it remains extremely challenging to employ a completely quantum mechanical treatment to this situation, due to the large number of photons involved~\cite{Mandel2008}. Often, a semiclassical approach is used instead, where the photon fields are assumed to be high-energy coherent states, and their dynamics is neglected. This leads to an effective Floquet description of the atomic dynamics, which can be employed to engineer materials with novel emergent properties \cite{hubener2021engineering,wang2019cavity,schlawin2022cavity,li2020manipulating,kiffner2019manipulating,gulacsi2015floquet,schafer2018ab,eckhardt2022quantum,Rokaj2022,lloyd20212021}. Unfortunately, while this semiclassical treatment is very powerful in modeling the matter subsystem, it fails to describe the photonic driving field.

\begin{figure*}
	\includegraphics[width=0.99\linewidth]{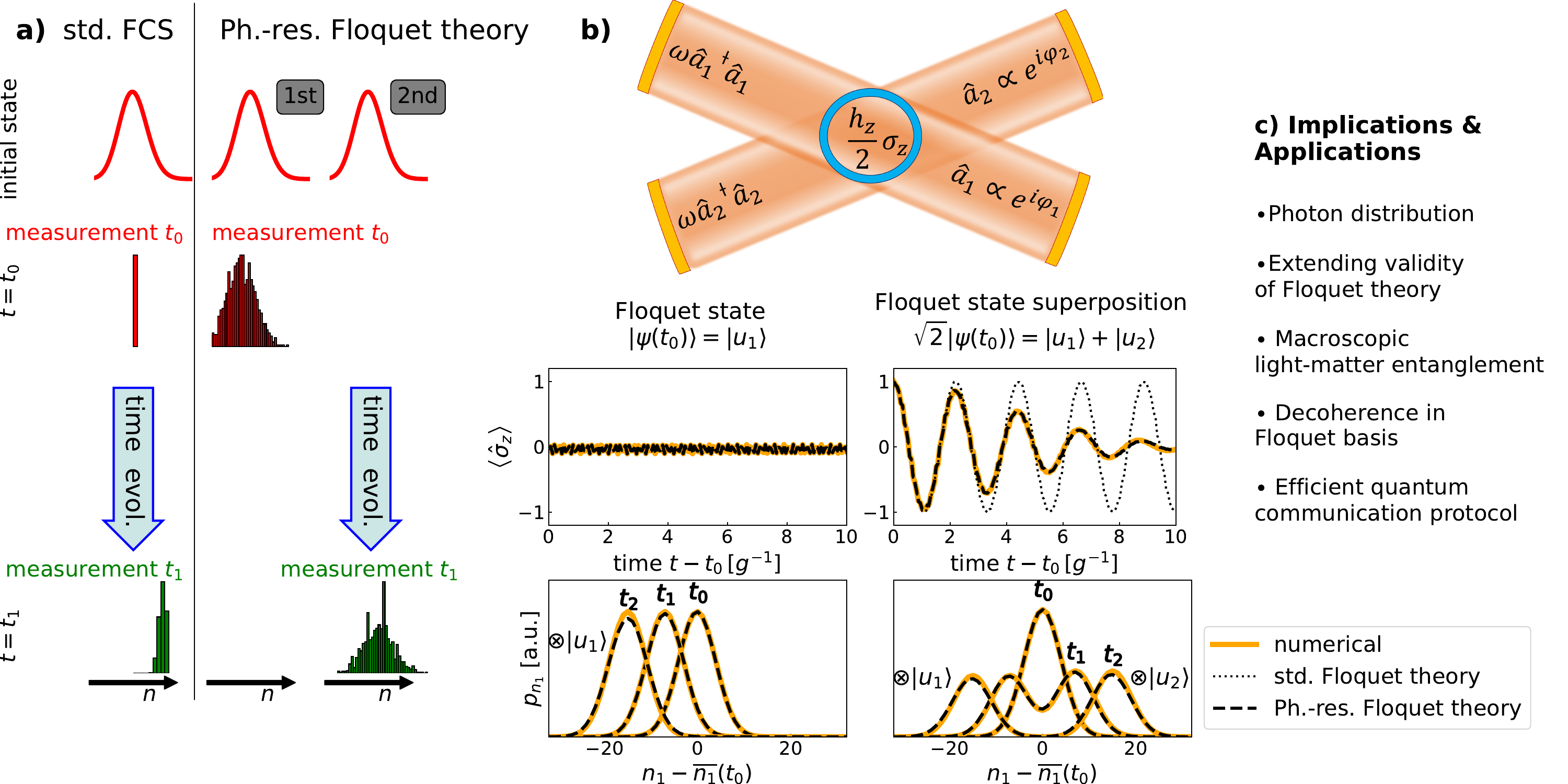}
	\caption{  Schematic illustration of the fundamental difference between the FCS and the PRFT. (a) The standard FCS extracts the statistical information based on two-point projective measurements at times $t_0$ and $t_1$.  The  counting statistics in the PRFT relies on two-point tomographic measurements of two independent batches. One of which is used to determine the photon statistics $p_n(t)$ at time $t_0$, and the other determines the statistics of the time-evolved state $t_1$ without being measured at $t_0$. (b) Illustration of the light-matter entanglement in a two-mode Rabi model. As explained in details in Secs.~\ref{sec:FloquetStateAnalysis} and \ref{sec:twoModeRabiModel}, the two-level  system controls the photon transport between the two photonic modes, which leads to entanglement in the Floquet basis. Consequently, this entanglement gives rise to decoherence, which is incorrectly described by the standard Floquet theory, but correctly predicted by the PRFT. (c) Overview of some important implications and applications of the PRFT.}
	\label{figCountingStatSketch}
\end{figure*}

In this paper, we take a significant step beyond the semiclassical Floquet theoretic description of light-matter interactions by developing a   framework dubbed ``photon-resolved Floquet theory" (PRFT). The PRFT bridges  Floquet theory and quantum optics by introducing full-counting statistics (FCS) of photons in the semiclassical description of the quantum system. 
Originally developed in the context of quantum optics~\cite{mandel1995optical,mandel1979sub,cook1981photon}, FCS is a powerful method that has been employed to study mesoscopic transport~\cite{levitov1996electron,Brandes2008,Boehling2018,Xue2015,Engelhardt2019a,You2019,dasenbrook2014floquet,yadalam2016statistics}, quantum dots~\cite{groth2006counting,Schaller2018,kleinherbers2018revealing,benito2016fullCounting,Honeychurch2020}, spin chains \cite{stephan2017full,groha2018full}, spontaneous photon emission~\cite{pletyukhov2015quantum,zhang2018quantum,brange2019photon,nesterov2020counting}, thermodynamics~\cite{Restrepo2018,engelhardt2018maxwells},  and the entanglement entropy of noninteracting fermions~\cite{klich2009quantum,lacroix2019rotating,smith2021counting}. However, we must exercise caution in applying the standard FCS framework to  coherently laser-driven systems described in the previous paragraphs. This is because FCS is inherently based on  two-point projective measurements~\cite{esposito2009nonequilibrium}. Unfortunately, such measurements  destroy the coherent photonic states, thereby rendering the Floquet description of the matter system invalid. 

The PRFT developed here can track the photonic dynamics, without destroying the Floquet description of the matter system. This is achieved by introducing a framework based on two-point tomographic measurements of the photonic field, instead of the usual projective measurements. The distinction between two-point projective and two-point tomographic measurements is illustrated in Fig.~\ref{figCountingStatSketch}(a). Formally, the PRFT introduces counting fields into the semiclassical equations of motion, leading to a dynamical cumulant generating function. This in turn enables us to investigate the redistribution of photons amongst the Fock states and gain a clear picture of the photonic dynamics. We note that while some recent interesting works have investigated the photonic dynamics of driven systems in Sambe space~\cite{Long2021,Crowley2020,PhysRevB.99.064306}, these approaches scale exponentially with the number of frequency modes; the PRFT does not suffer from this limitation.  The PRFT thus enables us to go beyond the previous approaches that connects Floquet theory to cavity dressed states as in Ref.~\cite{Guerin1997}, since those approaches do not provide correct results for the photon statistics.

Based on analytical derivations and extensive numerical benchmarking, we demonstrate that the PRFT is valid for   coherent and number squeezed states with a moderate mean photon number $\overline n >500$ and photon standard deviation as low as $\sigma =4$. The  PRFT thus covers all types of moderate and highly occupied  photonic fields in experiments, such as radio frequencies, microwaves and lasers, i.e., all driving fields for which standard Floquet theory is believed to be valid. In other words, the PRFT approach here assumes the same level of generic conditions as the standard Floquet theory, but is found remarkably capable of capturing  the dynamics of matter system as well of its photonic counterpart.

We employ the PRFT to analyze multimode driving, and discover that this leads to macroscopic light-matter entanglement at long times due to the matter-system-controlled transport of photon between distinct modes. This light-matter entanglement causes a complete decoherence of the matter system in the basis of the Floquet states. This effect is depicted for a pardigmatic two-mode Rabi model in Fig.~\ref{figCountingStatSketch}(b).  The PRFT thus provides a quantum-optical interpretation of Floquet states as the decohering basis for the matter system. As the standard Floquet theory is unable to describe this fundamental decoherence effect, it will incorrectly predict the dynamics of the matter system in general. The PRFT thus demonstrates that even in the semiclassical regime,  fundamental physical implications of the  standard Floquet theory are not understood and require further investigation. {\color{\markColorOne} In this context, it is worth noting that light-matter entanglement can arise even in single-mode models due to the photon shot noise~\cite{GeaBanacloche1990,GeaBanacloche1991,GeaBanacloche2002,Phoenix1991,Goldberg2020}. However,  this shot-noise induced entanglement has a much smaller impact on the photonic dynamics compared to the transport-related entanglement revealed by the PRFT.  A detailed analysis of these issues is presented in Sec.~\ref{sec:rabiModel}.}

Furthermore, the transport-entanglement-related decoherence effect has far-reaching experimental consequences. In particular, we demonstrate that  the quantum-optical coherence time is reasonably short (a few ms) for typical optical  fields used in experiments, but it can be very long for radio-frequency driving. This implies that the radio-frequency regime is optimal for realizing quantum memories and quantum operations. Furthermore, we argue that quantum time crystals provide a powerful platform for realizing quantum memories irrespective of the driving frequency. Intriguingly, we demonstrate that the light-matter entanglement described by the PRFT can be  deployed in a quantum-communication protocol that is intrinsically robust against photon loss. In particular, we demonstrate that using coherent light, it is possible for the quantum state transfer rate to reach the $0.1\, \text{KHz}$ regime over $500\,\text{km}$, thereby far exceeding the $\text{Hz}$ regime that is predicted in current theoretical protocols~\cite{Sangouard2011}. Our analysis thus demonstrates that the PRFT can play a pivotal role in the development of future quantum technologies.

This paper is organized as follows: In Sec.~\ref{sec:overview}, we introduce  the theoretical framework of the PRFT. In Sec.~\ref{sec:benchmarking}, we apply the PRFT to mulitmode  quantum Rabi models for benchmarking, and investigate the  light-matter entanglement. In Sec.~\ref{sec:experimentalTechnicalImpliations}, we discuss the experimental verification of the theory and implications for quantum memories, quantum time crystals, and other quantum applications.   In Sec.~\ref{sec:quantumCommunication}, we devise a quantum-communication protocol employing the PRFT framework. Finally, we summarize the main findings of this paper and discuss avenues for future research in Sec.~\ref{sec:conclusions}.

\section{Photon-resolved Floquet theory}
\label{sec:overview}

In this section, we introduce the basic ideas and main results of the PRFT. We emphasize that even though we primarily analyze Floquet systems in this paper, the formalism can also be used to analyze aperiodically driven systems. For a more detailed analysis, we refer the reader to Appendix~\ref{sec:photonResolvedFloquettheory}. 

\subsection{System}

\label{sec:system}

We consider the following generic Hamiltonian describing a matter system interacting with a multimode photonic field:
\begin{equation}
H_{\text{Q} }  = H_0 +  \sum_{k=1}^{R} \omega_k \hat a_k^\dagger  \hat a_k  +  \sum_{k=1}^{R} \tilde g H_k \left(  \hat a_k^\dagger   + \hat a_k \right),	
\label{eq:hamiltonian:quantum}
\end{equation}
where $k$ denotes the different photonic modes, $\hat a_k$ are  annihilation operators quantizing these modes and $H_k$ acts on the matter system. The light-matter interaction strength is parameterized by $\tilde g$. The dynamics of this system can be determined by representing the photonic operators with Fock states, which are the eigenstates of the occupation operators $\hat N_k = \hat a_k^\dagger \hat a_k$, i.e., $\hat N_k  \left| n \right>_k  = n   \left| n \right>_k   $. However, for typical laser fields, the photonic modes are highly occupied, such that an analytical or numerical treatment becomes infeasible for more than two modes.

Alternatively, one can employ a semiclassical description of the system by assuming that it is initially in the state
\begin{eqnarray}
\left|\psi(t_0) \right> = \left| \phi(t_0) \right> \otimes \prod_{k=1 }^{R}  \left|  \alpha_k e^{i\varphi_k} \right>,
\label{eq:initalState}
\end{eqnarray}
where $ \left|  \alpha_k e^{i\varphi_k} \right>$ are   coherent states of the photon operators $\hat a_k$ with real-valued amplitudes $\alpha_k> 0$ and phases $\varphi_k\in \left[ 0, 2\pi  \right)$, and the state $\left| \phi(t_0) \right> $ is the initial state of the matter system~\cite{sentef2020quantum,Shirley1965}.  In this semiclassical limit, we can substitute $\hat a_k \rightarrow \alpha_k e^{i\varphi_k- i \omega_k t} $ in Eq.~\eqref{eq:hamiltonian:quantum} such that we obtain the corresponding semiclassical Hamiltonian
\begin{equation}
\mathcal H(t) = H_0 +  \sum_{k=1}^{R}2 H_k  g_k  \cos (\omega_k t - \varphi_k),
\label{eq:def:hamiltonian:floquet}
\end{equation}
where we have introduced the effective light-matter interactions $g_k =\tilde{g} \alpha_k$. This description is valid as long as the back action of the quantum system on the photonic field is negligible, i.e., if $\tilde g\ll \omega_k\alpha_k $. Nevertheless the impact of the photonic field on the matter system can be large because of the product $\tilde g\alpha_k$. The semiclassical approach  is thus valid for large mean occupation numbers $\langle \hat N_k\rangle =\alpha_k^2 $.

For  a single  photonic frequency $\omega_k$ or  for commensurate frequencies (where all $\omega_{k'}/\omega_{k}$ are rational numbers), the semiclassical Hamiltonian is time periodic $\mathcal H(t) = \mathcal H(t + \tau)$ with a driving period $\tau$. Under these conditions one can apply the celebrated Floquet theory to analyze the system~\cite{Shirley1965}.
Unfortunately, this effective semiclassical description loses the microscopic information about the photonic field. The PRFT resolves the problem by introducing  counting fields into the semiclassical description. \\

{ \color{\markColorOne} Before proceeding further, we would like to point out that the transition from the quantum Hamiltonian in Eq.~\eqref{eq:hamiltonian:quantum} to the mean-field Hamiltonian in Eq.~\eqref{eq:def:hamiltonian:floquet} corresponds to the transition from Fock space to  Sambe space. In the latter, the photonic operators are replaced by their unbounded counterparts
\begin{eqnarray}
	\hat a_k^\dagger \hat a_k &\rightarrow& \sum_{n_k=-\infty}^{\infty}  n_k \left| n_k\right> \left< n_k\right|, \nonumber \\
	\hat a_k^\dagger   &\rightarrow&    \sum_{n_k=-\infty}^{\infty} g_k \left| n_k+1\right> \left< n_k\right| ,
	\label{eq:def:sambeSpace}
\end{eqnarray} 
which can be formally derived by a Fourier transformation of the Schr\"{o}dinger equation determined by the Hamiltonian in Eq.~\eqref{eq:def:hamiltonian:floquet}. It is worth noting that the photon number dependence of the matrix elements of $\hat a_k^\dagger $ in Fock space can lead to a light-matter entanglement even for coherent states, which has been extensively investigated~\cite{GeaBanacloche1990,GeaBanacloche1991,GeaBanacloche2002,Phoenix1991,Goldberg2020}. This is a consequence of the photon-number dependent light-matter interaction $g_k(n_k)\propto \sqrt{n_k}$, which results in a photon-number dependent dynamics of the matter system. As this effect is induced by the finite photon number uncertainty of the coherent light fields, we will refer to it as photon shot-noise entanglement. We emphasize that this form of light-matter entanglement has a minor effect on the photonic dynamics.}

\subsection{Photon-resolved time evolution}

The time-evolution operator $\mathcal U (t)$ corresponding to the Hamiltonian in Eq.~\eqref{eq:def:hamiltonian:floquet} does not contain the information about the microscopic state of the photonic fields. In order to track the photonic dynamics, we introduce real-valued counting fields   $\chi_k  \in \left[ 0, 2\pi\right)$, and define the generalized time-evolution  operator,
\begin{eqnarray}
U_{\boldsymbol \chi} (t) = \exp(-i \sum_k \chi_k \hat{N}_k) U(t) \exp(i \sum_k \chi_k \hat{N}_k),
\end{eqnarray}
where $U(t)$ is the time-evolution operator corresponding to the Hamiltonian in Eq.~\eqref{eq:hamiltonian:quantum}.
This transformation implies that the annihilation (creation) operators transform as $\hat{a}_k \rightarrow \hat{a}_k e^{i \chi_k}$ ($\hat{a}^{\dagger}_k \rightarrow \hat{a}^{\dagger}_k e^{-i \chi}_k$), leading to the following generalized time-evolution  operator in the semiclassical limit,
\begin{eqnarray}
\mathcal U_{\boldsymbol  \chi} (t) &\equiv &  \mathcal  T e^{-i \int_{0}^{t}\mathcal H_{\boldsymbol  \chi}(t')dt'} , \label{eq:def:generalizyedTimeEvolutionOperator} \\
\mathcal H_{\boldsymbol \chi}(t) &=&  H_0 +\sum_j ^R H_k \alpha_k \left( e^{i\omega_k  t - i \chi_k  } +  e^{-i\omega_j t + i \chi_k  }  \right) ,
\label{eq:def:generalizyedHam}
\end{eqnarray}
where $\boldsymbol  \chi = \left(\chi_1 ,\dots,\chi_R \right) $  is a vector of counting fields  (see Appendix~\ref{sec:photonResolvedFloquettheory} for a detailed derivation).
Based on the generalized time-evolution operator in  Eq.~\eqref{eq:def:generalizyedTimeEvolutionOperator}, we define the photon-resolved time-evolution operators 
\begin{equation}
\mathcal U^{(\boldsymbol m)} (t) =  \frac{1} {\left( 2\pi\right)^R}  \int_{0}^{ 2\pi } d^R\boldsymbol \chi \,\mathcal  U_{\boldsymbol \chi} (t) e^{i\boldsymbol m \cdot \boldsymbol \chi} ,
\label{eq:photonResolvedEvolutionOperatorSCmultiSum}
\end{equation}
where  $\boldsymbol m = (m_1, \dots ,m_R )$ is a vector of photonic transition numbers. In terms of the photon-resolved time-evolution operator, we can now express arbitrary system observables. For example, the probability that the photonic modes are in the Fock states $\boldsymbol n = (n_1, \dots ,n_R ) $ is given as
\begin{eqnarray}
\left< \hat P_{\boldsymbol  n} \right>_t  &=& \sum_{\boldsymbol m, \boldsymbol m' }   \left<    \mathcal U_{}^{(\boldsymbol m )\dagger} (t)  \mathcal U_{}^{(\boldsymbol m')} (t)    \right>_{t_0} \nonumber \\ 
&\times& \prod_{k=1}^{R}   a_{n_k-m_{k}}^{(k)*} a_{n_k-m_{k}'}^{(k)},
\end{eqnarray}
where $\hat P_{\boldsymbol  n} $ denotes the projector onto the Fock states with quantum numbers $\boldsymbol  n$,  $ a_{m}^{(k)}  $ are the expansion coefficients of the photonic initial state $\left|  \alpha_k e^{i\varphi_k} \right> = \sum_{m} a_{m}^{(k)} \left| m \right>_k   $, and $\langle \hat O \rangle_t \equiv  \langle \psi(t)  \mid \hat O \mid \psi(t)\rangle $, where  $\left| \psi(t)\right>$ can be either a state in the matter system or the  composite light-matter system depending on the enclosed operator $\hat O$. Thus, by semiclassically calculating the generalized time-evolution operators in Eq.~\eqref{eq:def:generalizyedTimeEvolutionOperator}, we can evaluate genuine quantum properties of photonic observables.

\subsection{Full-counting statistics}

While the analytical (or numerical) evaluation of the photon-resolved time-evolution operators in Eq.~\eqref{eq:photonResolvedEvolutionOperatorSCmultiSum} is an infeasible task in many cases, it is relatively easy to compute the moments and cumulants of the photonic modes $\hat a_k$. Importantly, the counting statistics within the PRFT is fundamentally different from the standard FCS formalism.  As shown in Fig.~\ref{figCountingStatSketch}(a), the standard FCS framework is based on  two-point projective measurements, where the state is formally projected to the Fock state basis at the beginning  $t_0$ and at the end $t_1$ of each experimental run~\cite{esposito2009nonequilibrium,Brandes2004,levitov1996electron} (see Appendix~\ref{sec:standardFCS} for more details). However, performing a projective photon number measurement at the beginning of the experiment would completely destroy the coherent photonic state. To circumvent this issue, we propose using two-point tomographic measurements, which are performed at the beginning and the end of the time evolution. The tomography is  independently carried out for two  batches of experimental runs with the same initial states. As illustrated in Fig.~\ref{figCountingStatSketch}(b), the photon statistics is determined by photon-number measurements at $t=t_0$ for the first batch, and at $t=t_1$  for the second batch. This alternative approach to FCS has been investigated for heat transport between incoherent baths in Ref.~\cite{cerrillo2016nonequilibrium}.

Formally, the counting statistics of the photon modes can be calculated via the cumulant- or the moment-generating functions, which are defined by
\begin{eqnarray}
K_{\boldsymbol \chi}(t) &=& \log \left[ M_{\boldsymbol \chi}(t)\right], \\
M_{\boldsymbol \chi}(t) &=& \left<  e^{-i \sum_k \chi_k  \hat N_k  } \right>_t ,
\label{eq:def:cumulantMomentGenFct}
\end{eqnarray}
respectively. The associated $n$th cumulant and moment of  mode $k$ are determined via
\begin{eqnarray}
\kappa_n^{(k)}(t) &=& \frac{d^n}{d\,(-i\chi_k)^n} K_{\boldsymbol \chi}(t)  ,\nonumber \\
m_n^{(k)}(t) &=&  \frac{d^n}{d\,(-i\chi_k)^n} M_{\boldsymbol \chi}(t)  .
\label{eq:def:cumulants}
\end{eqnarray}
We are interested in the time evolution of the  cumulant- and moment-generating functions. To this end, we define the dynamical cumulant-generating function $ K_{\text{dy},\boldsymbol \chi}( t_1 )$ via
\begin{eqnarray}
K_{\boldsymbol \chi}( t_1 ) 
&=& K_{\text{dy} ,\boldsymbol \chi}( t_1 ) +  K_{\boldsymbol \chi}( t_0 ) ,
\label{eq:def:dynamicalcumulantGeneratingFunction}
\end{eqnarray}
where $K_{\boldsymbol \chi}( t_0 ) $ and $K_{\boldsymbol \chi}( t_1 ) $ can be determined by independent tomographies at the beginning and the end of the time evolution~\cite{cerrillo2016nonequilibrium}.  
As shown in detail  in Appendix~\ref{sec:fullCountingStatistics}, the dynamical cumulant-generating function can be  expressed as
\begin{equation}
K_{\text{dy},\boldsymbol \chi}( t)  =  \log \frac{1}{2} \left<      \mathcal U_{ \boldsymbol \varphi }^{\dagger} (t)  \mathcal U_{\boldsymbol \varphi +\boldsymbol \chi}(t)  +     \mathcal U_{ \boldsymbol \varphi -\boldsymbol \chi}^{\dagger} (t)  \mathcal U_{\boldsymbol \varphi}(t)   \right>_{t_0} ,
\label{eq:res:dynamicalcumulantGenFct}
\end{equation}
where $\boldsymbol \varphi =\left(\varphi_1 , \dots \varphi_R \right)$ is the vector of phases of the  photonic states. 
Employing Eq.~\eqref{eq:def:dynamicalcumulantGeneratingFunction}, we can now obtain the change of the cumulants,
\begin{equation}
\kappa_{\text{dy},n}^{(k)}(t)   \equiv \left. \frac{d^n}{d(-i\chi_k)^n} 	K_{\text{dy},\boldsymbol \chi}( t) \right|_{\boldsymbol \chi=0}= \kappa_{n}^{(k)}(t)  - \kappa_{n}^{(k)}(t_0)  ,
\label{eq:dynCummulants}
\end{equation}
which describes the change of the photon statistics in the two-time tomographic measurement  sketched in Fig.~\ref{figCountingStatSketch}(b). We can see for instance that the first and second dynamical cumulants
\begin{eqnarray}
\Delta \langle\hat N_k (t) \rangle &=& \kappa_1^{(k)}(t) - \kappa_1^{(k)}(t_0) ,\nonumber \\
\Delta \sigma_k^2 (t) &=& \kappa_2^{(k)}(t) - \kappa_2^{(k)}(t_0) 
\label{eq:meanAndVariancePRFT}
\end{eqnarray}
correspond to the change of the mean photon number $\langle \hat N_k  \rangle $ and the variance $\sigma_k^2$, respectively. In the rest of this paper, we will analyze Eq.~\eqref{eq:res:dynamicalcumulantGenFct} in a variety of contexts to gain a transparent picture of the photonic dynamics. The derivation of Eq.~\eqref{eq:res:dynamicalcumulantGenFct} assumes a separable initial state of the form in Eq.~\eqref{eq:initalState}. In Appendix \ref{sec:generalizations}, we explain how the PRFT can be generalized to more general initial states such as entangled states in a similar fashion.

It is worthwhile to point out that the well-defined phase in the PRFT is in contrast to the well-defined particle number in the standard FCS. This is the origin of the difference in the measurement protocols for the two cases. As shown in Appendix~\ref{sec:standardFCS}  the cumulant-generating function is $\tilde K_{\boldsymbol \chi}(t) =\log \langle \mathcal U_{ \boldsymbol \varphi -\boldsymbol \chi/2}^{\dagger} (t)   \mathcal U_{ \boldsymbol \varphi +\boldsymbol\chi /2} (t)  \rangle_{t_0} $ according to the standard FCS, where the phases $\boldsymbol \varphi$ have been evaluated heuristically. This results is a profound difference with the predictions of the PRFT, and highlights the crucial role played by two-point tomographic measurements in the PRFT. We note that the  first cumulant in Eq.~\eqref{eq:meanAndVariancePRFT}  agrees with the standard semiclassical calculation as shown in Appendix~\ref{sec:relationClassicalDriving}.

\subsection{Probability redistribution}

Using the moment-generating function of the final state given in Eq.~\eqref{eq:def:cumulantMomentGenFct}, we can calculate the probability distribution of the Fock states  at time $t$ via
\begin{eqnarray}
p_{\boldsymbol n}(t)  &=& \frac{1}{(2\pi)^R}\int_{0}^{2\pi }d^R\boldsymbol \chi\, M_{\boldsymbol \chi}( t)e^{i\boldsymbol \chi \cdot \boldsymbol n }  \nonumber \\
&=&  \sum_{\boldsymbol m} q_{\boldsymbol n- \boldsymbol m}(t) p_{\boldsymbol m}(t_0),
\label{eq:tra:pseudoProb}
\end{eqnarray}
where we have introduced the \textit{quasiprobabilities } as
\begin{equation}
\label{eq:def:pseudoProbilities}
q_{\boldsymbol n}(t) =\frac{1}{(2\pi)^R}\int_{0}^{2\pi }d^R\boldsymbol \chi\, M_{\text{dy},\boldsymbol \chi}( t) e^{i\boldsymbol \chi \cdot \boldsymbol n } ,
\end{equation}
where $M_{\text{dy},\boldsymbol \chi}( t) = \exp \left[ K_{\text{dy},\boldsymbol \chi}( t) \right]$. Akin to regular probabilities, these quasiprobabilities are  real valued  and fulfill 
\begin{equation}
\sum_{\boldsymbol n}  q_{\boldsymbol n} = 1.
\end{equation}
However, due to interference effects, $q_{\boldsymbol n}$ may also be negative.     
Such negativity is a characteristic signature of nonclassical temporal correlations~\cite{hofer2016negative}. We note that according to Eq.~(\ref{eq:tra:pseudoProb}), the quasiprobabilities may be interpreted as the kernel of the probability redistribution.

\subsection{Full-system state }

The PRFT not only predicts the photonic probability distribution, but also the state of the matter system, from which we can even reconstruct the  state of the total system.
We can express  Eq.~\eqref{eq:tra:pseudoProb} in the form
\begin{equation}
p_{\boldsymbol n} (t)  = \bra{\phi(t_0)}  \hat P_{\boldsymbol n} (t) \ket{\phi(t_0)},
\label{eq:probabityOperator_ExpectationValue}
\end{equation}
where we have defined the probability operators
\begin{eqnarray}
\hat P_{\boldsymbol n} (t)  &=& \sum_{\boldsymbol m}  \hat Q_{\boldsymbol m}(t)  p_{\boldsymbol n-\boldsymbol m}  ,\nonumber\\
\hat Q_{\boldsymbol m} &=& \frac{1}{2}  \int_{0}^{2\pi} \frac{ d^R\boldsymbol\chi}{(2\pi)^R} \,  \mathcal U_{ \boldsymbol\varphi}^{\dagger} (t)  \mathcal U_{\boldsymbol\varphi+\boldsymbol\chi}(t)  e^{i {\boldsymbol m}\cdot  \boldsymbol\chi} + \text{H.c.}\;
\label{eq:probabilityOperator}
\end{eqnarray}
It is easy to show that the set of probability operators fulfill
\begin{equation}
\sum_{n=0}^{\infty} \hat P_{\boldsymbol n} (t)  =    \mathbbm 1.
\end{equation}
Moreover, in the parameter regime and time scope  in which  the PRFT is valid, we have  $0\leq \hat P_{\boldsymbol n} (t)\leq 1 $, which is equivalent to $0\leq p_{\boldsymbol n}(t)\leq 1$ in Eq.~\eqref{eq:probabityOperator_ExpectationValue}. Thus, the operators $\hat P_{\boldsymbol n} (t)$ define a positive-operator-valued  measurement (POVM)~\cite{Hayashi2017}, which consistently describes the photon measurement process sketched in Fig.~\ref{figCountingStatSketch}. 

According to the theory of quantum measurements~\cite{Wilde2017,Hayashi2017}, the reduced density matrix of the matter system after the measurement is given by 
\begin{eqnarray}
\rho_{\text{M} }(t)   &=&   \sum_{\boldsymbol n} p_{\boldsymbol n}(t) \rho_{\boldsymbol n}(t) ,\nonumber \\
\rho_{\boldsymbol n}(t)   &=&   \frac{1}{p_{\boldsymbol n}(t)} \mathcal U_{ \boldsymbol\varphi}(t)\sqrt{\hat P_{\boldsymbol n} (t)} \rho(t_0) \sqrt{\hat P_{\boldsymbol n} (t)}\mathcal U_{ \boldsymbol\varphi}^{\dagger}(t),
\label{eq:matterDensityMatrix}
\end{eqnarray}
where  $\rho(t_0) = \left| \phi(t_0) \right>  \left< \phi(t_0) \right|$   is the initial density matrix, and $\rho_{\boldsymbol n}$ denotes the reduced density matrix conditioned on the measured photon number $n$. As the POVM  $\hat P_{\boldsymbol n} (t)$ acts on the initial state, the time evolution operator $\mathcal U_{ \boldsymbol\varphi}(t)$ has been added heuristically to account for the dynamics of the matter system. Equation \eqref{eq:matterDensityMatrix} thus defines a valid set of Krauss operators describing photon-number-dependent quantum channels.  If $\rho(t_0)$ is a pure state, then  $\rho_{\boldsymbol n}(t) \equiv  \ket{ u_{\boldsymbol n} }    \bra{ u_{\boldsymbol n} } $ will be also pure as $\hat P_n (t)$ is positive semidefinite. A purification of  $\rho_{\text{M} }$ is given by
\begin{equation}
\ket{\Psi(t)}  = \sum_{\boldsymbol n}  \sqrt{p_{\boldsymbol n}  }  e^{-i(\boldsymbol \omega t - \boldsymbol\varphi)\cdot \boldsymbol n}  \ket{ u_{\boldsymbol n} }  \ket{\boldsymbol n},
\label{eq:state:totalSystemState}
\end{equation}
where we have assigned the phases $ e^{-i(\boldsymbol \omega t - \boldsymbol\varphi)\cdot\boldsymbol n} $ in terms of the frequency vector $ \boldsymbol \omega = \left(\omega_1,\dots,\omega_{R}\right)$, such that we can interpret the states $\ket{\boldsymbol n}$ as the Fock states of the photonic system.  Tracing over the light system, we can verify that indeed $\text{tr}_{\text{L}} \left[  \ket{\Psi(t)} \bra{\Psi(t)}\right] = \rho_{\text{M} }(t) $. Thus, we can identify the $\ket{\Psi(t)}$ as the state of the total light-matter system.

\subsection{Floquet-state analysis}

\label{sec:FloquetStateAnalysis}
We now proceed to analyze systems subjected to multimode driving where the photonic field is composed of multiple commensurate photonic frequencies $\omega_k$.  In this scenario, the matter subsystem is still described by a Floquet Hamiltonian in the semiclassical limit. However, we find that the PRFT leads to some important insights about such systems (like light-matter entanglement), that are beyond the reach of a standard Floquet analysis and imply a fundamental limitation on its validity.

According to Floquet theory, the generalized time-evolution operator can be written as 
\begin{equation}
\mathcal U_{\boldsymbol  \chi}(t) = \mathcal U_{\text{kick},\boldsymbol \chi}(t)  \exp\left(-i H_{\text{Fl},\boldsymbol  \chi}t\right),
\end{equation}
where the generalized Floquet Hamiltonian can be expanded as  
\begin{equation}
H_{\text{Fl},\boldsymbol  \chi} = \sum_\mu E_{\mu, \boldsymbol \chi} \left| u_{\mu,\boldsymbol \chi} \right>  \left<  u_{\mu,\boldsymbol \chi} \right|.
\label{eq:defFloquetHamiltonian}
\end{equation}
As the kick operator  $\mathcal U_{\text{kick},\boldsymbol\chi}(t)   = \mathcal U_{\text{kick},\boldsymbol \chi}(t+\tau)  $ is time periodic, it accounts only for  periodic changes in the photon redistribution.   
Thus,  the asymptotic dynamics of the cumulant-generating function in Eq.~\eqref{eq:res:dynamicalcumulantGenFct} is  determined by the generalized quasienergies $E_{\mu,\boldsymbol \chi } $. The stroboscopic dynamics of the matter system is characterized by the Floquet states $\left| u_{\mu,\boldsymbol \chi} \right> $, that generalize the common eigenstates in time-independent systems.    When expanding the initial state in the Floquet-state basis $ \left| u_{\mu,\boldsymbol \varphi}\right>$ 

\begin{equation}
\left| \Psi (t_0)\right> = \sum_\mu c_\mu \left| u_{\mu,\boldsymbol \varphi}\right> \otimes \left| A(t_0) \right>,
\label{eq:genericInitialState}
\end{equation}
where $\left|  A(t_0) \right>$ is the initial state of the photonic field, the asymptotic dynamical  cumulant- and moment-generating functions read as
\begin{align}
K_{ \text{dy}, \boldsymbol \chi}( t )   &\rightarrow   \log M_{ \text{dy}, \boldsymbol \chi}( t ), \label{eq:stroboscopicKummulantGenFunction} \\
M_{ \text{dy}, \boldsymbol \chi}( t )   &\rightarrow 
\qquad \qquad \nonumber
\\ \sum_\mu  & \frac{\left|c_\mu \right|^2}{2}\left(  e^{i \left( E_{\mu, \boldsymbol  \varphi} - E_{\mu ,\boldsymbol  \varphi + \boldsymbol  \chi} \right) t } +  e^{i \left( E_{\mu, \boldsymbol  \varphi-\boldsymbol  \chi}- E_{\mu, \boldsymbol  \varphi} \right) t } \right).
\nonumber 
\end{align}
This derivation is rigorously explained in Appendix~\ref{sec:app:periodicallyDrivenSystems}.
Thus, $M_{ \text{dy} , \boldsymbol \chi}$ is a weighted average of the dynamical  moment-generation functions of the Floquet states, with the weights given by the expansion coefficients $\left|c_\mu \right|^2$.
We can now obtain the mean photon number change,
\begin{equation}
\Delta \langle \hat N_k\rangle =  - \sum_\mu  \left|c_\mu \right|^2 \frac{d E_{\mu ,\boldsymbol\varphi}}{d\varphi_k} \, t =  \sum_\mu  \left|c_\mu \right|^2 \Delta \langle \hat N _k \rangle \mid_{\mu } \, t,
\label{eq:meanPhotonNumberExpansion}
\end{equation}
where we have evaluated the quasienergies at the phases of the photonic fields $\boldsymbol \varphi$, and we have denoted the mean-photon change in a specific Floquet state by $\Delta \langle \hat N _k \rangle \mid_{\mu  }$. 
Intriguingly, when preparing the system initially in an arbitrary Floquet state $\mu$, we find
\begin{equation}
\left. \Delta \sigma_k^2 \right|_{\mu } = 0,
\label{eq:vanishingVariance}
\end{equation}
i.e., the variance change $\Delta \sigma_k^2 $ vanishes.

For a superposition of Floquet states, Eqs.~\eqref{eq:def:pseudoProbilities}  and ~\eqref{eq:stroboscopicKummulantGenFunction} imply that the quasiprobabilities are given by a weighted average of Floquet-state dependent quasiprobabilities $ q_{n\mid \mu}$, i.e., $q_n  = \sum_\mu  \left|c_\mu \right|^2  q_{n\mid \mu}$. Similarly, from Eq.~\eqref{eq:tra:pseudoProb}, we can infer that
\begin{equation}
p_{\boldsymbol n} = \sum_\mu  \left|c_\mu \right|^2  p_{\boldsymbol n\mid \mu}\quad.
\label{eq:probabilityExpansion}
\end{equation}
Thus, $p_{n\mid \mu}$ defines a conditional probability distribution, and the dynamics of the photonic state is controlled by the Floquet states.

Our results  lead to a quantum-optical interpretation of Floquet states. To see this, we consider the time-evolution of the  generic initial state in Eq.~\eqref{eq:genericInitialState}. According to Eq.~\eqref{eq:probabilityExpansion}, the light-matter state becomes entangled in the course of the time-evolution:
\begin{equation}
\left| \Psi (t)\right> = \sum_\mu c_\mu e^{-iE_{\mu,\boldsymbol \varphi}(t-t_0)} \left| u_{\mu,\boldsymbol \varphi}(t) \right> \otimes \left| A_\mu (t)\right>,
\label{eq:lightMatterEntanglement}
\end{equation}
where $\left| u_{\mu,\boldsymbol \varphi}(t) \right>  = \mathcal U_{\text{kick},\boldsymbol\varphi}(t) \left| u_{\mu,\boldsymbol\varphi } \right>  $, and the photonic wave functions for long times is given by
\begin{equation}
\left| A_\mu (t)\right> = \sum_{\boldsymbol n} \sqrt{p_{\boldsymbol n\mid\mu}} e^{-i(\boldsymbol \omega t - \boldsymbol \varphi)\cdot \boldsymbol n}\ket{n},
\end{equation}
which is in agreement with Eq.~\eqref{eq:state:totalSystemState}.
Thereby, the mean photon number of each $\left| A_\mu \right>$  changes linearly in time [see Eq.~\eqref{eq:meanPhotonNumberExpansion}], while its variance remains unchanged [see Eq.~\eqref{eq:vanishingVariance}]. This means that the conditional probability can be approximated by
\begin{equation}
p_{\boldsymbol n\mid \mu} (t) = p_{\boldsymbol n - \Delta\boldsymbol n_\mu(t)}(t_0)
\label{eq:photonRedistributionSimplified},
\end{equation}
where $ \left[ \Delta\boldsymbol n_\mu(t)\right]_k   = -{d E_{\mu ,\boldsymbol\varphi}}/ {d\varphi_k}  t $, which can be numerically efficiently evaluated.

Consequently, the photonic states $\left| A_\mu  (t)\right>$ will become mutually orthogonal for sufficiently long times, and the reduced density matrix of the matter system becomes,
\begin{eqnarray}
\rho_{\text{M}} =   \sum_\mu  \left| c_\mu \right|^2  \left| u_{\mu,\boldsymbol \varphi}(t) \right>\left< u_{\mu,\boldsymbol \varphi}(t) \right|.
\label{eq:rhoMatterLongTimes}
\end{eqnarray}
Thus, from a quantum-optical point of view, Floquet states act as the decohering basis. We note that this interpretation holds as long as there is a nonvanishing photon flux between  distinct photonic modes. This effect is illustrated in Fig.~\ref{figCountingStatSketch}(b)  and will be further analyzed in Sec.~\ref{sec:twoModeRabiModel} for a two-mode Rabi model.

We emphasize that a large-scale photon flux cannot develop when the quasienergy does not depend on the counting field, as we see from Eq.~\eqref{eq:stroboscopicKummulantGenFunction}. This situation occurs for single-mode systems, where the  single counting field can be transformed away by a counting-field-dependent shift in time $t\rightarrow \chi/\omega$, which can be seen in Eq.~\eqref{eq:def:generalizyedHam}. {\color{\markColorOne} Consequently, the transport-induced light-matter entanglement can strictly appear only in two- or higher-mode  systems. } We recall however that there are always two  polarization modes of light, such that this  decoherence effect discussed here has practical importance.

At this point, it is instructive to compare our results with Ref.~\cite{Guerin1997}, which has computed photonic observables  by considering a static phase for photonic coherent states. However, the static-phase assumption leads to unphysical predictions such as the diverging higher-order moments and cumulants and, thus,  is not suitable to quantitatively capture the  dynamics of the photon field or the decoherence of the matter system. { \color{\markColorOne} It is worth mentioning that the PRFT formally operates in the Sambe space rather than in Fock space.  Thus, the PRFT cannot account for the shot-noise induced entanglement discussed in Sec. II A. However, as we demonstrate in Sec.~\ref{sec:rabiModel}, this effect has a minor influence on the photonic dynamics.}

\subsection{Error analysis}

To describe the deviation from the exact time evolution  quantitatively, we specify  the initial state to be
\begin{equation}
\left| \psi(t_0) \right> =  \left| \phi(t_0) \right> \bigotimes_{k=1}^{R} \left| \overline n_k,\sigma_k,\varphi_k \right>,
\label{eq:InitalStateBenchmarking}
\end{equation}
where $\left| \phi(t_0) \right>$ is the initial state of the matter system, and the  photonic state of mode $k$ is given by 
\begin{equation}
\left|\overline n_k,\sigma_k , \varphi_k\right>  = \mathcal N \sum_{n} e^{-\frac{(n -\overline n_k)^2}{4\sigma_k} +i\varphi_k n}\left| n\right>_k.
\end{equation}
Thereby, $\overline n_k$ is the mean photon number, $\sigma_k$ is the standard deviation, $\varphi_k$ is the mean  phase, and  $ \mathcal N$ is a normalization factor. For $\sigma_k< \sqrt{\overline n_k}$ ($\sigma_k> \sqrt{\overline n_k}$) the system is in a number (phase) squeezed state, while for $\sigma_k= \sqrt{\overline n_k}$ it is in a coherent state.
As discussed in details in Appendix~\ref{sec:errorAnalysis}, the deviation of the probabilities $ p_{\boldsymbol n}   $   in Eq.~\eqref{eq:tra:pseudoProb} from the exact ones $  p_{\boldsymbol n}^{(\text{Ex})}$ scales as
\begin{eqnarray}
\Delta  p_{\boldsymbol n}(t) &\equiv &  p_{\boldsymbol n}(t)- p_{\boldsymbol n}^{(\text{Ex})}(t) \nonumber  \\ 
&=& \mathcal F \left[\left\lbrace\frac{g_k t}{\sigma_k^2},\frac{\sigma_k}{\overline n_k},\frac{g_k t}{\overline n_k} \right\rbrace_k  \right]\label{eq:errorScaling},
\end{eqnarray}
where $ \mathcal F\left[ x\right]$ denotes the scaling function, which depends on the set of ratios $\frac{g_k\cdot t}{\sigma_k^2}$, $\frac{\sigma_k}{\overline n_k} $, and $\frac{g_k\cdot t}{\overline n_k}  $ of all photonic modes $k$.
This shows that the PRFT performs well for photonic states with a large standard deviation $\sigma_k$ and large mean  photon number $\overline n_k$. Large $\sigma_k$ makes sure that the phase is well defined, which is reflected by the first argument of the function $\mathcal F $ in Eq.~\eqref{eq:errorScaling}. A large $\overline n_k$ guarantees that the matrix elements of the photonic operators $\hat a_k$ do not depend on the photon number, which is described by the second and third argument of $\mathcal F$. { \color{\markColorOne} Clearly, the latter source of error is absent when considering the Floquet theory in Sambe space [introduced in Eq.~\eqref{eq:def:sambeSpace}].}
All requirements are naturally fulfilled for coherent states with $\sigma_k = \sqrt{\overline n_k}\gg 1$ in the thermodynamic limit $\overline n_k \rightarrow \infty$. 
Intriguingly, the PRFT makes accurate predictions, even when $\sigma_k$ is small; we discuss this in  Sec.~\ref{sec:twoModeRabiModel}.

The linear change of the mean photon number in Eq.~\eqref{eq:meanPhotonNumberExpansion}  is a consequence of the vanishing photon-number dependence of the  matrix elements $\left< n+1 \right| \hat a_k^\dagger \left| n\right> =\sqrt{n+1} \approx \sqrt{\overline n_k}$ for $-\sqrt{\overline n_k} < n-\overline n_k < \sqrt{\overline n_k} $. This establishes a `translational invariance' in  Fock space. In each unit of time, the matter system can pump a certain number of photons  from one driving mode into another   independent of the photon number. When the mean photon number change reaches the order $\overline n_k$, the translational invariance in Fock space is lost, and the PRFT theory breaks down.

\section{Applications}
\label{sec:benchmarking}

\begin{figure*}
	\includegraphics[width=\linewidth]{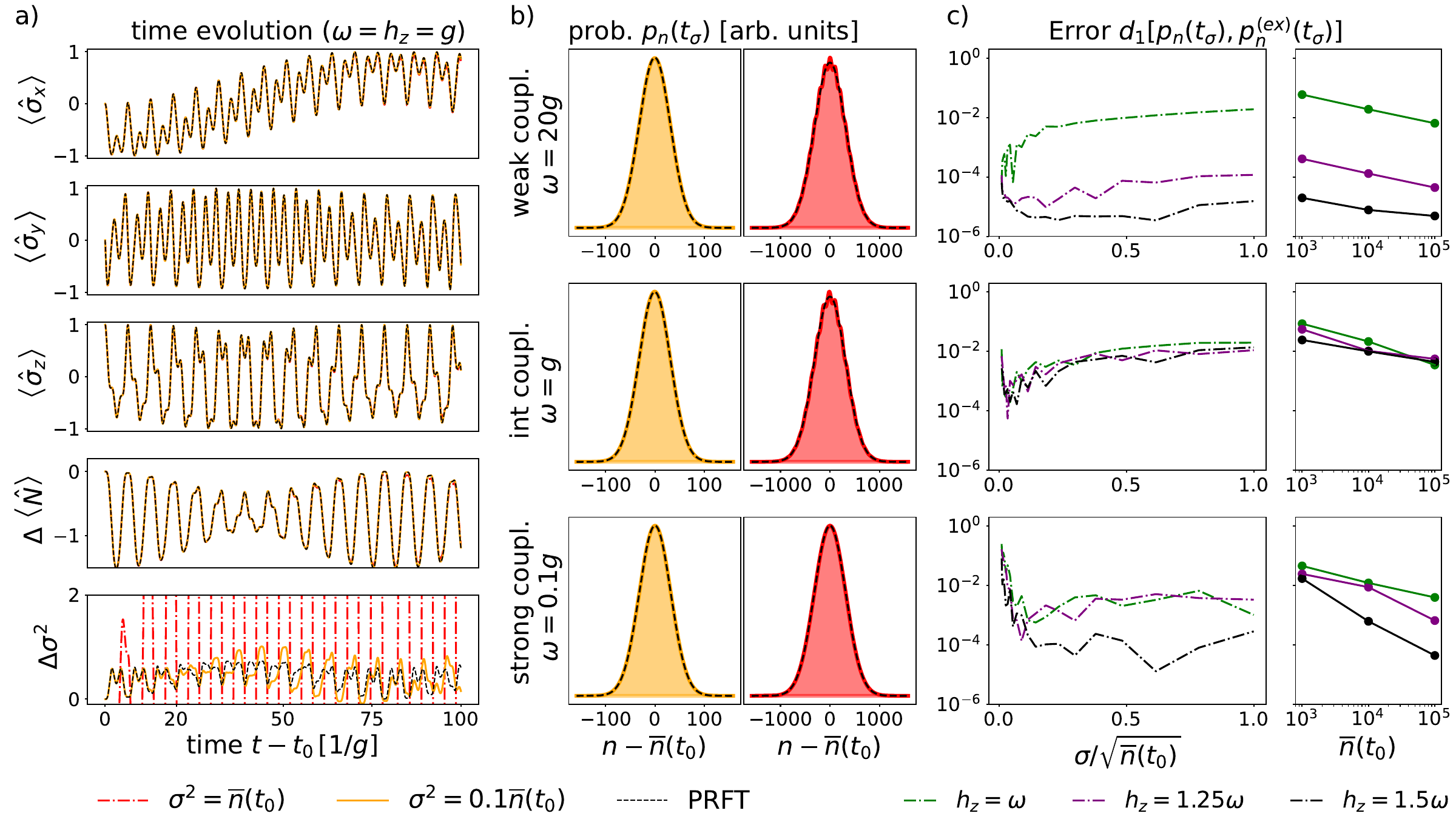}
	\caption{ \color{\markColorOne} (a) Time evolution of the spin components and the first two photonic cumulants in the quantum Rabi model in the intermediate-coupling regime $\omega=h_z=g$. The initial condition is given in Eq.~\eqref{eq:InitalStateBenchmarking} with $\left| \phi(t_0) \right> =(  \left| \downarrow \right> + \left| \uparrow \right>)/\sqrt{2}$ for $\overline n(t_0) = 10^5$ photons, as well as for $\sigma = \sqrt{\overline n(t_0)}$ (red, dash-dotted) and $\sigma = 0.1\sqrt{\overline n(t_0)}$ (yellow,solid). (b) Probability distribution at the scaled time $t_\sigma \equiv  (2\pi/g) 6\sigma$ for the same coherent (right) and number-squeezed (left) initial states as in (a) and for the resonance condition $h_z=\omega$. (c) Analysis of the trace distance in Eq.~\eqref{eq:variationalDistance} as a measure of the error at the scaled time $t_\sigma$ as a function of noise $\sigma$ (left) and mean photon number $\overline n(t_0)$ with scaled $\sigma = 0.3\sqrt{\overline n(t_0)}$ and $t_\sigma$ (right). }
	\label{fig:RabiModel}
\end{figure*}

We  apply the PRFT to three versions of the  quantum Rabi model, and demonstrate that the framework is accurate in the semiclassical limit. The Hamiltonian describing the system is
\begin{equation}
H_{\text{QR}} =   \frac{h_z }{2}\hat \sigma_z + \sum_{k=1}^{R} \omega_k \hat a_k^\dagger\hat a_k  +\sum_{k=1}^{R}  \tilde  g_k \hat \sigma_x   \left( \hat a_k   +  \hat a_k^\dagger    \right)   ,
\label{eq:ham:quantumRabi}
\end{equation}
where the two-level system  is described  by the common Pauli matrices $\hat \sigma_{\alpha}$ with $\alpha = \left\lbrace x, y, z\right\rbrace $, and $h_z$ denotes level splitting of the two-state model. We denote the eigenstate of $\hat \sigma_z$ with eigenvalue $1$ ($-1$) by $\left| \uparrow\right>$ ($\left| \downarrow\right>$).  The photonic system operators and parameters have been described in Sec.~\ref{sec:system}.

To compare the photonic probability distributions predicted by the numerical exact quantum calculation $p_{n_k}^{(\text{Ex})}$ and by the PRFT $p_{n_k}$ of mode $k$, we employ the  trace distance
\begin{equation}
d_{1}\left(p_{n_k},p_{n_k}^{(\text{Ex})} \right) =  \sum_{n_k} \left| p_{n_k}-p_{n_k}^{(\text{Ex})} \right|.
\label{eq:variationalDistance}
\end{equation}
 We quantify the  entanglement of the light and matter systems in terms of the purity of the two-level system $\mathcal P \equiv \text{tr} \left(\rho_{\text{M}}^2 \right)$, where $\rho_{\text{M}}$ is the reduced density matrix of the two-level system. For  a pure state $\mathcal P =1$, while for a completely mixed, i.e., maximally entangled, state $\mathcal P =0.5$.

\subsection{Rabi model}

\label{sec:rabiModel}

First, we investigate the paradigmatic single-mode quantum Rabi model with $R=1$. In the semiclassical limit, the atomic dynamics is described by $  \mathcal H_R = h_z \hat \sigma_z/2 + 2 g_1 \hat \sigma_x \cos(\omega_1 t).$ In the following discussions, we neglect the index $k=1$. 
{\color{\markColorOne} Since there is no possibility for large scale photon transport in the single-mode Rabi model, the error will be mainly determined by the photon shot-noise of the initial states. As we will see, this error is small as compared to the transport dynamics introduced in Sec.~\ref{sec:FloquetStateAnalysis} and discussed in Sec.~\ref{sec:twoModeRabiModel}. In Appendix~\ref{sec:app:benchmarkCalculations}, we also benchmark the PRFT against the dynamics in the Sambe space [introduced in Eq.~\eqref{eq:def:sambeSpace}], where the accuracy of the PRFT is even further improved.

\textit{Time evolution.} In Fig.~\ref{fig:RabiModel}(a), we depict the dynamics for $\overline n =10^5  $ photons and two different  $\sigma = \sqrt{ \overline n} $ (red, dash-dotted) and $\sigma = 0.1 \sqrt{ \overline n}  $  (yellow, solid), i.e., for a coherent and a number squeezed photonic state. The PRFT results are depicted by  a black dashed line.  We observe that the PRFT results of the spin observables $\left< \hat \sigma_\alpha\right>$ with $\alpha \in \left\lbrace \text{x}, \text{y},\text{z} \right\rbrace$ agree perfectly to the numerical results for the coherent and  number-squeezed initial conditions. Likewise, we do not observe differences in the mean photon number change $\Delta \left< \hat N \right>$. For the photon variance change $\Delta \sigma $, we observe that the PRFT prediction agrees reasonable well with the exact calculation for the number squeezed state for short times. Yet,  the PRFT strongly deviates for the coherent photonic initial state. These findings are in agreement with the error analysis in Eq.~\eqref{eq:errorScaling}, which shows that the error scales with $\sigma_k/\sqrt{\overline n_k}$ due to the photon shot noise. 
Yet, we note that  deviations in the probability distributions are heavily enhanced by the definition of the variance, which  disproportional weights  photon numbers away from the mean value with $\propto (n-\overline n)^2$. Intriguingly, the PRFT agrees perfectly with the numerical results when the dynamics is simulated in the Sambe space instead of the  Fock space as demonstrated in Appendix~\ref{sec:app:benchmarkCalculations}. For these reasons, we continue to analyze the photon statistics in terms of  the trace distance defined in Eq.~\eqref{eq:variationalDistance}. Moreover, we mention that the observed variance values in Fig.~\ref{fig:RabiModel} are small compared  to  the rapidly diverging variance, which can appear in multimode systems due to photon transport.

\textit{Photon probabilities.}  In Fig.~\ref{fig:RabiModel}(b), we benchmark the photon probability distribution in the weak- (upper row), intermediate- (middle row), and strong-coupling (bottom row) regimes for number-squeezed (left) and the coherent (right) initial states, respectively. To test the validity of the PRFT, we choose a scaled time $t_\sigma \equiv  (2\pi/g) 6\sigma$, which agrees with the analytically predicted validity according to the first argument in Eq.~\eqref{eq:errorScaling}. The scaled time $t_\sigma$ is a sufficient time scale to observe the transport entanglement effect explained in Sec.~\ref{sec:FloquetStateAnalysis} and demonstrated in Sec.~\ref{sec:twoModeRabiModel}. In the weak- and the intermediate-coupling regime, we observe that the PRFT agrees well with the exact numerical calculations. For the coherent photonic state, we observe some minor deviations exhibiting an oscillating dependence  with photon number, which will be explained below. 

\textit{Error scaling.}   In Fig.~\ref{fig:RabiModel}(c), we investigate the error quantified by Eq.~\eqref{eq:variationalDistance} in the weak-, intermediate- and strong-coupling regimes  at the scaled time $t_\sigma$. Thereby, we investigate the error for three different level splittings $h_z$. In the left column, we investigate the error as a function of $\sigma$ for $\overline  n = 10^5$. In the weak- and intermediate-coupling regimes we find that the error increases  with $\sigma$  in agreement with the second and third argument in Eq.~\eqref{eq:errorScaling}. For very small $\sigma \approx 0.05 \sqrt{\overline n }$, we also observe a rapid error increase, which is due to the first argument in Eq.~\eqref{eq:errorScaling}.  In contrast, the error exhibits an overall decaying behavior as a function of $\sigma$ in the strong-coupling regime, suggesting that the first argument in Eq.~\eqref{eq:errorScaling} plays a more prominent role.

In the right column of Fig.~\ref{fig:RabiModel}(c) we  investigate the error as a function of $\overline n(t_0)$ while simultaneously scaling $\sigma =0.3 \sqrt{\overline n(t_0)} $. In the weak- and intermediate-coupling regime, we find that the error scales approximately as $d_1 \propto \overline n(t_0)^{-0.5} \propto \sigma/\overline n  $, i.e., according to the second argument in Eq.~\eqref{eq:errorScaling}. While it might appear that the error scales as $d_1 \propto \overline n(t_0)^{-1} \propto 1/\sigma^2  $ in the strong-coupling regime, we interpret this as a consequence of  numerical fluctuations, which are caused by the $\sigma$-dependent evolution time $t_\sigma$.  Similar observations can be also found for other ratios of $\sigma / \sqrt{\overline n(t_0)} $ and matter initial states (not shown). 

Overall, we find that the error is smaller than $d_1\lessapprox 0.01$ for the Rabi model at resonance $h_z \approx\omega $. Away from the resonance condition $h_z > \omega$, we find that the error even decreases. We explain this by a reduced shot-noise entanglement, which occurs most efficient at resonance.}

\textit{Analytical analysis.} When $\left| h_z-\omega\right|\ll g$, we can neglect the counter-rotating terms  $\sigma_{+}\hat a^\dagger$, $\sigma_{-}\hat a$ and the Hamiltonian reduces to the Jaynes-Cummings model  $  \mathcal H_{\text{JC} } = h_z \hat \sigma_z/2 +  g \hat \sigma_+ \hat a + g \hat \sigma_- \hat a_1^\dagger$.
In this case, the photon-resolved time evolution operators in Eq.~\eqref{eq:photonResolvedEvolutionOperatorSCmultiSum} are,
\begin{eqnarray}
\mathcal U_{}^{(0)} (t) &=& e^{-i \frac{\omega }{2}\hat \sigma_z t}   \left[ \cos\left( Et \right) \mathbbm 1 +i \sin\left( Et \right)  \cos\theta \hat \sigma_z  \right] \nonumber, \\
\mathcal U_{}^{(\pm 1)} (t) &=&   i e^{-i \frac{\omega }{2} \hat \sigma_z t}   \sin\left( Et \right) \sin\theta \hat \sigma_{\mp} ,
\label{eq:res:photonResolvedTimeEvolutionOperators}
\end{eqnarray}
where $E=\frac 12  \sqrt{ \left( h_z  -\omega\right)^2   + 16 g^2  }$ is the  energy of the excited eigenstate (see Appendix ~\ref{sec:jaynesCummingsModel} for a detailed derivation).

This form of the photon-resolved operators encodes the conservation of the quantity,
$
N_{\rm tot} =\hat  a^{\dagger}\hat a + \hat \sigma_{z},
$
which is a salient feature in the Jaynes-Cummings model. To illustrate this, let us consider the  initial state
$
\left| \psi (t_0) \right> = \left| \uparrow  \right> \otimes  \left| n  \right> ,
$
that gives rise to the following photonic occupations of the Fock states $n$ and $n+1$
\begin{eqnarray} 
\left< \hat P_{n} \right>   &=&   \left [ \cos\left( Et \right) \right]^2 + \left [\cos\theta \sin\left( Et \right) \right]^2 \nonumber, \\
\left< \hat P_{n+1} \right>   &=&   \left [\sin\theta \sin\left(Et \right) \right]^2,
\label{eq:res:photonOccupations}
\end{eqnarray}
which maintains the probability
$
\langle \hat P_{n} \rangle +  \langle \hat P_{n+1} \rangle = 1.
$
Intriguingly, the occupations in Eq.~\eqref{eq:res:photonOccupations} are identical to the exact time evolution of the quantum Jaynes-Cummings model. Thus, we have determined the dynamics of a genuine quantum model by employing semiclassical methods of the PRFT. 

{\color{\markColorOne}  Analysis of Eq.~\eqref{eq:res:photonOccupations} explains the minor oscillating deviations in the weak- and intermediate-coupling regimes in Fig.~\ref{fig:RabiModel}(b) for the coherent photonic state. The PRFT assumes a fix Rabi frequency $g =\tilde g \sqrt{\overline n}$. Taking a more microscopic perspective,  each initial Fock state $\left| n \right> $ determines its specific oscillation frequency $g(n) = \tilde g \sqrt{n}$, such that the probabilities in Fig.~\ref{fig:RabiModel}(b)  oscillate slower (faster) for photon numbers $n$ below (above) the mean photon number, leading to the observed minor derivations. This interpretation is further underpinned by the improved accuracy of the PRFT in Sambe space, which is analyzed in Appendix~\ref{sec:app:benchmarkCalculations}. }

\subsection{Two-mode Rabi model}
\label{sec:twoModeRabiModel}

{ \color{\markColorOne}  We found for the $R=1$ Rabi model that the shot-noise entanglement causes a small error for the photonic probabilities ($d_1\lessapprox0.01$ in Fig.~\ref{fig:RabiModel}), which is even vanishing in the thermodynamic limit $\overline n \rightarrow \infty$, $\sigma  \rightarrow \infty$ or when working in the Sambe space defined in Eq.~\eqref{eq:def:sambeSpace}.} 
As a more advanced example, that still allows for numerical benchmark calculations, we consider now the two-mode quantum  Rabi model in Eq.~\eqref{eq:ham:quantumRabi} with $R=2$. This model allows for a large-scale photon transport between the two photonic modes, which leads to a  more prominent photonic dynamics as in the $R=1$ model, and a light-matter entanglement effect, that persists in the thermodynamic limit.

The initial state is given in Eq.~\eqref{eq:InitalStateBenchmarking} for  $\overline n_k=5000$ and three different photonic distribution widths $\sigma_k\in \left\lbrace 2,3,4\right \rbrace$ for both modes $k=1,2$. Larger values for $\sigma_k$ cannot be numerically simulated. According to the error scaling in Eq.~\eqref{eq:errorScaling}, the PRFT requires large values of $\sigma_k$. However, we find that the PRFT is already very accurate for $\sigma_k =4$ and improves quickly. In  Appendix~\ref{sec:app:benchmarkCalculations}, we also benchmark the PRFT for various mean photon numbers $\overline n_k$, and in the  dephasing, adiabatic, and high-frequency driving regimes. 

\begin{figure*}
	\includegraphics[width=\linewidth]{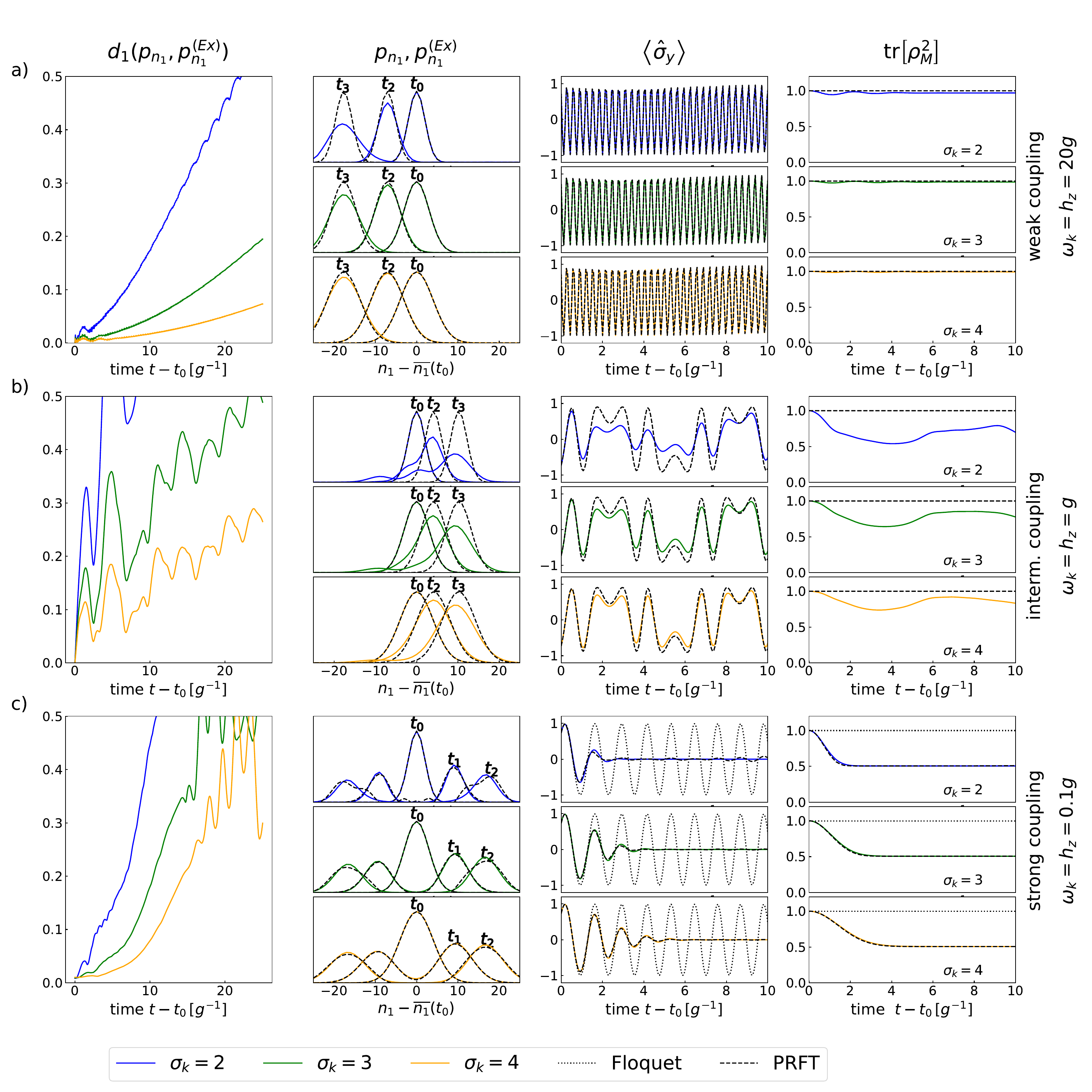}
	\caption{ Benchmark calculations of the PRFT for an initial Floquet  state and various photon number widths $ \sigma_1=\sigma_2$ in the two-mode Rabi model. The first column depicts the trace distance $d_1(p_{n_1}  , p^{\text{(Ex)}}_{n_1}   ) $ as a function of time.  The second column shows the photonic probability distributions for selected times $t_1g = 4$,  $t_2g = 8$, and $t_3g = 20$ as labeled in the panels. The third column depicts the expectation value of $\left<\hat \sigma_y \right>$ as a function of time. The fourth column shows the purity.  (a), (b), and (c)  depict the dynamics in the weak-,  intermediate-, and strong-light-matter-coupling regime, respectively. Colored lines depict the  numeric quantum calculations. We choose a symmetric coupling of both modes to the two-level system $g_k =g$ and $\overline n_k(t_0) =5000$ for $k=1,2$.  Other parameters are specified in the figure.  }
	\label{fig:benchmarkTwoModeRabiFloquetState}
\end{figure*}

\begin{figure*}
	\includegraphics[width=\linewidth]{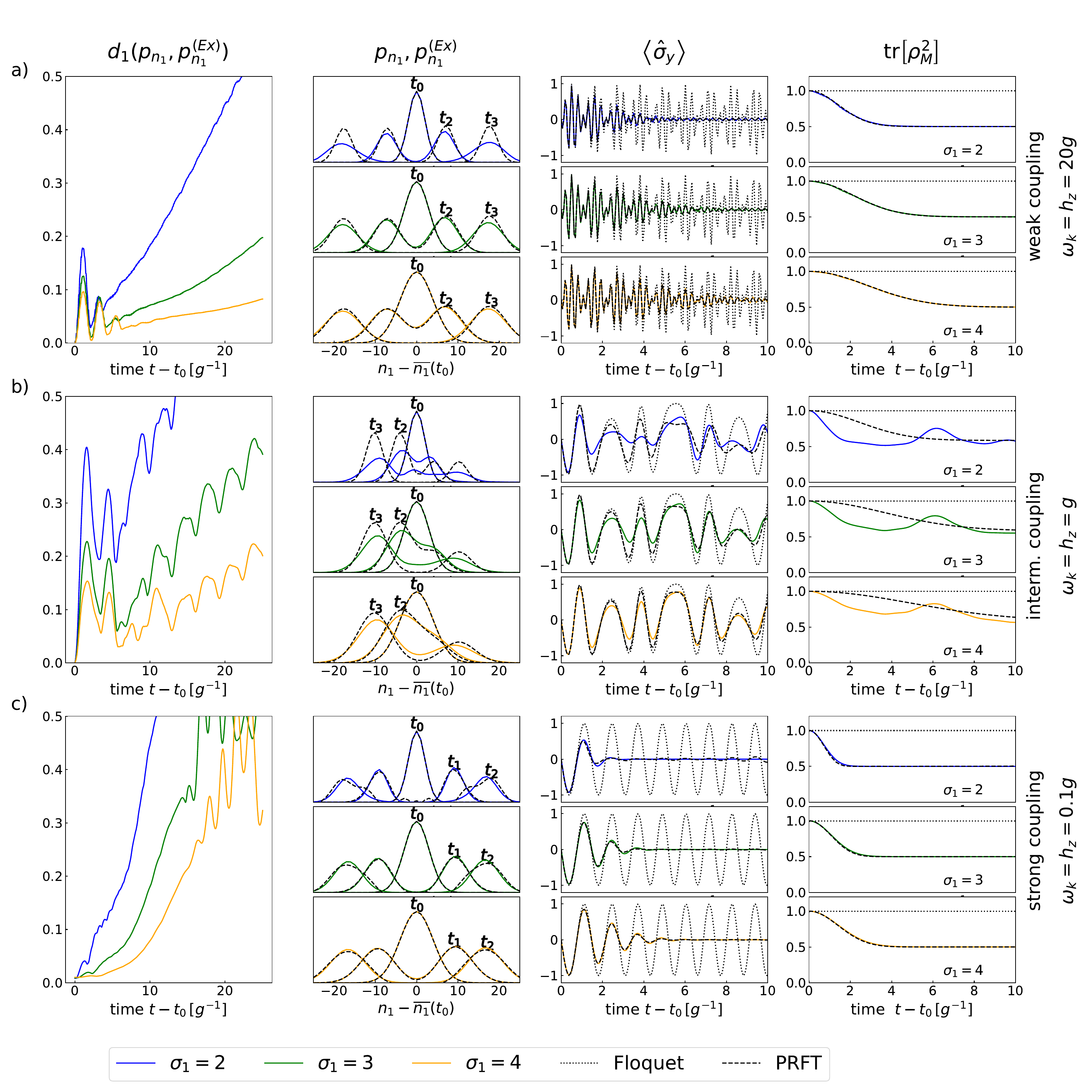}
	\caption{  Same as in Fig.~\ref{fig:benchmarkTwoModeRabiFloquetState}, but for the initial state $\ket{\phi(t_0) } =\ket{\uparrow}   $.   }
	\label{fig:benchmarkTwoModeRabiSpinUp}
\end{figure*}

As explained in Sec.~\ref{sec:FloquetStateAnalysis}, the time evolution sensitively depends on the initial state.   For this reason, we carry out benchmarking for an initial Floquet state $\ket{\phi(t_0)} = \ket{u_\mu}$, and an initial spin-up state $\ket{\phi(t_0)} = \ket{\uparrow}$ in Figs.~\ref{fig:benchmarkTwoModeRabiFloquetState} and \ref{fig:benchmarkTwoModeRabiSpinUp}, respectively. In the first column of Figs.~\ref{fig:benchmarkTwoModeRabiFloquetState} and \ref{fig:benchmarkTwoModeRabiSpinUp} we depict the variational distance $d_1$ defined in Eq.~\eqref{eq:variationalDistance}.  The second column depicts the probability distribution at selected times. The third column shows the expectation value $\left< \hat \sigma_y\right>$ of the two-level system, while the forth column depicts the purity of the matter system.

For short simulation times $t<2\pi/\omega$, we use Eq.~\eqref{eq:tra:pseudoProb} to determine the photon probability distribution, while for longer times $t>2\pi/\omega$ we employ Eq.~\eqref{eq:photonRedistributionSimplified}.
To calculate $\left< \hat \sigma_y\right> $ and $ \mathcal P $  within the PRFT, we use Eq.~\eqref{eq:matterDensityMatrix} for short simulation times. For long times, we employ the reduced density  matrix of the state in Eq.~\eqref{eq:lightMatterEntanglement}.
For the spin observables, we also depict the predictions of the standard Floquet theory. In the following, we discuss the performance of the PRFT in the weak-, intermediate-, and strong-light-matter-coupling regimes.

\textit{Weak coupling.} In the weak-coupling regime  for an initial Floquet state depicted in Fig.~\ref{fig:benchmarkTwoModeRabiFloquetState}(a), we observe that the photonic probability distribution is shifted to smaller photon numbers $n_1$ with increasing time. For small $\sigma_1 =2$, the compact initial photon distribution significantly diffuses with time. However, when increasing to $\sigma_1 =4$, we already find a very good agreement with the exact probabilities. The improvement with increasing $\sigma_1$ can be also clearly seen  in  the trace distance $d_1$. This improvement with increasing $\sigma$ is according to the first argument in Eq.~\eqref{eq:errorScaling}, while the two other arguments have a negligible influence on the dynamics. An excellent agreement of both the standard Floquet theory and PRFT to the exact calculation can be also observed for  $\left<\hat \sigma_y \right>$ and the purity.

As we investigate here the resonant system with $h_z = \omega_1 =\omega_2$ in the weak-coupling regime, we can apply the rotating-wave approximation, and investigate the corresponding two-mode Jaynes-Cummings model. As explained in Sec.~\ref{sec:FloquetStateAnalysis}, the asymptotic dynamics is defined by the counting-field depend quasienergies, which in this case can be exactly calculated and read as
\begin{equation}
E_{\boldsymbol \chi,\mu} = \pm 2 \left| G(\boldsymbol \chi )  \right|  
\end{equation}
with $\mu=1,2$, where we have defined $ G(\boldsymbol \chi ) = \sum_{k=1,2}   g_k e^{- i \chi_k   } $. The corresponding Floquet states are given by $\sqrt{2}\ket{u_{\mu, \boldsymbol \chi}} = e^{i\phi_{\boldsymbol \chi}} \ket{ \downarrow }   \pm   e^{-i\phi_{\boldsymbol \chi}}  \ket{ \uparrow } $  with $ \phi_{\boldsymbol \chi} = \text{arg}\, G(\boldsymbol \chi )  $. Interestingly, the quasienergies depend only on the difference $\chi = \chi_1 -\chi_2$. Evaluating the first dynamical cumulant, we obtain
\begin{eqnarray}
\left. \Delta\langle \hat   N_1 \rangle \right|_{\mu }&=&  (-1)^\mu \frac{2 g_1g_2}{ E_{ \varphi} }  \sin(\varphi)  (t-t_0)  +\mathcal O \left[(t-t_0)^0\right], \nonumber \\
\label{eq:res:mean_JcModel}
\end{eqnarray}
where $\varphi = \varphi_1 - \varphi_2$ is the phase difference of the two photonic states. For $\varphi\neq \left\lbrace 0,\pi \right\rbrace $, there is a net photon flux between the photon modes.  The variance change $\Delta \sigma_1$ vanishes according to Eq.~\eqref{eq:vanishingVariance}. The current  vanishes for  $\varphi= \left\lbrace 0,\pi \right\rbrace $ due to the symmetry of the initial condition. In Fig.~\ref{fig:benchmarkTwoModeRabiFloquetState}(a), we consider Floquet state $\mu=1$, such that photons are transported from mode $k=1$ to $k=2$, as can be inferred from Eq.~\eqref{eq:res:mean_JcModel}. Likewise, photons would be transported from $k=1$ to $k=2$ when the system is initialized in Floquet state $\mu=2$.

Figure~\ref{fig:benchmarkTwoModeRabiSpinUp}(a) depicts the same as Fig.~\ref{fig:benchmarkTwoModeRabiFloquetState}(a) but for the initial state  $\ket{\phi(t_0)} = \ket{\uparrow}\approx (\ket{u_{1,\boldsymbol\varphi } } + \ket{u_{2,\boldsymbol\varphi }  })/\sqrt{2}$, which is here a  balanced superposition of the two Floquet states. The overall accuracy is similar for both initial conditions. According to Eq.~\eqref{eq:meanPhotonNumberExpansion}, the photon flow is controlled by the Floquet state, such that the photon redistribution in the long-time limit becomes entangled with the two-level system. This effect can be clearly observed for $\sigma_1 =4$, where the left peak is entangled with the Floquet state $\ket{u_{1,\boldsymbol\varphi } }$, while the right peak is entangled with $\ket{u_{2,\boldsymbol\varphi} }$ [see also Fig.~\ref{figCountingStatSketch}(c)].

The light-matter entanglement leads to decoherence as is clearly manifested in the time evolution of $\left< \sigma_y\right>$.
It is noteworthy that the PRFT agrees almost perfectly with the exact quantum calculation, while the standard Floquet theory strongly deviates. In particular, the PRFT accurately predicts the evolution of an initial pure state (with purity $\mathcal P = 1$) to a maximally entangled state (purity $\mathcal P = 0.5$). In contrast, the purity remains 1 in standard Floquet theory.

\textit{Intermediate coupling.} In Fig.~\ref{fig:benchmarkTwoModeRabiFloquetState}(b) we depict the dynamics for an initial Floquet state in the intermediate coupling regime for $h_z = \omega =g_k=g$. As in the weak-coupling regime, we observe that the PRFT calculation rapidly approaches the exact probabilities for increasing $\sigma_1$. The mean photon number $\overline n_1(t)$ increases linearly, while the width $\sigma_1(t)$ stays almost constant, which can be seen for $\sigma_1 =4$.
We note that even though the difference between exact numerics and the PRFT is larger than in the weak coupling case, the convergence can be clearly anticipated even for small $\sigma_1$.
In general, large couplings $g$ result in more Fourier components in the periodic dynamics of the Floquet states, that require an initial probability distribution with larger $\sigma_1$ to be smoothed out (see also Appendix~\ref{sec:errorAnalysis}). 

Furthermore, similar to the dynamics in Fig.~\ref{fig:benchmarkTwoModeRabiSpinUp}(a), we observe that  the probability distribution eventually splits into two peaks in Fig.~\ref{fig:benchmarkTwoModeRabiSpinUp}(b), each being entangled with  a Floquet state. The height of each peak is thereby determined by the amplitude of the expansion coefficients in the Floquet basis according to Eq.~\eqref{eq:probabilityExpansion}.

\textit{Strong coupling.}  In Figs.~\ref{fig:benchmarkTwoModeRabiFloquetState}(c) and \ref{fig:benchmarkTwoModeRabiSpinUp}(c) we analyze the  dynamics in the strong-coupling regime. As in the weak- and intermediate-coupling regime, we observe that the PRFT calculations for both the photonic and spin observables approach the exact calculation for increasing $\sigma_1$. In contrast, the standard Floquet theory clearly fails to reproduce the correct dynamics of the two-level system. 

Due to numerical limitations, the probability distribution is depicted only for times $t$ smaller than the driving period $t_i \ll \tau = 2\pi/\omega$ in Fig.~\ref{fig:benchmarkTwoModeRabiFloquetState}(c). As the findings in Sec.~\ref{sec:FloquetStateAnalysis} are only valid at stroboscopic times $t = n\tau$, we do not observe a steady photon flux towards a higher (lower) photon number as in Fig.~\ref{fig:benchmarkTwoModeRabiFloquetState} (a) [Fig.~\ref{fig:benchmarkTwoModeRabiFloquetState}(b)]. However, we expect that  both peaks will merge at stroboscopic times $t = n\tau$ at a mean photon number as predicted by Eq.~\eqref{eq:meanPhotonNumberExpansion}. Similarly, we cannot associate the two peaks with the two Floquet states in Fig.~\ref{fig:benchmarkTwoModeRabiSpinUp}(c) for the depicted times $t_j \ll \tau$.

\subsection{Three-mode Rabi model }
\label{sec:treeModeRabiModel}

\begin{figure}
	\includegraphics[width=\linewidth]{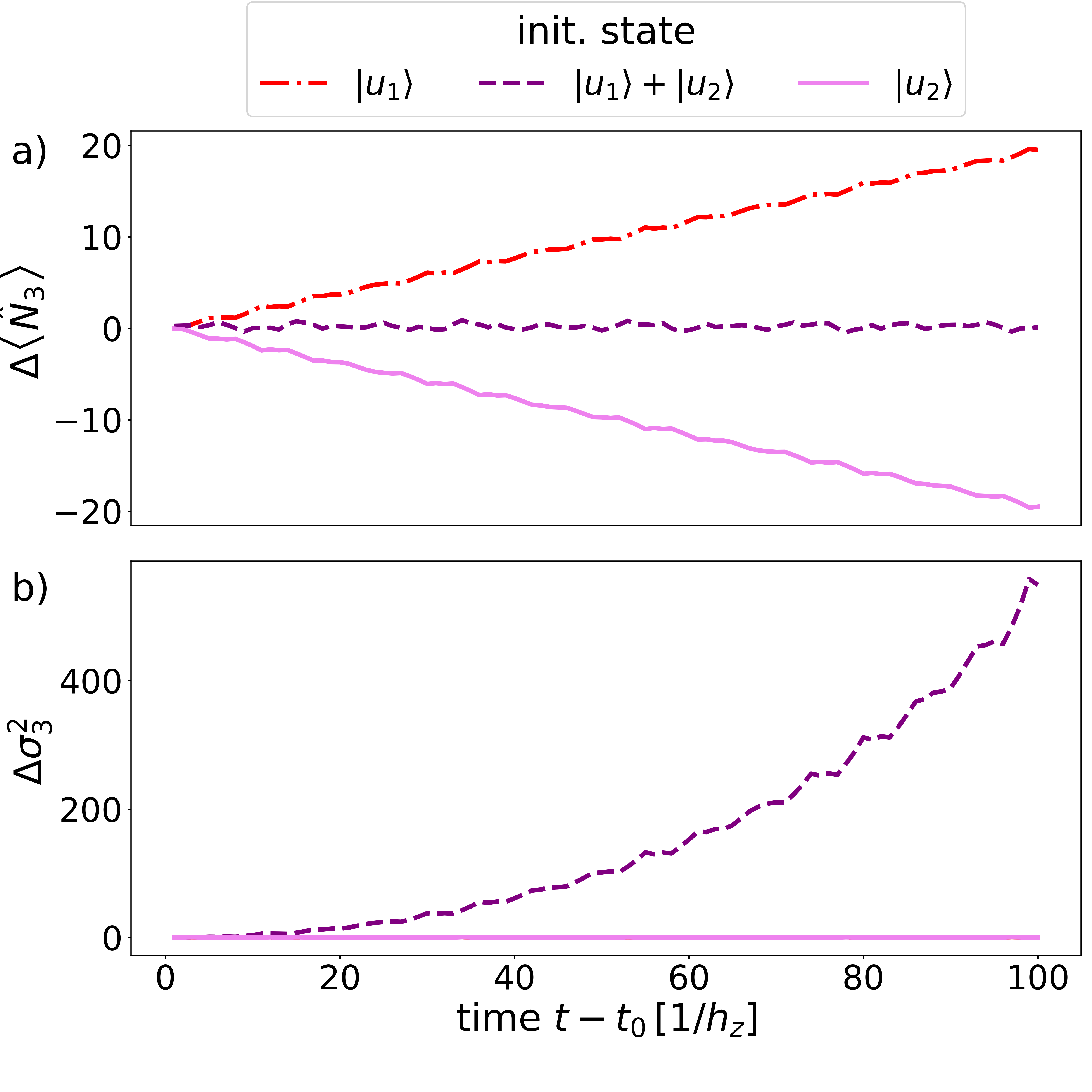}
	\caption{Analysis of the photon statistics in the Rabi model in Eq.~\eqref{eq:ham:quantumRabi} for $R=3$.  (a) Change of the mean photon occupation of mode $k=3$. (b) Change of the corresponding variance. Parameters are $  h_z  = 2.1 \omega_1$, $  \omega_2  = 2 \omega_1$, $  \omega_3  = 3 \omega_1$, and $g_k = \alpha_k\tilde g_k= \omega_1 $, $\varphi_1 =\varphi_3 = 2 \varphi_2 =0.5$.    }
	\label{fig:ThreeModeRabiModel}
\end{figure}

To illustrate that the light-matter entanglement explained in Sec.~\ref{sec:FloquetStateAnalysis} is a generic effect, we now investigate a three-mode Rabi model with distinct commensurate photonic frequencies $\omega_k$. For concreteness, we choose $\omega_2 = 2\omega_1$ and $\omega_3 =3\omega_1$. In this case, the  semiclassical Hamiltonian is a periodically-driven Rabi model with period $2\pi/\omega_1$, and we compute the photon-mean and photon-variance changes using the PRFT. Noteworthy, a simulation of the full quantum Rabi model  in Eq.~\eqref{eq:ham:quantumRabi} is numerically hardly tractable due to the three photonic modes. When representing each mode with $m$ states, the Hilbert space has dimension $D = 2\times m^3$, i.e., $D= 2\times 10^6$ states for a moderate $m=100$.
Similar to the analysis for the two-mode Rabi model,  we consider the three initial conditions $\left| \phi(t_0)\right>  = \left| u_{1,\boldsymbol\varphi } \right>$,  $\left| \phi(t_0)\right>  = \left| u_{2,\boldsymbol\varphi } \right>$, and  $\left| \phi(t_0)\right>  = \frac{1}{\sqrt{2}}  \left( \left| u_{1,\boldsymbol\varphi } \right> +  \left| u_{2,\boldsymbol\varphi } \right> \right) $, where the $\left| u_\mu \right>$ are the Floquet states of the two-level system, i.e., the eigenstates of Eq.~\eqref{eq:defFloquetHamiltonian} for $\boldsymbol \chi =0$.

In Fig.~\ref{fig:ThreeModeRabiModel}, we depict  the photonic mean and variance change of mode $k=3$ for  each initial condition.  When the initial state is the  Floquet state  $\left| u_1 \right> $ ($\left| u_2 \right> $), the mean photon number grows (decreases) linearly in time, while the variance remains almost unchanged. This is analog to the probability distributions  in Fig.~\ref{fig:benchmarkTwoModeRabiFloquetState} (a) in the two-mode model, where the variance remains constant for a Floquet initial state.

In contrast, the mean photon number change is close to zero for the superposition state, while the variance increases rapidly. This corresponds to the probability distributions for an initial superposition in Fig.~\ref{fig:benchmarkTwoModeRabiSpinUp}(a). Since  we have chosen a balanced superposition of the two Floquet states with $c_1 =c_2=1/\sqrt{2}$, the mean photon number remains unchanged according to Eq.~\eqref{eq:meanPhotonNumberExpansion}. However, as the Floquet state $\mu=1$ linearly increases the photon number, while Floquet state $\mu=2$ simultaneously linearly decreases the photon number, the variance diverges quadratically.

\section{Experimental and technological  implications}
\label{sec:experimentalTechnicalImpliations}

In this section, we discuss the light-matter induced decoherence  for coherently driven systems. This effect can be detected in experiments and has a significant impact on the design of quantum memories and  quantum operations. {\color{\markColorOne} The following analysis therefore focuses on the transport-induced entanglement and neglects the shot-noise entanglement.}

\subsection{Quantum-optical coherence time}

\label{sec:quantumMemories}

To estimate the quantum-optical coherence time for periodically driven systems, we consider  the  mean photon-number changes  $\left. \Delta \langle \hat N _k \rangle \right|_{\mu}  = - E'_\mu (\varphi_k) t  $   for two distinct Floquet states $\mu_1$ and $\mu_2$, where $E'_\mu (\varphi_k) = d E_\mu (\boldsymbol \varphi) / d\varphi_k$. The system is completely decohered at time $t_c$ when  the difference in the mean photon number for these two Floquet states exceeds the width of the photon distribution in Fock space, i.e.,
\begin{equation}
t_c \left| E'_{\mu_1} (\varphi_k) -E'_{\mu_2}  (\varphi_k)    \right| = \sigma_k,
\end{equation} 
where $\sigma_k^2$ is the initial variance of photon mode $k$. We recall that the variance does not change with time for  Floquet states as shown in Sec.~\ref{sec:FloquetStateAnalysis}. To estimate $\sigma_k$ in terms of physical quantities, we distinguish  two cases: (i) a closed light-matter system, where the photonic field is confined in a cavity; (ii) an externally driven quantum system, where the photonic field is a traveling wave. Each case gives a different scaling behavior for $t_c$.

\textit{ Closed  light-matter systems:} For simplicity, we consider coherent states, whose  mean and variance are equal  $  \overline n_k = \sigma_k^2$. The mean photon number in a cavity mode is given by $\overline n_k = \epsilon_0 E^2  V/(2 \hbar \omega_k)$ where $E$ is the electromagnetic field, $V$ is the cavity volume, and $\epsilon_0$ is  the dielectric constant. This implies that  the quantum-optical coherence time is given as
\begin{equation}
t_{\text{c}}  = \min_{k,\mu_1,\mu_2} \frac{\sqrt{ \frac{ \epsilon_0 E^2  V  }{2 \hbar \omega_k} } }{ \left|E'_{\mu_1}  (\varphi_j) -E'_{\mu_2} (\varphi_j)    \right|  }  .
\label{eq:coherenceTime}
\end{equation}
We note that the coherence time in Eq.~\eqref{eq:coherenceTime} is an approximation, since the above arguments have assumed that the photon modes decohere independently. 

\textit{Externally-driven quantum systems:} For the experimentally more relevant situation, in which the matter system is driven by a traveling wave, the estimate for the coherence time has to be modified. Here we establish a connection to a cavity setup to get an estimate for the coherence time. Consider that the light field is a pulse of duration $t_{\text{p} } $ with  central frequency $\omega_k$ and spectral width $\Delta\omega\propto 1/t_{\text{p} }$. For long times $t_{\text{p} } $, the spectral width vanishes and we assign all pulse photons to the central frequency. The mean photon number in a pulse is $\overline n_k = P(\omega_k) t_p$, where $P(\omega_k) $ is the power of the electromagnetic field at frequency $\omega_k$. Now, we model the pulse as a cavity mode  with initial occupation $\overline n_k$.  Consequently, we find that the quantum-optical coherence time is,
\begin{equation}
t_{c}  = \min_{k,\mu_1,\mu_2} \frac{ P(\omega_k)  }{  \hbar \omega_k\left|E'_{\mu_1}  (\varphi_k) -E'_{\mu_2} (\varphi_k)    \right|^2  }  .
\label{eq:coherenceTimeOpen}
\end{equation}
We note that since each photonic mode  contributes to the decoherence effect, the coherence time in Eq.~\eqref{eq:coherenceTimeOpen} can be considered as an upper bound.

\subsection{Experimental verification}  

\label{sec:experimentalVerification}

The decoherence effect discussed above provides a route to verify the PRFT without measuring photon statistics, which may be a challenging task. We explain our approach for the two-mode Rabi model for illustration. The key idea is to determine the  decoherence effect by purity measurements of the atomic system. For this, it is crucial to isolate the quantum-optical decoherence from other decoherence sources. 
We achieve this by varying the amplitudes of the coherent states $\alpha_k$ and the light-matter interactions $\tilde g_k$ such that the effective parameters  $g_k =\tilde g_k \alpha_k$ remain constant. In doing so, the spin system experience the  same semiclassical driving fields [see Eq.~\eqref{eq:generalizedHamiltonianMMJCmodel}] and, thus, is subject to the same decoherence sources other than the quantum-optical decoherence. As for a coherent driving field  $ \sigma_k \propto \alpha_k  \propto  P(\omega_k)^{1/2}$,  the quantum-optical decoherence dynamics can be accessed by measuring the purity decay for various driving-field powers $P(\omega_k) $.

To get  an estimate for the coherence time in the optical regime, we evaluate Eq.~\eqref{eq:coherenceTimeOpen} for the  two-mode Rabi model, where the photonic frequency is assumed to be $\omega =400\,\text{THz}$. The quasienergy difference in the denominator in Eq.~\eqref{eq:coherenceTimeOpen} can be approximated by a typical Rabi frequency of atomic systems   $\Omega_{\text{R} } = 40\,\text{MHz}$, generated by   a laser  with power $P(\omega) = 10\,\mu\text{W}$. For these parameters, the  quantum-optical coherence time is $t_{\text{c} } =5\,\text{ms}$, which is comparable to the duration of typical cold-atom experiments~\cite{wang2021observation,Wang2022,Viermann2022}, but significantly shorter than quantum information storage times achieved with trapped ions~\cite{harty2014highFidelity}. 

We obtain the order of magnitude of the coherence time for radio frequencies using the following parameters: $\omega =10 \,\text{MHz}$, $\Omega_{\text{R}} = 10\,\text{kHz}$, and $P(\omega) = 10\,\text{W}$, which are typical in current experiments~\cite{zhong2015optically}. In this regime, the quantum-optical coherence time is very long ($t_{\text{c} } \approx 3\times  10^{18}\,\text{s}$). This is a consequence of the high photon occupation of radio-frequency modes for realistic experimental parameters. Thus, in the radio-frequency regime, the electromagnetic field can be considered as completely classical, and the decoherence can be neglected. 

\subsection{Quantum memories and quantum operations} 

The quantum-optical decoherence in driven systems can have a significant impact on the design of quantum memories and  quantum operations. To achieve long quantum information storage times,  sophisticated control protocols have been developed, that typically involve time-periodic electromagnetic fields. Using dynamical decoupling, quantum information could be conserved for more than six hours in rare-earth atoms embedded in a crystal structure~\cite{zhong2015optically}, and more than $50\,\text{s}$ in trapped ions~\cite{harty2014highFidelity}. In both cases, the quantum information is stored in the hyper-fine levels of the ground-state manifold, which are energetically separated in the radio-frequency regime.

Depending on the control protocol, the quantum-optical coherence time in Eq.~\eqref{eq:coherenceTime} influences the performance of quantum memories. The quantitative considerations in Sec.~\ref{sec:experimentalVerification} suggest that driving protocols involving optical-frequencies should be avoided, while radio-frequency control protocols are optimal. We note that quantum information storage and retrieval protocols often employ optical frequencies~\cite{fleischauer2005electromagnetically}. Even though the pulse duration in these cases may be rather short, $t_{\text{p} }<1\;\mu \text{s}$, an inappropriately adjusted pulse sequence might lead to a degradation of the quantum information via the quantum-optical decoherence effect. Similar considerations  also apply to other quantum operations based on periodic driving, such as two-qubit gates. As quantum error correction usually requires high fidelities $> 99 \%$~\cite{lidar2013quatum}, even a minor quantum-optical decoherence can have a significant effect on quantum operations. Furthermore,  inspection  of the coherence time in Eq.~\eqref{eq:coherenceTimeOpen} reveals that the coherence time can be enhanced by choosing control protocols for which the difference of quasienergies are not sensitive to the driving phases. 

 As the coherence time depends on the specific system, no general statements can be made about how to eliminate the quantum-optical decoherence effect.
An alternate intriguing approach to mitigate this decoherence would be to employ quantum time crystals~\cite{nayak2019review,khemani2019review,yao2017prl}.  Discrete time crystals, that are driven by an external driving field with period $\tau = 2\pi/\omega $, exhibit subharmonic response with a frequency $\omega/n$, where $n>1$ is an integer. This intriguing subharmonic response arises from the structure of the eigenspectrum which is composed of Floquet eigenstates that are separated by a quasienergy of $\omega/n$. Since  time crystals are robust to generic perturbations, the difference of quasienergies would have little dependence on the driving phases, leading to stable quantum memories.

\section{Quantum communication}
\label{sec:quantumCommunication}

The light-matter entanglement discussed in the previous section can  be employed in a quantum communication protocol that is robust against photon loss. To this end, we consider that Alice and Bob---the two participants in the communication---successively carry out the light-matter entanglement process described for the two-mode Rabi model and postselect the measurement results. We now proceed to explain how Greenberger-Horne-Zeilinger (GHZ) states can be employed to speedup the light-matter entanglement generation, before delving into the details of the communication protocol in Sec.~\ref{sec:RemoteEntanglementGeneration}.

\subsection{Rapid generation of light-matter entanglement}

\label{sec:speedingUpEntanglementProcess}

Typically, the light-matter interaction between a single atom and the light-field is relatively weak, which slows down the  generation of maximally-entangled light-matter states as depicted in Fig.~\ref{figCountingStatSketch}(b). To enhance this effect, we again employ the setup in Fig.~\ref{figCountingStatSketch}(b), but with $N_{\text{A}}$ atoms. The corresponding Hamiltonian reads as
\begin{equation}
H = \sum_{j}\frac{h_z }{2} \hat \sigma_z^{(j)} +  \tilde g \sum_{j,k}  \hat \sigma_{+}^{(j)} \hat a_{k} + \hat \sigma_{-}^{(j)} \hat a_{k}^\dagger + \sum_k \omega \hat a_k^\dagger \hat a_k    ,
\end{equation}
where the Pauli operators $\hat \sigma_z^{(j)} , \hat \sigma_{+}^{(j)} ,\hat \sigma_{-}^{(j)}  $ act on the atoms $j = 1,\dots,  N_{\text{A}}$. This is a rotating-wave approximation version of the Hamiltonian in Eq.~\eqref{eq:ham:quantumRabi} with many two-level systems, i.e., the Tavis-Cummings model,  in which the many-body interaction is mediated via the quantized electromagnetic field.  As in Sec.~\ref{sec:twoModeRabiModel}, we consider two photonic modes $k=1,2$ which are initially in  coherent states $\left|  \alpha_k  e^{i\varphi_k}  \right>$. 

The PRFT is a powerful tool to analyze this system when the atom numbers are large and exact numerical calculations are very expensive.
Upon introducing the counting fields $\chi_1$ and $\chi_2$, the corresponding semiclassical Hamiltonian reads as
\begin{eqnarray}
\mathcal H(t) &=& \sum_{j=1}^{N_
	{\text{A} } } \mathcal H^{(j)}(t)\nonumber, \\
\mathcal H^{(j)}(t) &=& \frac{h_z }{2} \sigma_z^{(j)} +    \sum_{k=1}^{2}  \sigma_{+}^{(j)} g_k  e^{-i\omega t+i\chi_k }+ \text{H.c.},
\end{eqnarray}
where $\mathcal H^{(j)}(t)$ denotes the semiclassical Hamiltonian of atom $j$.
The atoms are formally decoupled in the semiclassical description, yet, the interaction is still implicitly encoded in the counting fields. The quasienergy of the total system can be written as the sum of the quasienergies of the individual atoms
\begin{equation}
E_{\boldsymbol \mu,\boldsymbol  \chi}^{(N_{\text{A}} )} = \sum_{j=1}^{N_{\text{A} }} E_{\mu_j,\boldsymbol  \chi},
\end{equation}
where $E_{\mu_j,\boldsymbol  \chi}$ is the quasienergy of  atom $j$ in Floquet state $\mu_j$. We have introduced the vector notation $\boldsymbol \mu  = \left( \mu _1  ,\dots ,\mu_{N_A}\right)$, that contain the quantum numbers $\mu_j$ of the $N_A$ atoms. The corresponding Floquet states read as
\begin{equation}
\left| \Psi_{\boldsymbol \mu,\boldsymbol \chi} \right> = \bigotimes_j  \left| u^{(j)}_{\mu_j ,\boldsymbol \chi} \right>,
\label{eq:collectiveFloquetStates}
\end{equation}
where  $\left|  u^{(j)}_{\mu_j, \boldsymbol \chi} \right>$ denotes the Floquet state of   atom $j$ with quantum number $\mu_j$.

To drastically enhance the number of photons transported from mode $k=1$ to $k=2$ (or vice versa), we prepare the system in either of the Floquet states characterized by the $N_{\text{A}}$-component quantum numbers
$\boldsymbol \mu^{(1)} =(1,\dots,1) $  or $\boldsymbol \mu^{(2)} =(2,\dots,2) $, i.e., the state in which all atoms $j$ are  in the same Floquet state $\mu_j =\alpha \in \left\lbrace1,2\right\rbrace$. The quasienergy of the atom ensemble in either of these states  is then
\begin{equation}
E_{\boldsymbol \mu^{(\alpha)} ,\boldsymbol \chi} =    N_{\text{A} }  E_{ \alpha  , \boldsymbol \chi} .
\label{eq:quasienergyEnhancement}
\end{equation}
According to Sec.~\ref{sec:FloquetStateAnalysis}, this leads  to an enhancement of the photon transport proportional to the atom number.
To be more specific, we choose $h_z =\omega$ and the driving phases  $\varphi_1 = \pi/4, \varphi_2 = - \pi/4$. In this case, the Floquet states of the matter system $\left|u_\mu \right>$ for each atom are $ \left| u_1 \right> = \left| -\right> = \left(  \left| 0  \right> - \left| 1  \right> \right) /\sqrt{2} $  and $\left| u_2\right> = \left| + \right> = \left(  \left| 0  \right> + \left| 1  \right> \right) /\sqrt{2} $, and the Floquet states of the atom ensemble in Eq.~\eqref{eq:collectiveFloquetStates}   are
$ 	\left| \Psi_{\boldsymbol \mu^{(1)}} \right>  = \left| -\cdots - \right> \equiv \left| \pmb - \right> $
and
$ 	\left| \Psi_{\boldsymbol \mu^{(2)}} \right>  = \left| +\cdots + \right> \equiv \left| \pmb + \right> $.

\begin{figure}
	\includegraphics[width=\linewidth]{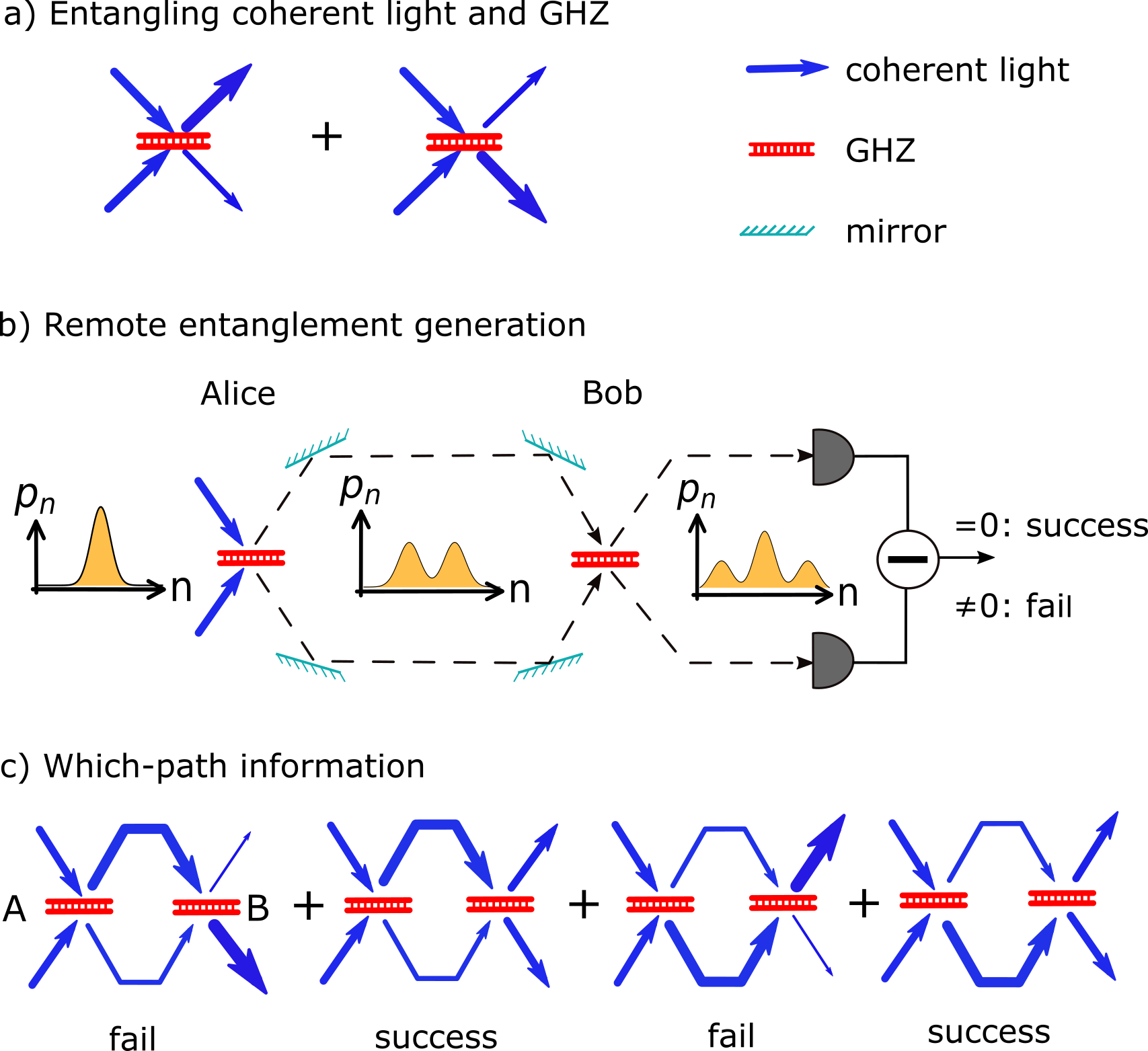}
	\caption{ Remote entanglement generation protocol. (a) Illustration of the local light-matter entanglement process. After the interaction of two coherent light fields at a GHZ state, the final state is a superposition of states where either coherent field is enhanced (thick arrow) while the other is reduced (thin arrow).  (b) Entanglement protocol outlined in Sec.~\ref{sec:RemoteEntanglementGeneration}. After light-matter entanglement with Alice's GHZ state, the light is transmitted to Bob, where it interacts with Bob's GHZ state. The dashed lines sketch the light paths. After Bob's light-matter entanglement, the coherent fields are measured (gray half circles). A vanishing signal difference heralds success of the protocol.   The three insets show the photon distributions $p_n$  before and after Alice's entanglement process, and after Bob's process. (c) Analysis of the \textit{which-path} information. After interaction with Bob's GHZ state, the system is in a superposition of four states. In two of which, the two output fields have changed their amplitude which reveals the path of quantum information and results in  failing of the protocol. In the other two states, the \textit{which-path} information remains hidden and results in success of the protocol.   }
	\label{fig:quantumCommnuication}
\end{figure}

By constraining the system  to the collective states $ \left| \pmb - \right> $  and  $ \left| \pmb + \right> $, we restrict the model to an effective two-level system with a renormalized quasienergy, for which the findings in the Jaynes-Cummings model in Appendix~\ref{sec:multiModeJaynesCummings} are valid. To  create   light-matter entanglement in the  basis  $ \left| \pmb + \right> $  and  $ \left| \pmb - \right> $, we use the initial condition
\begin{equation}
\left| \Psi(0) \right>  = \hat U_{\text{H} }^{N_{\text{A}  } }\left| \text{GHZ} \right>,
\end{equation}
where $ \left| \text{GHZ} \right> $ is the celebrated GHZ state. This state is defined by a superposition $ \left| \text{GHZ} \right>  = \left( \left| \pmb 0  \right> + \left| \pmb 1 \right> \right) /\sqrt{2} $, where $\left| \pmb 0  \right> \equiv \left|0 \dots 0 \right>$  and  $\left| \pmb 1  \right> \equiv \left|1 \dots 1 \right>$. The Hadamard gate $\hat U_{\text{H} } = \exp\left[-i\pi \hat \sigma_y /4 \right] $ that locally rotates the state of each atom, is independently applied to all atoms. 
For later purpose, we illustrate the  light-matter entanglement generation  in Fig.~\ref{fig:quantumCommnuication}(a) showing a superposition of states, where either driving field is enhanced and the other is reduced. 
We emphasize that even though we take the Tavis-Cummings model as an example, this enhancement effect is valid for  general Floquet systems according to the PRFT.

\subsection{Remote entanglement generation}

\label{sec:RemoteEntanglementGeneration}

The most crucial task of the quantum communication protocol is the generation of remote entanglement between two atomic ensembles possessed by Alice and Bob. Quantum state transfer can then be carried out via quantum teleportation~\cite{Wilde2017}.  As schematically sketched in Fig.~\ref{fig:quantumCommnuication}(b), this is achieved by repeating the  light-matter entanglement process using the same driving fields. The three steps of the protocol are as follows:

\begin{enumerate}
	\item \textit{State preparation:} For an efficient light-matter entanglement generation, we assume that both Alice and Bob have prepared a GHZ state such that the initial state
	$
	\left| \Psi_{\text{AB} }(t_0) \right> = \left| \text{GHZ}_{\text{A} } \right> \otimes  \left| \text{GHZ}_{\text{B} } \right> 
	$
	is separable.  Alice and Bob carry out  local Hadamard gates such that   their states  become
	\begin{equation}
	U_{H,X}^{N_{\text {A}}} \left| \text{GHZ} \right>_X = \left|\pmb - \right>_X + \left|\pmb + \right>_X,
	\end{equation}
	where $X= \text{A,B}$. The choice of the basis $ \left|\pmb - \right>$ and $\left|\pmb + \right>$ is thereby determined by the driving phases, which we assume to be $\varphi_1 = -\pi/4$ and $\varphi_2 = \pi/4$.
	
	\item \textit{Light-matter interaction:} Alice impinges two coherent light beams onto her atom ensemble leading to the generation of a light-matter entangled state according to the explanations in Secs.~\ref{sec:FloquetStateAnalysis} and \ref{sec:twoModeRabiModel}:
	\begin{equation}
	\left( \left|\pmb -\right>_{\text{A} }   \left| A_{-}  \right> + \left|\pmb + \right>_{\text{A} } \left| A_{+} \right>   \right)\left( \left|\pmb - \right>_{\text{B} }  + \left|\pmb + \right>_\textbf{B} \right) ,
	\end{equation}
	where $\left| A_{+} \right>  $ ($\left| A_{-} \right>  $) denotes a photonic state with enhanced (diminished) amplitude.
	The output light fields are transmitted to Bob, where they interact with Bob's atom ensemble. The resulting state can be written as
	\begin{eqnarray}
	&&\left|\pmb -\right>_{\text{A} }  \left|\pmb -\right>_{\text{B} }   \left| A_{2-}  \right> + \left|\pmb + \right>_A\left|\pmb - \right>_{\text{B} }  \left| A_{0} \right>   \nonumber  \\
	&&+  \left|\pmb - \right>_{\text{A} } \left|\pmb + \right>_{\text{B} }  \left| A_{0} \right> 
	+  \left|\pmb + \right>_{\text{A} }  \left|\pmb + \right>_{\text{B} }  \left| A_{2+} \right> ,\nonumber
	\end{eqnarray}
	where the photonic states $ \left| A_{2-} \right> $,   $ \left| A_{0} \right> $,   $ \left| A_{2+} \right> $  are close to coherent states  with amplitudes smaller, comparable and larger compared to the initial coherent states. The change of amplitude refers to mode $k=1$, while mode $k=2$ will conversely have a larger, comparable or smaller amplitude. The photon distribution   of the modes at different stages of the protocol is sketched in Fig.~\ref{fig:quantumCommnuication}(b).
	
	\item \textit{Measurement and postselection:} 
	Bob makes a projective measurement defined by the operator $\left|A_{0} \right> \left<A_{0} \right|$. If the measurement is successful, Alice and Bob carry out  Hadmard gates (and other local operations to correct for phases accumulated during the light-matter interactions). Finally, Alice and Bob hold a share of the entangled state
	\begin{equation}
	\left|\pmb 0 \right>_{\text{A} }  \left|\pmb  1\right>_{\text{B} }    +  	\left|\pmb 1 \right>_{\text{A} }  \left|\pmb  0 \right>_{\text{B} }  
	\end{equation}
	in the  basis of collective excitations $\left|\pmb 0 \right>_X $ and $\left|\pmb 1 \right>_X $. The projective measurement  can be implemented by measuring the intensity difference  of both output fields. If the difference is close to zero, then the two atoms have been successfully entangled. Otherwise, the process has failed and must be repeated.
\end{enumerate}

The  working principle of the protocol is based on the \textit{which-path} information, that is illustrated in Fig.~\ref{fig:quantumCommnuication}(c). After interacting with Alice ensemble, either of the output fields will be enhanced. This effect can be either repeated or reversed after interaction with Bob's ensemble. If the field $k=1$ is two times enhanced, the intensity measurement reveals that it is in the photonic state $\left| A_{2+} \right>  $. Consequently,  Alice's and Bob's ensemble both are in the state $\left| \pmb +\right>$ and thus not entangled. A similar reasoning applies to   the state $\left| A_{2-} \right>  $.
If the effect is reversed and the intensities of the output modes are equal, the \textit{which-path} information of the photonic fields remains hidden, such that  both ensembles preserve their uncertainty and become entangled. From Fig.~\ref{fig:quantumCommnuication}(c) we  find that the success probability is $50\%$. 
{ \color{\markColorThree} We note that the photon probability distribution of the coherent states decay exponentially from the mean value such that the states $\left| A_{2+} \right>  $, $\left| A_{0} \right>  $, and $\left| A_{2+} \right>  $ can be distinguish with only a small error probability in the intensity difference measurement. }

\subsection{Quantum state transfer rate}

\label{sec:transmissionRate}

The major obstacle in  quantum communication is photon loss. A typical damping rate of optical fibers is $\gamma \approx 0.051\, 1/\text{km} $, leading to a loss of more than $99\%$ of photons after $100 \,\text{km}$. This  heavily  limits the reach of  quantum state transfer protocols based on few photons. Current theoretical  transfer protocols predict transmission rates of up to $1\, \text{Hz}$ over $500\,\text{km}$~\cite{Sangouard2011}. These protocols typically employ  quantum repeaters strategically placed  between the transmission endpoints.

The coherent-light protocol introduced in Sec.~\ref{sec:RemoteEntanglementGeneration} is naturally robust against photon loss. The information of Alice's qubit is encoded as an enhanced or diminished light amplitude during transmission [see inset in Fig.~\ref{fig:quantumCommnuication}(b)]. When a photon is lost, it is hardly possible to determine from which transmission peak it originated. Still, while the \textit{which-path} information is preserved, photon loss has a detrimental effect, as it leads to a broadening of the photonic probability distribution. When the broadening exceeds the distance of the two peaks, i.e.,  $ \langle \Delta \hat N_1\rangle \approx \Delta \sigma_1   $  the quantum information is lost. 

Along the same arguments for photon loss, the quantum information is also robust against classical amplification.  To compensate  the photon loss, we assume that it will be amplified with rate $\gamma_{\text{Amp} } =\gamma$, such the mean photon number is conserved during transmission. Yet, the amplification will lead to a broadening of the probability distributions. Modeling  photon loss and amplification as independent Poissonian processes with rate $\gamma$, the width of each peak increases by  $\Delta \sigma_1  = \sqrt{ 2\gamma d P t_{\text{p}}/(\hbar \omega) } $, where $P$ is the power of the transmitted pulse, $t_{\text{p}}$ is the pulse duration, and $d$ the transfer distance. We recall from Sec.~\ref{sec:quantumMemories} that the separation between the two peaks is given as $\langle \Delta \hat N_1\rangle = | E_{1,\varphi}' -   E_{0,\varphi}' |  t $. In atomic systems, the quasienergy splitting can be associated with the Rabi frequency $\Omega_R$.  Using the GHZ amplification in Eq.~\eqref{eq:quasienergyEnhancement}, the probability peaks are separated by      $\langle \Delta \hat N_1\rangle = N_{\text{A}} \Omega_{\text{R}}  t_{\text{p}}$. The quantum state transfer rate  can be thus estimated as
\begin{equation}
f= 1/t_{\text{p} } = \frac{\mathcal S N_{\text{A} }^2 \Omega_{\text{R} } ^2 \hbar \omega} {2 \gamma P} \frac{1}{d},
\label{eq:transmissionFrequency}
\end{equation}
where we introduced $\mathcal S \equiv \sigma / \sqrt{\overline n} $  as the ratio of the photonic distribution widths of a number-squeezed state $\sigma$ and a coherent state $\sqrt{\overline n}$.
For instance, we assume a photonic frequency $\omega =400\,\text{THz}$ and a typical  Rabi  frequency of atomic systems   $\Omega_R = 40\,\text{MHz}$, corresponding to a laser power $P = 10\,\mu\text{W}$. Recent experiments  of the Lukin  group have successfully created GHZ states with $N_{\text{A} } =12$ Rydberg atoms~\cite{Bluvstein2022}. When assuming a coherent state with $\mathcal S =1$, the  transfer rate is $122\, \text{Hz} $ over $500 \,\text{km}$, thus exceeding typical few-photon protocol by two orders of magnitude.

\subsection{Discussion}

Even though  strongly idealized, the proposed protocol  merits a thorough discussion. As quantum information is often stored in the ground-state manifold of atoms, that are coupled via Raman transitions, a more realistic modeling in terms of a three-or-more level system is required. We emphasize that the proposed protocol is not restricted to the Tavis-Cummings model used here for illustration, as the light-matter entanglement is a generic effect appearing in all Floquet systems as predicted by the PRFT.

There is a series of points that can be discussed independent of the concrete physical implementation. Along with future progress in quantum control, the suggested protocol has enormous \textit{development potential}:
\begin{itemize}
	\item \textit{Atom number.} The quantum state transfer rate in Eq.~\eqref{eq:transmissionFrequency} scales quadratically in the number of atoms $N_{\text{A} }$. It has been estimated that for fault-tolerant quantum computation $N_{\text{A}} =1000$ physical qubits in a highly entangled state are required. This would increase the transfer rate by a factor of $10^4$ compared to the estimate in Sec.~\ref{sec:transmissionRate}.
	\item \textit{Transmission power.} The transfer rate is inversely proportional to the laser power $P$. When Alice deamplifies the beam after interaction with her GHZ state prior to transmission by a factor of $\alpha $,  the transfer rate can be increased by that factor $\alpha$. Bob must amplify the received signal before interaction with his GHZ state. Noiseless deamplification and amplification can be implemented with  additional atom  ensembles which are prepared in  Floquet states as  discussed in Sec.~\ref{sec:FloquetStateAnalysis} and shown in Figs.~\ref{figCountingStatSketch}(b) and \ref{fig:benchmarkTwoModeRabiFloquetState}(a). This might enhance the transfer rate by an additional factor of $\alpha =10$. 
	\item  \textit{Number squeezing.}  According to Eq.~\eqref{eq:transmissionFrequency}, the transfer frequency scales linearly with the number squeezing parameter $\mathcal S$. Moreover,  a moderate $\mathcal S$ also minimizes the shot-noise entanglement discussed in Sec.~\ref{sec:rabiModel}. We consider a number squeezing of $\mathcal S=0.1$ for the following estimation.
\end{itemize}
Taking these points into account, the transfer rate will be on the order of $f \approx 100 \text{MHz}$, and thus commercially relevant.
Compared to  few-photon quantum protocols, the coherent-state protocol comes with a serious of \textit{advantages}:
\begin{itemize}
	\item \textit{Distance dependence.} The transfer rate scales inverse proportional with the distance $d$. This is a more favorable scaling than for few-photon protocols whose quantum state transfer rate typically decrease exponentially due to   intensive postselection in  quantum repeaters~\cite{Sangouard2011}. 
	
	\item \textit{Simple implementation.} The implementation is based on the  light-matter entangling process. It does not require sophisticated encoding and decoding schemes to protect the quantum information that are experimentally and computationally challenging.
	
	\item \textit{On-demand light fields.}
	Presently, the absence of on-demand single photon sources pose a major challenge to several quantum communication protocols.
	This problem is circumvented in the proposed protocol, as  coherent pulses of light can be easily produced and controlled.
	
	\item \textit{Photon detection.} The efficiency of single-photon detectors strongly  influences the transfer rate  of few-photon protocols. As the success flag in the coherent-light protocol is determined by measuring the intensity difference [c.f., Fig.~\ref{fig:quantumCommnuication}(b)], a standard photon multiplier will be sufficient.
\end{itemize}

Compared with  few-photon quantum communication protocols, the coherent-state entanglement protocol has two major \textit{disadvantages}, though these can be overcome:
\begin{itemize}
	\item \textit{GHZ generation.} To establish an efficient protocol,  Alice and Bob must generate high-fidelity GHZ states, that might be technically challenging for large atom numbers. However, quantum information will be always stored in an encoded form. While the GHZ state is the basis of Shor's quantum repetition code~\cite{Shor1995}, other quantum error correction codes are based on graph states, that are generalizations of the GHZ states~\cite{lidar2013quatum}. Thus, the generation of  GHZ states will only lead to negligible overhead. 
	
	\item \textit{Noise and decoherence.}  As the quantum information is encoded in the photon number, the transmitted photonic state will be sensitive to phase noise and decoherence  related to the operators $\hat a_k^\dagger \hat a_k$ for $k=1,2$. Fortunately, as the ratio of   $\Delta \langle  \hat N_k \rangle$ and $\langle \hat N_k \rangle$  is on the order $10^{-5}$,  the environment can learn only little about the quantum state, such that the \textit{which-path} information is not leaked. To further enhance the protection of the quantum information, dynamical decoupling  can be employed by periodically switching $\left|\pmb - \right>_X \leftrightarrow \left|\pmb + \right>_X  $  for $X=\text{A},\text{B}$, leading to a periodic variation of  $ \Delta\langle \hat N_k \rangle$. 
\end{itemize}

The entanglement generation based on coherent light can be interpreted as a physical encoding of the quantum information, which is in contrast to the logical encoding of quantum information typically deployed in quantum communication protocols~\cite{lidar2013quatum}.  Assuming a pulse length of $t_{\text{p} }  = 20\, \text{ms}$, the two peaks are separated by $\Delta\langle  \hat N_1\rangle \approx 10^7$ photons which corresponds to a Hamming distance of $ \log_2 (\Delta \langle  \hat N_1\rangle )\approx 24 $ bits. As  the total number of photons is on the order of $\langle \hat N_1 \rangle = Pt/\hbar\omega \approx 10^{12}$ corresponding to $40$ qubits, the physical encoding presented here is thus comparable to a $\left[ \left[40 ,1,24 \right] \right]$ quantum error correction code.

\section{Summary and outlook}
\label{sec:conclusions}

\subsection{Summary}

In this subsection, we compare the PRFT with other well-established methods for analyzing light-matter systems. The PRFT combines important features of  established frameworks, while avoiding their shortcomings:\\

\textit{Floquet theory.} The PRFT introduces counting fields into the semiclassical equation of the light-matter system to track the quantum dynamics of the photonic driving field, thereby making an important advancement to the framework of Floquet theory.  Crucially, the PRFT defines a quantum channel for the dynamics of the driven matter system, that describes the decoherence induced by the light-matter interaction in the Floquet basis.
This inherently quantum effect is completely neglected in the standard Floquet theory, which treats the matter subsystem as an effectively closed quantum system. We note that while this eventual decoherence may be anticipated from other semiclassical techniques [53], a quantitative calculation is generally beyond the reach of these methods. Our investigations clearly demonstrate that Floquet theory suffers from fundamental limitations in describing light-matter-coupled systems. The  PRFT provides a semiclassical approach to address these issues, and emphasizes the need to carefully investigate the standard Floquet theory even in parameter regimes, in which it has been thought to be valid.  
{{\color{\markColorTwo} The PRFT renders the Floquet theory  as an open quantum system framework, by providing a microscopic derivation of the Kraus operators.}

We emphasize that the PRFT has the same computational complexity as the standard Floquet theory, since it requires the integration of semiclassical equations.
As a consequence, the PRFT  has significant computational advantages over Sambe space methods that investigate photonic dynamics by effectively requantizing the semiclassical driving field~\cite{Long2021,Crowley2020,PhysRevB.99.064306}. Finally, we note that, unlike other approaches, the PRFT can distinguish between modes with commensurate frequencies, thereby extending its reach over a wide class of driven systems.}   The phase representation approach in Ref.~\cite{Guerin1997} derives photonic observables by formally considering the photon phase as a dynamical variable. However, this approach becomes problematic when specifying to coherent photonic states, for which the phase has been considered as static leading to unphysical prediction such as  diverging higher-order moments and cumulants. For this reason, the method in  Ref.~\cite{Guerin1997} comes short to describe the light-matter entanglement, which takes a prominent role within the PRFT.   \\

\textit{Quantum optics.} An extremely appealing feature of the PRFT is that it makes accurate predictions about  photonic observables, even though it only relies on semiclassical equations. Our framework thus provides a drastic computational advantage over established methods in quantum optics. In particular, we note that while it is extremely difficult to numerically investigate systems with more than two photon modes using traditional methods such as phase-space frameworks~\cite{Mandel2008,Scully1997,Gardiner2004}, the numerical effort  is independent of the mode number in the PRFT. Intriguingly, the PRFT predicts the generation of light-matter entangled states in general multimode driven systems, which is essentially infeasible to describe with  standard quantum optical methods.  
  This entanglement effect is a fundamental consequence of the photon transport between different modes, which is controlled by the matter quantum system. 
It is thus distinct from the entanglement resulting from the photon-number uncertainty (i.e., the shot-noise) of a single mode in a coherent state~\cite{GeaBanacloche1990,GeaBanacloche1991,GeaBanacloche2002,Phoenix1991,Goldberg2020}. Our analysis has shown that the transport induced entanglement has a profound impact on the photon dynamics in contrast to the more well-studied shot-noise induced entanglement. 
Furthermore, the PRFT is nonperturbative in nature, and thus capable of making predictions beyond methods from nonlinear spectroscopy~\cite{Mukamel1995}.

Before proceeding further, we note that there are several straightforward applications of the PRFT that could not be addressed in this work. For instance, in the benchmarking of the quantum Rabi model, it has been assumed that  the photon field is switched on in a quantum quench. A more realistic situation is a smooth switching of the photon density, that could be modeled as a coherent superposition in continuum mode photon field, or by a time-dependent light-matter coupling $\tilde g(t)$. In all of these cases, the overall framework remains valid.
\\

\textit{Full-counting statistics.} 
The PRFT provides a stochastic description of  photonic fields in terms of probabilities, moments and cumulants in a manner analogous to the standard FCS~\cite{levitov1996electron,Schoenhammer2007}. The standard FCS is based on two-point projective measurements, and  can describe spontaneous photon emission~\cite{mandel1995optical,mandel1979sub,cook1981photon}. However, the coherences in the photon number basis  are formally destroyed in this method, thereby leading to wrong predictions for photon modes in coherent states. In contrast, the photon counting statistics in the PRFT is obtained by two-point tomographic measurements, that are compatible with coherent states.  The formalism provides an exact expression for the change of the  dynamical moment- and cumulant-generating functions, that describe the change in the photonic statistics of the  photon modes. Finally, the \textit{ quasiprobabilities}  capture the redistribution of the initial probability distribution and therefore describe the photonic dynamics.

\subsection{Discussion and outlook}

The PRFT has far-reaching implications for quantum science and technologies. In particular, the
quantum optical decoherence predicted by the PRFT has serious consequences for quantum memories and quantum operations.

Quantum memories often use sophisticated Floquet control protocols, where the control fields  can  unintentionally induce quantum-optical decoherence. In the optical frequency regime, the quantum-optical coherence time predicted by the PRFT is reasonably short ($\approx$ ms), which is orders of magnitude smaller than the targeted storage time of quantum information. This analysis suggests that optical frequencies should be avoided in quantum memories and quantum operations, as even a small decoherence is detrimental to maintain the high fidelity required for quantum error correction. Moreover, we have argued that quantum time crystals are ideal candidates for quantum memories. Finally, we have employed the PRFT to propose a quantum communication protocol  that is robust against photon loss. Our proposal employs coherent photonic states instead of single photons to establish remote entanglement. The robustness of this protocol originates from the fact that the  width of the photonic probability distribution increases only with the square root of the number of lost photons. Consequently, the  quantum state transfer rate scales inversely with distance, thereby outperforming few-photon protocols based on quantum repeaters, that typically decay exponentially with distance. We plan to investigate detailed implementations of this protocol in future work.\\

 The PRFT developed in this paper can potentially have a significant impact on various research directions. 
Some promising applications include thermodynamics~\cite{Strasberg2017}, heat engines~\cite{Restrepo2018,Kolodrubetz2018,Nathan2021} and quantum phase transitions in  interacting spin chains in a cavity~\cite{li2020fast,li2021long,bastidas2021nonequilibrium}, and control of many-body localization~\cite{Koshkaki2022,Ng2021} in the presence of external driving. Thereby, the PRFT has a significant computational advantage over quantum optical methods, where accurate numerical calculations are extremely challenging in this regime. An extension to open quantum systems can also clarify the compatibility with the standard FCS. We further speculate that a suitable development of the PRFT will have important implications in spectroscopy and metrology.    The PRFT can be applied in the analysis of highly occupied Fock-state lattices, which have been shown to exhibit an intriguing collapse and revival dynamics~\cite{Saugmann2023}.
 Methods developed for noise suppression in electron transport can be also combined with the PRFT to control the photonic counting statistics~\cite{Xu2023}.

\section*{Acknowledgments}

This work is supported by the Guangdong Provincial Key Laboratory (Grant No.2019B121203002), the  AFOSR Grand No. FA9550-23-1-0598, the MURI-ARO Grant No. W911NF17-1-0323 through UC Santa Barbara, and the Shanghai Municipal Science and Technology Major Project through the Shanghai Research Center for Quantum Sciences (Grant No. 2019SHZDZX01). The authors acknowledge inspiring discussions with Victor Manuel Bastidas, Georgious Sivilogou, Zhigang Wu, JunYan Luo, Yiying Yan, and Gloria Platero.

\appendix
\allowdisplaybreaks

\section{Photon-resolved Floquet theory: Derivations}
\label{sec:photonResolvedFloquettheory}
In this Appendix, we study the PRFT in more details. In the following derivations, we focus on the special case of a single counting field to enhance the readability. The generalization to multimode photon fields  works along the same lines. 

\subsection{Unraveling of the quantum dynamics} 

\label{sec:unravelingQuantunDynamics}

To reveal the quantum properties of the Floquet theory, we start from the underlying quantum Hamiltonian in Eq.~\eqref{eq:hamiltonian:quantum} with a single photonic mode $k=1$. We unravel the corresponding  time-evolution operator in a perturbation picture defined by $H_{\text{S} } =H_0 + \omega \hat a^\dagger  \hat a $. In doing so, we find
\begin{eqnarray}
U(t) &=&  e^{-i H_{\text{Q} } t } \nonumber \\
&=&  \sum_{n=0}^{\infty}\int_0^{t}dt_n   \cdots      \int_0^{t_2}dt_1 \tilde U^{(n)}(t),
\label{eq:trfm:perturbationExpansion}
\end{eqnarray}
where we have defined the perturbation-resolved propagation operator
\begin{eqnarray}
\tilde U^{(n)}(t) =  e^{-i H_{\text{S} } (t - t_n)  } \cdots e^{-i H_{\text{S} } (t_2 - t_1)  }  H_1 \left(  \hat a^\dagger   + \hat a \right)  	  e^{-i H_{\text{S} } t_1 }\nonumber ,\\
\end{eqnarray}
in which the term $H_1 \left(  \hat a^\dagger   + \hat a \right)$ appears $n$ times.  Next, we introduce the counting field $\chi$ into the expansion in Eq.~\eqref{eq:trfm:perturbationExpansion} by  replacing $\hat a \rightarrow \hat a e^{i \chi}$  and $\hat a^\dagger  \rightarrow \hat a^\dagger  e^{-i \chi}$. This  generalizes the time-evolution operator $U(t) \rightarrow U_\chi(t) $, which can be formally written as
\begin{equation}
U_{\chi}(t) =   e^{-i\chi \hat N  }  U(t)e^{i\chi \hat N  }.
\label{eq:def:genTimeEvOpeartorQuantum}
\end{equation}
We can expand  the generalized time-evolution  operator as
\begin{eqnarray}
U_\chi(t) &=&   \sum_{m=-\infty}^{\infty} U^{(m)}(t) e^{-i \chi m},
\label{eq:trfm:photonExpansion} 
\end{eqnarray}
where  the photon-resolved time-evolution operators $ U^{(m)}$ contain all terms in which the difference of the number of creators  $m_{C}$ and the number of annihilators $m_{A}$ is  $m = m_C - m_A$.  Each $U^{(m)}(t) $  can be thus represented as  a  polynomial of  $\hat a$ and $\hat a ^\dagger$.  The non-unitary  $  U^{(m)}$ can be obtained from $U_\chi(t) $ by performing a Fourier transformation with respect to $\chi$, i.e.,
\begin{equation}
U^{(m)} (t) =  \frac{1} {2\pi }  \int_{0}^{2\pi } d\chi U_\chi (t) e^{im \chi} .
\label{eq:photonResolvedEvolutionOperator}
\end{equation}
Using these photon-resolved time-evolution operators, we can express expectation values of arbitrary observables  $\hat O$ (can be either matter-like, photonic, or mixed observables) as 
\begin{equation}
\left< \psi(t)  \right| \hat O \left|\psi(t) \right>  = \sum_{m_1,m_2} \left< \psi(t_0)  \right| U^{(m_1)\dagger} (t)  \hat O  U^{(m_2)} (t) \left|\psi(t_0) \right>.
\end{equation}
We assume a product state as initial state 
\begin{eqnarray}
\left|\psi(t_0) \right> = \left| \phi(t_0) \right>\otimes \sum_{n=0}^{\infty} a_n \left| n \right> ,
\label{eq:generalInitalState}
\end{eqnarray}
where the expansion coefficients $a_n $ shall fulfill the conditions (i) and (ii) explained later in Appendix.~\ref{sec:errorAnalysis}. For illustration, we consider the projector onto a particular Fock state
\begin{equation}
\hat P_{n} =  \left| n \right> \left< n \right| ,
\end{equation}
but  more complicated operators can be treated accordingly. For this specific operator, the expectation value can be written in terms of the photon-resolved time-evolution operators as
\begin{eqnarray}
\langle \hat P_{n} \rangle &\equiv&	\left< \psi(t)  \right| \hat P_{n} \left|\psi(t) \right>  \label{eq:result:expectionValue}\nonumber \\
&=&  \sum_{m_1,m_2}   \left<   \left[ U^{(m_1)\dagger} (t)\right]_{n-m_1,n}     \left[  U^{(m_2)} (t)\right]_{n,n-m_2}    \right>_{t_0} \nonumber  \\
&&\times a_{n-m_1}^{*} a_{n-m_2} ,
\end{eqnarray}
where we have defined the  photon-resolved propagation matrices by
\begin{equation}
\left[  U^{(m)\dagger} (t)\right]_{n_1,n_2} \equiv  \left< n_1 \right|   U^{(m)\dagger}\left| n_2 \right>,
\end{equation}
which act on the states in the matter system. The expectation value in Eq.~\eqref{eq:result:expectionValue} is taken with respect to the matter initial state $\left|\phi (t_0)\right>$.

\subsection{Transition to the photon-resolved Floquet theory}

\label{sec:transitionToPhoton-resolved Floquet theory}

The derivations in Appendix~\ref{sec:unravelingQuantunDynamics} have been carried out in the Fock space.
The connection of the photon-resolved quantum time-evolution and the PRFT is established via Eq.~\eqref{eq:trfm:photonExpansion}. We recall that  $U_\chi (t)$ can be represented as a polynomial of  $\hat a^\dagger$ and  $\hat a$ operators. As in the standard semiclassical approximation, we now replace
\begin{equation}
e^{i \omega \hat a^\dagger \hat a t } \hat a  e^{-i \omega \hat a^\dagger \hat a t }\rightarrow \alpha e^{- i\omega t+i\varphi},
\end{equation}
where $\alpha\gg 1$  and $\varphi$ are the amplitude and the phase of  the photon field.   We denote the resulting operator as $\mathcal U_\chi (t) $, which acts on the matter system. The transition to the PRFT is readily done by realizing that
\begin{equation}
\mathcal U_\chi (t)  =  \mathcal  T e^{-i \int_{0}^{t} \mathcal H_\chi(t')dt'} ,
\end{equation}
where $\mathcal  T$ is the time-ordering operator and the generalized time-periodic Hamiltonian on the right-hand side  is defined as
\begin{equation}
\mathcal	H_\chi(t) =  H_0 +\tilde g H_1 \alpha \left( e^{i\omega t - i \chi  } +  e^{-i\omega t + i \chi  }  \right).
\label{eq:ham:singleModeQuantumHam}
\end{equation}
The calculation of $\mathcal U_\chi (t)$ can be thus performed by analytically or numerically solving the time-dependent Schr\"{o}dinger equation with $\mathcal H_\chi(t)$ for all $\chi \in \left[ 0 , 2\pi \right) $.
The final step in the transition is done by evaluating
\begin{equation}
\mathcal U^{(m)} (t) =  \frac{1} {2\pi }  \int_{0}^{2\pi } d\chi \, \mathcal U_\chi (t) e^{im \chi} 
\label{eq:photonResolvedEvolutionOperatorSC}
\end{equation}
and replacing
\begin{equation}
\left[  U^{(m)\dagger} (t)\right]_{n_1,n_2}  \rightarrow 	\mathcal U^{(m)} (t)
\label{eq:rel:semi-classicalReplacement}
\end{equation}
in  Eq.~\eqref{eq:result:expectionValue}. This replacement is well justified as the matrix elements of the terms such as  $\hat a^{^\dagger n } \hat a^{m } $ depend only weakly on  the photon number for large  $n_1,n_2$.  The expectation value of $\hat P_n$  in the PRFT thus reads as
\begin{equation}
\left< \hat P_{n} \right>  = \sum_{m_1,m_2}   \left<  \mathcal U^{(m_1)\dagger} (t)  \mathcal U^{(m_2)} (t)     \right>_{t_0}   a_{n-m_1}^{*} a_{n-m_2},
\label{eq:res:photonProjectorExpV}
\end{equation}
where the expectation value is  taken in the matter initial state.

\subsection{Full-counting statistics}

\label{sec:fullCountingStatistics}

In the following, we derive the dynamical cumulant-generating function of the photon field given in Eq.~\eqref{eq:res:dynamicalcumulantGenFct}. As the physical background and interpretation has been already explained in Sec.~\ref{sec:overview}, we focus here on the merely technical details. For simplicity, we  count only the photons in  mode $k=1$. Nevertheless, we still implicitly allow for other photon modes, which are not explicitly counted. In this case,  the cumulant- and moment-generating functions are given as
\begin{eqnarray}
K_\chi(t) &=& \log \left[ M_\chi(t)\right] , \nonumber \\
M_\chi(t) &=& \left<  e^{-i\chi \hat N  } \right>_t ,
\label{eq:def:cumAndMomGenartingFunctions}
\end{eqnarray}
where  the time evolution is calculated with the full quantum Hamiltonian in Eq.~\eqref{eq:def:hamiltonian:floquet}.
The moment-generating function in Eq.~\eqref{eq:def:cumAndMomGenartingFunctions} can be rewritten as
\begin{eqnarray}
M_\chi(t)
&=& \frac{1}{2} \left<  U^{\dagger}  (t) U_{\chi}(t)e^{-i\chi \hat N  }  + e^{-i\chi \hat N   } U_{-\chi}^{\dagger}(t)   U(t)  \right>_{t_0} , \nonumber \\
\label{eq:tra:symmetrization}
\end{eqnarray}
where $U_{\chi}(t)$ is the generalized time-evolution operator in Eq.~\eqref{eq:def:genTimeEvOpeartorQuantum}. Here, the moment-generating function is presented in a symmetric way. At this stage, an unsymmetrical representation would be also correct, however, this will get problematic when taking the semiclassical limit later. In doing so, we make sure that the essential transformation property $M_\chi^{*}(t) = M_{-\chi}(t)$ is maintained  in the semiclassical limit, which guarantees that all moments and cumulants are real valued.

We assume that the initial state is separable 
\begin{equation}
\left| \Psi (t)\right> =   \left| \phi(t_0) \right> \otimes \left| A(t_0)\right>,
\label{eq:InitialState}
\end{equation}
where $\left| \phi(t_0) \right>$ is the initial state of the matter system. We can expand the initial state of the light field in the Fock basis
\begin{equation}
\left| A (t_0)\right>  =   \sum_{n } a_{ n}  \left|  n  \right>.
\end{equation}
Using  now the photon-resolved time-evolution operator in Eq.~\eqref{eq:photonResolvedEvolutionOperator}, we can expand Eq.~\eqref{eq:tra:symmetrization} in terms of  photon processes such that
\begin{widetext}
	\begin{eqnarray}
	M_\chi(t)   &\equiv& \frac{1}{2}\sum_{n, m_1,  m_2 }   \left< \phi(t_0) \right|       \mathcal U^{(  m_1 )\dagger} (t) e^{-i \chi   m_2   }   \mathcal U^{( m_2)} (t)   e^{-i \chi   \left(n -m_2  \right) }   \left| \phi(t_0) \right>  
	   a_{ n-m_{1}}^{*}  a_{ n-m_{2}}  + \left( \text{c.c.}, \chi\rightarrow -\chi \right)\nonumber  \\
	&\equiv& \frac{1}{2}\sum_{ m_1,  m_2,n }   \left< \phi(t_0) \right|       \mathcal U^{(  m_1 )\dagger} (t) e^{-i \chi   m_2   }   \mathcal U^{( m_2)} (t)      \left| \phi(t_0) \right>  
	   a_{ n-m_{1}+ m_2}^{*}  a_{n}  e^{-i \chi  n  }   + \left(\text{ c.c.}, \chi\rightarrow -\chi \right)  \nonumber.  \\
	\label{eq:tra1:momGenFunction}
	\end{eqnarray}
\end{widetext}
 In this expression, we have already carried out the semiclassical replacement of the photon-resolved  time-evolution operators  introduced in Eq.~\eqref{eq:rel:semi-classicalReplacement}. The validity of this replacement will be analyzed in Appendix~\ref{sec:errorAnalysis}. To make progress, we apply a Fourier transform to the photonic expansion coefficients
\begin{eqnarray}
a_{  n} =  \frac{1}{2\pi}\int_0^{2\pi} d\varphi\;  a_{\varphi} e^{i   n \cdot   \varphi } .
\label{eq:rel:semiClassicalReplacement}
\end{eqnarray}
Moreover, using the representation of the photon-resolved time-evolution operators in Eq.~\eqref{eq:photonResolvedEvolutionOperatorSC}, the moment-generating function reads as
\begin{widetext}
	\begin{eqnarray}
	M_\chi  (t)
	&=& \frac{1}{2 (2\pi)^4}\int d\chi_1 d\chi_2  d\varphi_1 d\varphi_2 \sum_{ m_1, m_2,n }   \left< \phi(t_0) \right|     e^{ - im_{1}     \chi_1  }      \mathcal U_{\chi_1}^{\dagger} (t)  \mathcal U_{\chi_2 }(t)   e^{im_{2}  \left( \chi_2 -\chi \right)  }      \left| \phi(t_0) \right> \nonumber  \\ 
	&& \times    e^{-i(n-m_{1}+ m_2 ) \varphi_1 } a_{ \varphi_1}^{*} a_{ \varphi_2}  e^{in \varphi_2 } e^{-i \chi    n  }   + \left(\text{ c.c.}, \chi\rightarrow -\chi \right)  .\nonumber \\
	&=& \frac{1}{2 (2\pi)^4} \int d\chi_1 d\chi_2  d\varphi_1 d\varphi_2 \sum_{m_1, m_2,n }   \left< \phi(t_0) \right|     e^{  im_{1}  \left(    \varphi_1-\chi_1\right)  }      \mathcal U_{\chi_1}^{\dagger} (t)  \mathcal U_{\chi_2 }(t)   e^{im_{2}  \left( \chi_2 - \chi -\varphi_1 \right)  }       \left| \phi(t_0) \right> \nonumber  \\ 
	&& \times     a_{  \varphi_1}^{*} a_{\varphi_2}   e^{i n    (-\chi-\varphi_1 +\varphi_2 ) }  + \left(\text{ c.c.}, \chi\rightarrow -\chi \right) .\nonumber
	\label{eq:tra2:momGenFunction} \\
	&=& \frac{1}{2 (2\pi)^2}\int d\varphi_1 d\varphi_2 \sum_{ n }   \left< \phi(t_0) \right|     \mathcal U_{\varphi_1}^{\dagger} (t)  \mathcal U_{ \varphi_1+\chi}(t)      \left| \phi(t_0) \right>     a_{  \varphi_1}^{*} a_{\varphi_2}   e^{in \left(-\chi-\varphi_1+\varphi_2 \right)      } + \left(\text{ c.c.}, \chi\rightarrow -\chi \right).
	\label{eq:tra3:momGenFunction} 
	\end{eqnarray}
\end{widetext}

Albeit general, Eq.~\eqref{eq:tra3:momGenFunction} is inconvenient to evaluate both analytically and numerically. Motivated by the abundant use of lasers and other coherent electromagnetic fields in experiments, we focus in the following on Gaussian photonic states, for which the expansion coefficients are given by
\begin{equation}
a_{  n } = \frac{1}{(2\pi )^{\frac{1}{4} } \sqrt{\sigma} } e^{-\frac{1}{4\sigma^2}   \left( n- \overline n \right)^2  }  e^{i \varphi   n},
\end{equation}
where $n$ is considered as a continuous variable,  $\overline n\gg 1$   and $\sigma$ denote the mean photon number and width, respectively, and $\varphi$ is the  phase.  For $\sigma^2=\overline n$, the state is a coherent state, while for $\sigma^2<\overline n$ ($\sigma^2>\overline n$) it is denoted as a number-squeezed (phase-squeezed) state.  More general  photonic states, such as multimode squeezed states and light-matter entangled states will be discussed in Appendix~\ref{sec:generalizations}. Expressed as a function of  phase, the coefficients in Eq.~\eqref{eq:rel:semiClassicalReplacement} read as

\begin{eqnarray}
a_{ \varphi_1} 
&=& \frac{2\pi\sqrt{2 \sigma}}{(2\pi )^{\frac{3}{4} }  }    e^{-\left( \varphi_1  - \varphi \right)     \sigma^2   \left(  \varphi_1 -\varphi \right)  }e^{i\left(  \varphi_1      - \varphi \right)   \overline n }. \nonumber\\
\end{eqnarray}
For large $\sigma$ the coefficients $a_{\varphi_1 }$ quickly decay with $\left| \varphi_1-\varphi\right|$, such that it is justified to  expand the expectation value in Eq.~\eqref{eq:tra3:momGenFunction}  in a Taylor series as
\begin{eqnarray}
F_{t, \chi}( \varphi_1) &\equiv& \left< \phi \right|     \mathcal U_{ \varphi_1}^{\dagger} (t)  \mathcal U_{  \varphi_1 + \chi}(t)      \left| \phi \right> \nonumber  \\
&=&    F_{t, \chi} ( \varphi ) \nonumber   \\
&+ &   \partial_{ \varphi }  F_{t, \chi}( \varphi  )\cdot \left( \varphi_1-  \varphi \right)  \nonumber \\
&+ & \frac{1}{2} \partial^2_{\varphi}F_{t, \chi} ( \varphi ) \left( \varphi_1 - \varphi \right)^2  \nonumber \\
&+& \mathcal O \left[ \left( \varphi_1 - \varphi \right)^3 \right].
\label{eq:def:auxilaryFunction}
\end{eqnarray}
Accordingly, we can  expand the moment-generating function as 
\begin{eqnarray}
2M_{ \chi} (t) =&& M_{ \chi}^{(0)} (t) +  M_{ \chi}^{(1)} (t) + M_{\chi}^{(2)} (t)+\dots \nonumber \\
&&+   \left( \text{c.c.}, \chi\rightarrow -\chi \right).
\label{eq:momentGeneratingFkt_distribution}
\end{eqnarray}

For the first term, evaluation of the Gaussian integral gives
\begin{eqnarray}
M_{ \chi}^{(0)} (t)  &=&     \sum_{n} \int d \varphi_1 d \varphi_2   F_{t, \chi}( \varphi ) \nonumber \\
&\times&\frac{2 \sigma }{(2\pi )^{\frac{3}{2} }  } \prod_{j=1,2}      e^{- \varphi_j    \sigma^2     \varphi_j  }e^{(-1)^j i  \varphi_j   \left( n -   \overline n \right) }   e^{-i  n \chi      } \nonumber \\
&=&  F_{t, \chi}(\varphi)  \sum_{n} a_{ n}^* a_{ n} e^{-i  n \chi      } \nonumber \\
&=&F_{t, \chi}( \varphi)   M_{ \chi} (t_0),
\label{eq:momentGenFktZeroOrderContribution}
\end{eqnarray}
where in the last  equality we have identified the moment-generating function at time $t_0$, i.e.,  $ M_{ \chi} (t_0) =   \sum_{n} a_{ n}^* a_{ n} e^{-i  n \chi      } $. 

Evaluating  the Gaussian integrals of the second and third terms in Eq.~\eqref{eq:momentGeneratingFkt_distribution}, we obtain
\begin{eqnarray}
M_{ \chi}^{(1)} (t)  &=&    \sum_{ n}  \int d \varphi_1 d \varphi_2   ( \partial_{\varphi }F_{t, \chi}) \varphi_{1} \nonumber \\
&\times&\frac{2 \sigma }{(2\pi )^{\frac{3}{2} }  } \prod_{j=1,2}       e^{- \varphi_j     \sigma^2     \varphi_j    }e^{i (-1)^j \varphi_j   \left( n -   \overline n\right) }   e^{-i  n \chi      } \nonumber \\
&=&  \sum_{ n}  \int d \varphi_1 d \varphi_2   ( \partial_{\varphi }F_{t, \chi} ) \frac {2\sigma^2}{2\pi } \varphi_{1} \nonumber    \\
&&\times  \prod_{j=1,2} e^{-\left[    \varphi_j  + i  \left(  n - \overline  n \right)  \frac{1}{2\sigma^2}  \right]  \sigma^2  \left[    \varphi_j  +i \frac{1}{2\sigma^2}  \left(  n -  \overline  n \right) \right] }   \nonumber \\
&& \times   a_{n}^*  a_{n} e^{-i  n \chi      } \nonumber \\
%
%
%
%
%
&=&  \sum_{ n}  ( \partial_{\varphi }F_{t, \chi} )   \sigma^{-2} i \left( n - \overline  n \right) \cdot a_{ n}^{*} a_{  n} e^{-i  n \chi      } \label{eq:termM1}\nonumber, \\
\end{eqnarray}
and
\begin{eqnarray}
M_{ \chi}^{(2)} (t)  &=& \sum_n  \int d \varphi_1 d \varphi_2  ( \partial_{\varphi }^2 F_{t, \chi})  \varphi_{1}^2 \nonumber \\
&\times&  \frac{2 \sigma }{(2\pi )^{\frac{3}{2} }  }  \prod_{j=1,2}   e^{- \varphi_j     \sigma^2     \varphi_j    }e^{i (-1)^j \varphi_j  \left( n -   \overline  n \right) }   e^{-i  n \chi      } \nonumber \\
&=&\sum_{ n}   ( \partial_{\varphi }^2 F_{t, \chi})\nonumber  
  \sigma^{-4} \left( n - \overline  n \right)^2 \nonumber  
 a_{ n}^{*} a_{ n} e^{-i  n \chi  }\nonumber\\
&+&  \sum_{  n}  ( \partial_{\varphi }^2 F_{t, \chi} ) \sigma^{-2}  a_{  n}^{*} a_{  n} e^{-i  n \chi      } ,\label{eq:termM2}
\end{eqnarray}
respectively. 

As explained in the error analysis in Appendix~\ref{sec:errorAnalysis},  the terms $M_{ \chi}^{(1)} $, $M_{ \chi}^{(2)} $ and higher order terms  do not produce leading-order contributions to the moment-generating function for large $\overline n$ and $\sigma$ with $\overline n \gg \sigma$. Thus, the moment-generating function is mainly determined by  $M_{ \chi}^{(0)} $ in Eq.~\eqref{eq:momentGenFktZeroOrderContribution} and reads as
\begin{eqnarray}
M_{ \chi}(t) 
&=& M_{\text{dy} , \chi}(t)
M_{ \chi}(t_0)  
+ \mathcal F \left[\frac{\sigma}{\overline n},\frac{g t}{\overline n},\frac{gt}{\sigma^2} \right]
\label{eq:res:momenGenFunction},
\end{eqnarray}
where we have defined the dynamical moment-generating function by
\begin{equation}
M_{\text{dy}, \chi}(t) \equiv \frac{1}{2}   \left<        \mathcal U_{ \varphi}^{\dagger} (t)  \mathcal U_{\varphi+\chi}(t)  +   \mathcal U_{ \varphi- \chi}^{\dagger} (t)  \mathcal U_{  \varphi}(t)   \right>_{t_0}  ,
\label{eq:dynMomentGenFktFloquet_PRFT}
\end{equation}
and $\mathcal F \left[ x\right]$ denotes an appropriate scaling function.
Importantly, in this form, the moment-generating function fulfills the correct transformation properties under inversion of the counting field $\chi\rightarrow -\chi$. A non-symmetric representation, e.g., when only taking the first term in Eq.~\eqref{eq:res:momenGenFunction}, does violate this basic property. Using the  relation $K_{ \chi}(t) =\log M_{ \chi}(t)  $, and recalling the definition of the dynamical cumulant-generation function in Eq.~\eqref{eq:def:dynamicalcumulantGeneratingFunction},
\begin{eqnarray}
K_{ \chi}(t)  & \equiv  & K_{\text{dy}}(  \chi, t) + 	K_{  \chi}(t_0) ,
\end{eqnarray}
we find that  $ K_{\text{dy} }( \chi, t) =\log M_{\text{dy}, \chi}(t)   $, which is the expression given in Eq.~\eqref{eq:res:dynamicalcumulantGenFct} after generalization to multiple counting fields.

\subsection{Periodically driven systems}

\label{sec:app:periodicallyDrivenSystems}

The PRFT makes intriguing predictions for the important class  of periodically driven systems. According to Floquet theory, the time evolution operator can be written as
\begin{equation}
\mathcal U_{\chi}(t)  = \sum_\mu e^{ -i E_{\mu,  \chi}(t-t_0)} \left| u_{\mu, \chi} (t)\right>  \left<  u_{\mu, \chi}(t_0) \right|,
\label{eq:defFloquetTimeEvolutionOperator}
\end{equation}
where $ E_{\mu, \boldsymbol \chi}$ are the quasienergies and $ \left| u_{\mu,\chi} (t)\right> =  \left| u_{\mu, \chi} (t+\tau)\right>$ are the time-periodic Floquet states. Both depend on the counting field. 
Consequently, the probability operator in Eq.~\eqref{eq:probabilityOperator} can be expressed as
\begin{eqnarray}
\hat P_n (t)  &=& \sum_{m,\mu,m_1,\mu_1} \tilde q_{m\mid \mu }  \hat Q_{m_1}^{\mu_1,\mu}(t) e^{i (E_{\mu_1, \varphi} - E_{\mu, \varphi}) (t-t_0)} \nonumber \\
 && \qquad \qquad \times p_{n-m-m_1}(t_0) \nonumber \\
&& + \text{ H.c.}  , 
\label{eq:trf:probabilityOperator}
\end{eqnarray}
where
\begin{eqnarray}
\tilde q_{m\mid\mu} &=&  \frac{1}{4\pi}     \int_{0}^{2\pi}   e^{i \left( E_{\mu,  \varphi }  - E_{\mu,  \varphi  +   \chi} \right)(t-t_0)}      e^{i m \chi}d\chi \nonumber \\
\end{eqnarray}
describes the stroboscopic dynamics of the system, and
\begin{eqnarray}
\hat Q_{m_1}^{\mu_1,\mu} &=& \left| u_{\mu_1,\varphi} (t_0)\right>  \left<  u_{\mu_1,\varphi}(t)\right| \nonumber \\
&&\times  \frac{1}{2\pi}\int_{0}^{2\pi} \left|  u_{\mu, \chi} (t)\right>  \left<  u_{\mu, \chi}(t_0) \right|  e^{i m_1 \chi}d\chi \nonumber 
\end{eqnarray}
contains the information about the so-called micromotion. The later has the property
\begin{equation}
\sum_{m_1=-\infty}^{\infty} \hat Q_{m_1}^{\mu_1,\mu}  = \delta_{\mu_1,\mu }  \left| u_{\mu_1,\varphi} (t_0)\right>  \left<  u_{\mu_1,\varphi}(t_0)   \right|,
\end{equation}
which will become important later.

Equation~\eqref{eq:trf:probabilityOperator} is too complicated to allow for a clear physical picture and requires simplification. We observe that $\hat Q_{m_1}^{\mu_1,\mu} $ is constructed via a Fourier analysis of the operator $ \left| u_{\mu, \chi} (t)\right>  \left<  u_{\mu, \chi}(t_0) \right|$. Under physical reasonable conditions, the Fourier components are physically restricted by finite $ m_{\text{min}} \leq m_1 \leq  m_{\text{max}}   $. Thus, when the initial probability distribution varies only slowly with photon number, we can replace $p_{n-m-m_1}(t_0)  \rightarrow  p_{n-m}(t_0) $ in agreement with  the explanations in Appendix~\ref{sec:errorAnalysis} below. Consequently, Eq.~\eqref{eq:trf:probabilityOperator} simplifies to
\begin{eqnarray}
\hat P_{n}(t) &=& \sum_{\mu}p_{n\mid\mu}(t) \ \left| u_{\mu,\varphi} (t_0)\right>  \left<  u_{\mu,\varphi}(t_0) \right|      \\
p_{n\mid\mu}(t) &=& \sum_m q_{m\mid\mu} p_{n-m}(t_0)  \nonumber  \\
q_{m\mid\mu} &=&  \frac{1}{4\pi}     \int_{0}^{2\pi}   e^{i \left( E_{\mu, \varphi}  - E_{\mu, \varphi  +  \chi} \right)(t-t_0 )}      e^{i m \chi}d\chi + \text{c.c} \nonumber ,  \\
\end{eqnarray}
which shows that the photon redistribution is mainly determined by the quasienergies.
Moreover, the  state of the total system in Eq.~\eqref{eq:state:totalSystemState} can be specified as
\begin{equation}
\left| \Psi (t)\right> = \sum_\mu c_\mu e^{-i E_{\mu,\varphi}(t-t_0)}\left| u_{\mu,\varphi}(t) \right> \otimes \left| A_\mu (t)\right>,
\label{eq:app:lightMatterEntanglement}
\end{equation}
where we have identified  the Floquet-state conditioned photonic states  as
\begin{eqnarray}
\left| A_\mu (t)\right> = \sum_{n}  \sqrt{ p_{n\mid \mu}}  e^{-i(\omega t - \varphi)}    \ket{n}.
\end{eqnarray}
To bring the total system state in this form,  we took advantage of  Baye's theorem $ p_{\mu\mid n} p_{n}  =   p_{n\mid \mu} p_{\mu }$.

\subsection{Error analysis}
\label{sec:errorAnalysis}

Here we analyze the error of the semiclassical   moment-generating function
in Eq.~\eqref{eq:res:momenGenFunction}. To allow for a quantitative statement about  the error, we consider the generic photonic state in Eq.~\eqref{eq:InitalStateBenchmarking}. To simplify the notation, we consider here a single photon mode, while the generalization to a multimode system works along the same lines. 

Informally speaking, the derivations in the PRFT make the following assumptions for the expansion coefficients $a_{m}^{(k)}$ in the Fock basis of the photonic initial states $  \sum_{m} a_{m}^{(k)} \left| m \right>_k   $:   The amplitudes $ \left| a_{m}^{(k)} \right|$  vary slowly with the photon number $m$;  The phases are well defined in the sense that $  \arg a_{m}^{(k)}  =  \varphi_k m $ with  constant $\varphi_k$. These assumptions are naturally fulfilled for the coherent state in Eq.~\eqref{eq:initalState} when $\alpha_k$ is large, such that $\left| a_m^{(k)}\right|^2$ obeys the Poisson distribution.

To quantify the error, we have to investigate two approximations: (i)  the semiclassical replacement of the photon-resolved time-evolution operators in Eq.~\eqref{eq:tra1:momGenFunction}; (ii) the higher-order contributions  $M_{ \chi}^{(l\ge 1)} $ in Eq.~\eqref{eq:momentGeneratingFkt_distribution}. \\

\textit{(i) Semiclassical replacement.} In the semiclassical replacement in Eq.~\eqref{eq:tra1:momGenFunction}, we have assumed that  the matrix elements of the photonic operator $a^\dagger$ 
\begin{equation}
C(n) = \left< n+1\right| \hat a^\dagger \left| n\right>  =\sqrt{n+1}
\end{equation}
are independent of the photon number during the time evolution. To  estimate the validity of this approximation, we consider the ratio of matrix elements for  the  photon numbers $n=N$ and $n = N + \kappa_{\text{dy},1}t +\sigma$, where $\kappa_{\text{dy},1}t$ describes the change of the mean photon number and $\sigma$ is the initial photon number standard deviation. As the ratio scales as
\begin{equation}
\frac{C(\overline n+\kappa_{\text{dy},1}t+\sigma )}{C(\overline n)} = 1+  \frac{\kappa_{\text{dy} ,1}(t) }{\overline n }+ \frac{\sigma }{\overline n} +\mathcal O \left( \frac{1}{\overline n^2}\right),
\end{equation}
the semiclassical replacement is correct as long as the ratio $\kappa_{\text{dy},1}t / \overline n  $ is small. As $\kappa_{\text{dy},1} (t) $ is of the order of  light-matter interaction times time, we conclude that the semiclassical approximation gives an error of the order $g t/\overline n $ as indicated in Eq.~\eqref{eq:res:momenGenFunction}. Moreover, the standard deviation $\sigma$ is required to be small compared to the mean photon number $\overline n$ to ensure the photon number independence of the matrix elements. This analysis thus justifies the  first and second error scalings in Eq.~\eqref{eq:res:momenGenFunction}.\\

\textbf{(ii) Expansion contributions.}  Here, we examine the magnitude of the term  $M_{ \chi}^{(1)} $, which contributes the lowest order correction to the semiclassical moment-generating function in Eq.~\eqref{eq:res:momenGenFunction}. The analysis of the terms  $M_{ \chi}^{(l>1)} $ works along the same lines and gives the same estimate.  Transforming the sum over the photon  number $n$ into  an integral, we find
\begin{eqnarray}
M_{ \chi}^{(1)} (t)  
&\rightarrow &  \int dn    ( \partial_{\varphi }F_{t, \chi} ) \frac{ in }{\sigma^2}     \frac{1}{\sigma\sqrt{2\pi}}  \cdot e^{-\frac{n^2}{2\sigma^2}} e^{-i  (n+\overline{n})\chi      } \label{eq:termM1_evaluate}\nonumber \\
&=& ( \partial_{\varphi }F_{t, \chi} )  e^{-\frac{\sigma^2\chi^2}{2}} e^{-i  \overline{n}\chi  } \sigma^{-2} i (-i\sigma^2 \chi)
\label{eq:firstOrderEroor}.
\end{eqnarray}

To make progress, we represent the auxiliary function $F_{t, \chi}$ defined in Eq.~\eqref{eq:def:auxilaryFunction}  as an exponential
\begin{equation}
F_{t, \chi} (\varphi) =	 e^{i f_{t,\chi} (\varphi) }. \label{eq:longTimeLimitAuxFct}
\end{equation}
In doing so, the first-order correction of the moment-generating function can be written as
\begin{eqnarray}
M_{ \chi}^{(1)} (t)  
=  \partial_{\varphi }  e^{-i [\overline{n}-f_{t}^{\prime} (\varphi) ]\chi -  [\frac{\sigma^2}{2}- i f_{t}^{\prime\prime} (\varphi) ]\chi^2 +\mathcal O(\chi^3) }\chi , \nonumber \\
\label{eq:firstOrderErrorPeriodicallyDrivenSystems}
\end{eqnarray}
where $f_{t}^{\prime} (\varphi) $ and $ f_{t}^{\prime\prime} (\varphi) $ denote the first and second derivatives of $f_{t,\chi} (\varphi) $ with respect to the counting field at $\chi=0$.
Using Eqs.~\eqref{eq:firstOrderEroor} and~\eqref{eq:longTimeLimitAuxFct}, we can evaluate the contribution of $M_{ \chi}^{(1)} $ to the probability distribution to be
\begin{eqnarray}
p_{n}^{(1)}  &=& \frac{1}{2\pi} \int d\chi   \left[ M_{ \chi}^{(1)} (t) +   M_{ -\chi}^{(1)*} (t)   \right] e^{i\chi n} \nonumber \\
&=& \partial_{\varphi }    \frac{i}{\sqrt{2}}\frac{n-\overline{n}+f_{t}^{\prime} (\varphi) }{\sigma^2/2 - i f_{t}^{\prime\prime} (\varphi) } \frac {e^{-\frac{(n-\overline{n}+f_{t}^{\prime} )^2}{\sigma^2/2 - i  f_{t}^{\prime\prime} (\varphi) } }} {\sqrt{2\pi} \sqrt{ \sigma^2/2 - i  f_{t}^{\prime\prime} (\varphi) } }   \nonumber \\
&&+  \textit{c.c.} + \mathcal O \left( \frac{1}{\sigma^4} \right).
\end{eqnarray}
The terms $\mathcal O (\chi^3)$ in the exponent in Eq.~\eqref{eq:firstOrderErrorPeriodicallyDrivenSystems} generate  terms of order $\mathcal O \left( \frac{1}{\sigma^4} \right)$. 
Inspection of the probabilities $p_n^{(1)}$ reveals that they are small if  $ \sigma^2\ll  f_{t}^{\prime} (\varphi)  $ and $\sigma^2\ll    f_{t}^{\prime\prime} (\varphi)  t  $. 
Both $ f_{t}^{\prime} (\varphi) $ and $  f_{t}^{\prime\prime} (\varphi)$ are defined via the logarithm of $F_{t, \chi} (\varphi)$, which is defined in Eq.~\eqref{eq:def:auxilaryFunction}. As the time evolution operators are an exponential of the Hamiltonian,  we conclude  that both $ f_{t}^{\prime} (\varphi) $ and $  f_{t}^{\prime\prime} (\varphi)$  scale with the product of the light-matter coupling $g$ and time  $t$. Consequently, we can  estimate that the error magnitude scales as $p_{n}^{(l= 1)}  =\mathcal O \left( \frac{g t}{\sigma^2}  \right) $. Carrying out a  similar analysis for  $M_{ \chi}^{(l>1)} $, we  find the same error scaling, such that we can conclude
\begin{equation}
p_{n}^{(l\ge 1)}  =\mathcal F \left(\frac{g t}{\sigma^2} \right)  ,
\end{equation}
i.e., all terms $l\ge1$ can be neglected in the large $\sigma$ limit. As the moment-generating function  can be expressed as the Fourier transformation of the probabilities, we arrive at the third error scaling given in Eq.~\eqref{eq:res:momenGenFunction}.

\subsection{Generalizations}

\label{sec:generalizations}

The expression of the moment-generating function in Eq.~\eqref{eq:res:momenGenFunction} can be generalized to more general initial states along the same lines as in  Appendix~\ref{sec:fullCountingStatistics}. As the notation is tedious, we  state only the final result here.  

We consider a generic light-matter initial state of the form
\begin{equation}
\left| \Psi (t_0)\right> = \sum_\lambda c_\lambda \left| \phi_\lambda(t_0) \right> \otimes \left| A_\lambda (t_0)\right>,
\label{eq:GeneralInitialState}
\end{equation}
where we make no further assumption about the matter initial state $\ket{ \phi_\lambda(t_0) } $. The $c_\lambda$ are expansion coefficients. In contrast to the initial state in Eq.~\eqref{eq:InitialState} considered before, we here allow for an entangled initial state. The photonic states are Gaussian states  and parameterized as
\begin{equation}
\left| A_\lambda (t_0)\right>  =   \sum_{\boldsymbol n } a_{\lambda,\boldsymbol  n}  \left|\boldsymbol   n  \right>,
\end{equation}
where the expansion coefficients in the Fock basis $\left|\boldsymbol   n  \right>$ can be written as
\begin{equation}
a_{\lambda, \mathbf n } = \frac{1}{(2\pi )^{\frac{N_R}{4} } \sqrt{\det \boldsymbol \Sigma_\lambda} } e^{-\frac{1}{4}\left(\mathbf n-\overline{ \mathbf n}_\lambda   \right) \boldsymbol \Sigma_\lambda^{-2}  \left(\mathbf n-\overline{ \mathbf n}_\lambda  \right)^T   }  e^{i {\boldsymbol \varphi_\lambda} \cdot \boldsymbol n}.
\end{equation}
Thereby, $\overline{ \mathbf n}_\lambda = (\overline n_{\lambda, 1},\dots, \overline n_{\lambda, R})$ is the vector of the initial mean photon numbers, $\boldsymbol \Sigma_\lambda$ is a Hermitian matrix describing the corresponding standard deviation, and $\boldsymbol \varphi_\lambda = (\varphi_{\lambda, 1},\dots, \varphi_{\lambda, R})$ denotes the phases of the $R$ photon modes. 
This parametrization  covers coherent photonic states, number-squeezed photonic states, phase-squeezed photonic states, and multimode entangled photonic states.   A suitable linear combination  of the Gaussian states  also allows for  the description of Fock states. However, this would be numerically very expensive and eradicate the simplicity of the PRFT.

Generalizing the derivations in Appendix~\ref{sec:fullCountingStatistics} with regard to the initial state in Eq.~\eqref{eq:GeneralInitialState}, we finally arrive at the  generic moment-generating function
\begin{eqnarray}
M_{\boldsymbol \chi}(t) 
&=& \frac{1}{2} \sum_{ \lambda_1,\lambda_2 } \left[   \left<    \phi_{\lambda_1} (t_0 )\right|   \mathcal U_{ \boldsymbol   \varphi_{\lambda_1}}^{\dagger} (t)  \mathcal U_{ \boldsymbol  \varphi_{\lambda_1}+\boldsymbol  \chi}(t) \left|\phi_{\lambda_2}(t_0) \right> \right. \nonumber  \\
&+& \left. \left<    \phi_{\lambda_1}(t_0) \right| \mathcal U_{\boldsymbol  \varphi_{\lambda_2}-\boldsymbol  \chi}^{\dagger} (t)  \mathcal U_{\boldsymbol  \varphi_{\lambda_2 }}(t)  \left|\phi_{\lambda_2}(t_0) \right> \right] \nonumber\\ 
&&\times  M_{\boldsymbol \chi}^{\lambda_1,\lambda_2}(t_0)  \nonumber.\\
&+& \mathcal F \left[\left\lbrace \frac{1}{\sigma_{\lambda,k}^2},\frac{\sigma_{\lambda,k}}{\overline{n}_{\lambda,k}},\frac{g_{\lambda,k}\cdot t}{\overline{n}_{\lambda,k}} \right\rbrace_{\lambda,k} \right],
\label{eq:res:momenGenFunction_entangled}
\end{eqnarray}
where the initial moment-generating function is given by
\begin{equation}
M_{\boldsymbol \chi}^{\lambda_1,\lambda_2}(t_0)  =\sum_{\boldsymbol n} a_{\lambda_1,\boldsymbol  n}^*a_{\lambda_2,\boldsymbol  n} e^{i\boldsymbol \chi\cdot \boldsymbol  n}.
\end{equation}
The consequences of the initial light-matter entanglement in Eq.~\eqref{eq:GeneralInitialState} may lead to new dynamical effects, whose analysis would exceed the scope of this paper.

\subsection{Standard classical derivation}

\label{sec:relationClassicalDriving}

Here, we show that the cumulant-generating function in Eq.~\eqref{eq:def:dynamicalcumulantGeneratingFunction}  reproduces the standard semiclassical definition of the energy current as considered in, e.g., Refs.~\cite{Long2021,Crowley2020}. We first consider the cases of a single driving mode. Using the definition of the first dynamical cumulant in Eq.~\eqref{eq:def:cumulants},  we find
\begin{eqnarray}
\kappa_{\text{dy} ,1}(t)&=&  	\Delta  \left< \hat N(t)\right>_{t_0} \\
&=& \frac{-i}{2} \left.  \left<     \mathcal U_{\varphi}^\dagger (t)   \frac{d}{d \chi} \mathcal  U_{\varphi+  \chi} (t)   \right>_{t_0} \right|_{  \chi\rightarrow  0 } +\text{ c.c.} \nonumber 
\end{eqnarray}
To evaluate the derivative, we use the identity
\begin{eqnarray}
\mathcal  	U_{ \chi} (t) =  \mathcal U_{ \chi} (t,t_0) =  \mathcal  U_{0} (t-\chi/\omega ,t_0-\chi/\omega),
\end{eqnarray}
from which we can easily show that
\begin{eqnarray}
i \frac{d}{d \chi}  \mathcal 	U_{ \chi} (t)  =  \frac{1}{\omega } \left[ \mathcal   H(t) \mathcal  U_{ \chi} (t) -  \mathcal   U_{ \chi}(t)\mathcal  H  (t_0)  \right].
\end{eqnarray}
Consequently,
\begin{eqnarray}
\omega		\kappa_{\text{dy} ,1}(t)
&=& \left<\mathcal   H  (t) \right>_t - \left<\mathcal  H  (t_0) \right>_{t_0}.
\end{eqnarray}
Deriving with respect to time, we readily find the energy current from the matter system to the photon mode
\begin{eqnarray}
\omega \frac{d}{dt}    		\kappa_{\text{ dy} ,1}(t) 
&=& \left< \frac{d}{dt}  \mathcal  H  (t) \right>_{t_0} \equiv I(t)   ,
\label{eq:rel:classicalCurrent}
\end{eqnarray}
which is the expression of the semiclassical energy current commonly used in the literature.

The generalization of Eq.~\eqref{eq:rel:classicalCurrent}  to multiple modes can be easily performed by requantizing the photon modes, which are not counted: Assuming we are interested in the counting statistics of mode $k$, we only quantize the modes $k'\neq k$ in the semiclassical Hamiltonian in Eq.~\eqref{eq:def:hamiltonian:floquet}, such that
\begin{equation}
\mathcal H(t)= \mathcal H_k (t  ) + H_0 + \sum_{k'}\tilde g H_{k' }\left( \hat a_{k' } + \hat a_{k' }^\dagger \right) ,
\end{equation}
where $\mathcal  H_k (t  ) = g H_k \cos(\omega_k t +\varphi_k )$. This is just the Hamiltonian in Eq.~\eqref{eq:def:hamiltonian:floquet} with a complicated $H_0$ and a single semiclassical photon mode. Using the same reasoning as before, we can directly conclude that the photon flux into mode $k$ is given as
\begin{eqnarray}
I_k(t) \equiv \frac{d}{dt}     \omega_k	\left< \Delta \hat N_k \right> 
&=& \left< \frac{d}{dt} \mathcal   H_k  (t) \right>.
\end{eqnarray}
This implies that the photon-number change in mode $k$ is captured by the operator
\begin{equation}
\Delta\hat  N_k   = \int_{t_0}^{t} \left[  \frac{d}{dt} \mathcal  H_k  (t)\right] dt,
\end{equation}
which is thus an operator in space and time. Clearly, an analysis of the transport dynamics in terms of $ \Delta\hat  N_k  $  does not offer the convenience of the PRFT. \\

\section{Standard Full-Counting Statistics}
\label{sec:standardFCS}

Here we review the derivation of the standard FCS~\cite{levitov1996electron,Schoenhammer2007}. We assume that the system is described by Hamiltonian $H_{\text{Q} }$ in Eq.~\eqref{eq:hamiltonian:quantum}.  We want to count the change of photons in  photonic  mode $\hat a$ between  times $t_0$ and $t_1$. The system is assumed  to be initially in state $\rho(t_0)$. The photon number change shall be determined by two-point projective measurements as sketched in  Fig.~\ref{figCountingStatSketch}(a).

At time $t_0$, we perform  a  projective measurement defined by the projector $\hat P_{n_0} =\left| n_0 \right> \left< n_0 \right| $, where $ \left| n_0\right>$ denote the Fock states of $ \hat a$,  to determine the initial photon number $n_0$. The resulting density operator of the light-matter system is given as
\begin{eqnarray}
\rho_{\text{pr} }(t_0) &=& \sum_{n_0} \hat P_{n_0 }    \rho(t_0)  \hat P_{n_0 }  \nonumber \\
&= &   \sum_{n_0 }  p_{n_0 } (t_0)  \rho_{\text{pr},n_0},
\label{eq:projectedInitialState}
\end{eqnarray}
where $p_{n_0}(t_0) $ denotes the probabilities to find $n_0$ photons. The density matrix $\rho_{\text{pr} ,n_0}$  is the system state  conditioned on the measurement outcome $n_0$. 

We define $p_{n\mid n_0}(t_1)$  as the  probability distribution to measure $n$ photons by a projective measurement  at time $t_1$, given that there have been $n_0$ photons at time $t_0$.  This conditional probability distribution can be formally written as
\begin{eqnarray}
p_{n \mid n_0}(t) &=& \text{Tr} \left[ \rho_{\text{pr},n_0}(t) \sum_m \delta_{n,m} \hat P_{m} \right] \nonumber  \\
&=& \text{Tr} \left[ \rho_{\text{pr},n_0}(t) \frac{1}{2\pi } \int_{-\pi}^{\pi}d\chi e^{-i\chi n}   \sum_m e^{i\chi m}  \hat P_m \right] \nonumber \\
&=& \frac{1}{2\pi } \int_{-\pi}^{\pi} d\chi e^{-i\chi n}   \text{Tr} \left[ \rho_{\text{pr},n_0}(t)   e^{i\chi \hat N} \right],
\end{eqnarray}
where $\hat N = \hat a^\dagger \hat a$.
The time-evolved density matrix is given as $ \rho_{\text{pr},n_0}(t ) = \hat U(t)\rho_{\text{pr} ,n_0} \hat U^\dagger (t) $. For the density matrix conditioned on the first projective measurement, we can replace
\begin{eqnarray}
\rho_{\text{pr},n_0}  = \rho_{\text{pr} ,n_0} e^{i\chi \hat n_0}    e^{-i\chi \hat N}.
\end{eqnarray}
We note that this step is the  reason why we have applied a projective measurement at time $t_0$. In doing so, the conditional probability distribution can be written as
\begin{eqnarray}
p_{n \mid n_0}
&=&  \int_{-\pi}^{\pi} \frac{d\chi}{2\pi }  e^{i\chi (n-n_0)}   \text{Tr} \left[\hat U\rho_{\text{pr},n_0} e^{-i\chi \hat N}  \hat U^\dagger   e^{i\chi \hat N} \right] \nonumber.
\end{eqnarray}
We are interested in the probability distribution of the  photon number change $\Delta n = n-n_0$, that we denote as $\overline p_{\Delta n}$. This distribution is the average   of  the conditional probability distribution $p_{n_0 +\Delta n \mid n_0}$  weighted by the initial probability distribution, i.e.,
\begin{eqnarray}
\overline p_{\Delta n} &\equiv& \sum_{n_0} p_{n_0 +\Delta n \mid n_0}(t) p_{n_0}(t_0) \nonumber  \\
&=& \frac{1}{2\pi } \int_{-\pi}^{\pi} d\chi e^{i\chi \Delta n}   \text{Tr} \left[\hat U(t)\rho_{\text{pr}}(t_0)  e^{-i\chi \hat N}  \hat U^\dagger (t)  e^{i\chi \hat N} \right] \nonumber  \\
&=&  \frac{1}{2\pi } \int_{-\pi}^{\pi} d\chi e^{i\chi \Delta n}   M_{\text{dy} ,\chi}^{(\text{pr})}(t).
\end{eqnarray}
In the last step, we have introduced the dynamical moment-generating function of the two-point projective measurement, that can be alternatively expressed as
\begin{equation}
M_{\text{dy},\chi}^{(\text{pr})}(t)  = \text{Tr} \left[\hat U_{\chi/2}(t)\rho_{\text{pr}}(t_0)   \hat U_{-\chi/2}^\dagger (t)   \right] 
\end{equation}
in terms of the generalized time-evolution operator
\begin{eqnarray}
\hat U_{\chi}(t)   &=&  e^{-i\chi \hat N}  \hat U (t)e^{i\chi \hat N}  \nonumber \\
&=& e^{-i H_{\text{Q},\chi} t }.
\label{eq:generalizedTimeEvolutionOperator}
\end{eqnarray}
In the second equality, we have expressed the time-evolution operator in terms of the  Hamiltonian 
\begin{equation}
H_{Q,\chi} = e^{-i\chi \hat N}  H_{Q}  e^{i\chi \hat N} .
\end{equation}
We emphasize that the projective measurement in Eq.~\eqref{eq:projectedInitialState} destroys coherences in the photon basis and thus modifies the initial state.  If the initial state $\rho(t_0)$ is already diagonal in the photon basis, such as  the vacuum state or thermal states, it remains unchanged and the standard FCS makes correct predictions. For this reasons, the standard FCS correctly describes spontaneous photon emission. 

In contrast, coherences in the number basis are destroyed due to the projection at time $t_0$ in Eq.~\eqref{eq:projectedInitialState}. When one   blindly replaces  $ H_{\text{Q},\chi} \rightarrow \mathcal H_{\chi} (t)$ in Eq.~\eqref{eq:generalizedTimeEvolutionOperator} with the semiclassical Hamiltonian $ \mathcal H_{\chi} (t)$ to mimic a coherent field, the resulting expression is not equivalent to the dynamical cumulant generating function of the PRFT in Eq.~\eqref{eq:res:dynamicalcumulantGenFct} and thus makes wrong predictions about the photon statistics.


\section{Details to the application of the model calculations}
\label{sec:purityCalculation}

\subsection{Benchmark calculations}
\label{sec:app:benchmarkCalculations}

\begin{figure*}
	\includegraphics[width=\linewidth]{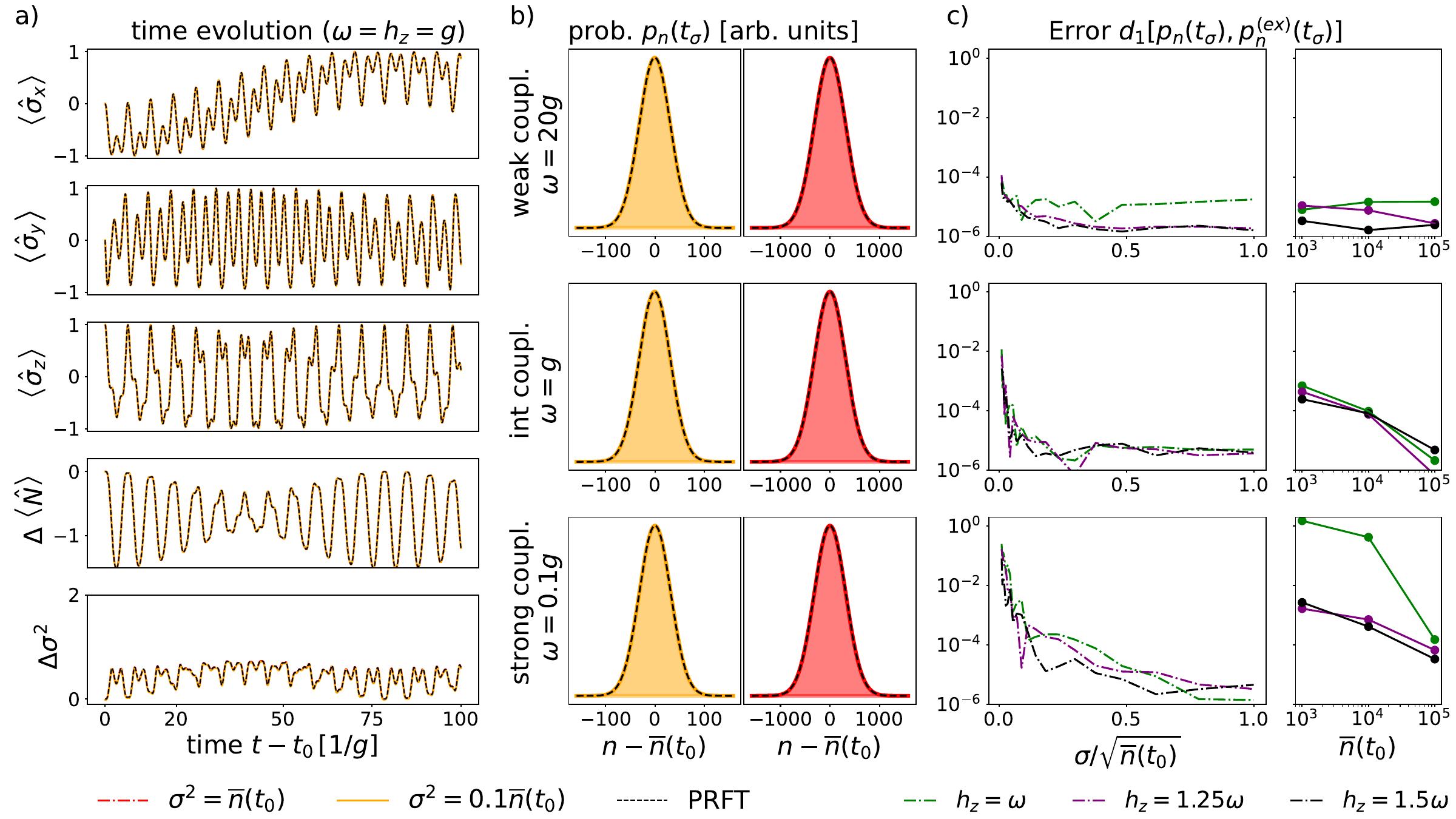}
	\caption{ \color{\markColorOne} Same as Fig.~\ref{fig:RabiModel}, but with the numerical time evolution simulated in the Sambe space instead of the Fock space. }
	\label{fig:RabiModelSambe}
\end{figure*}

\begin{figure*}
	\includegraphics[width=\linewidth]{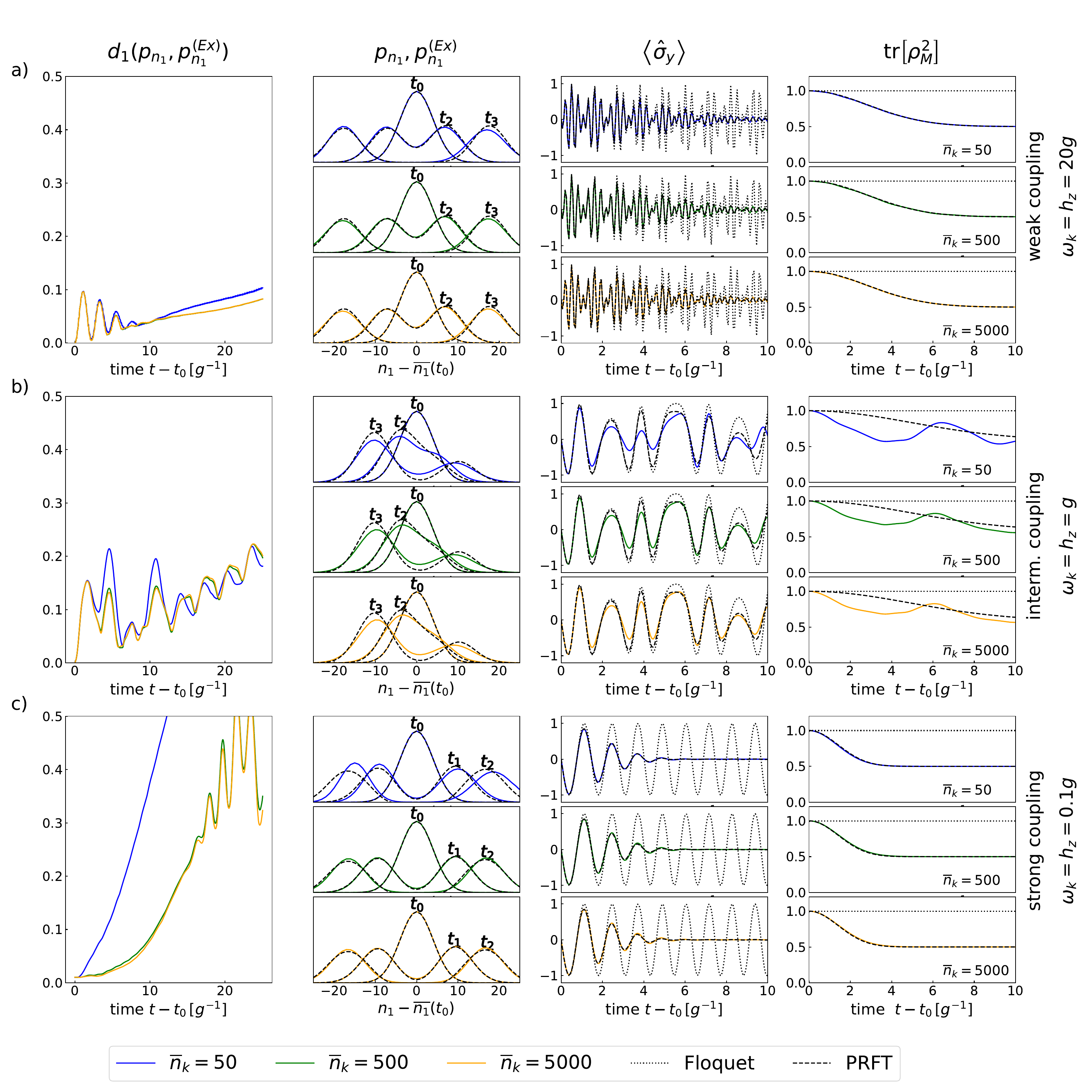}
	\caption{Same as in Fig.~\ref{fig:benchmarkTwoModeRabiFloquetState}, but for different initial mean photon numbers $\overline n_1 =\overline n_2 $  as indicated in the panels.   }
	\label{figTwoModeRabi-LmCoupling-meanPhotonNmb-spinUp}
\end{figure*}

\begin{figure*}
	\includegraphics[width=\linewidth]{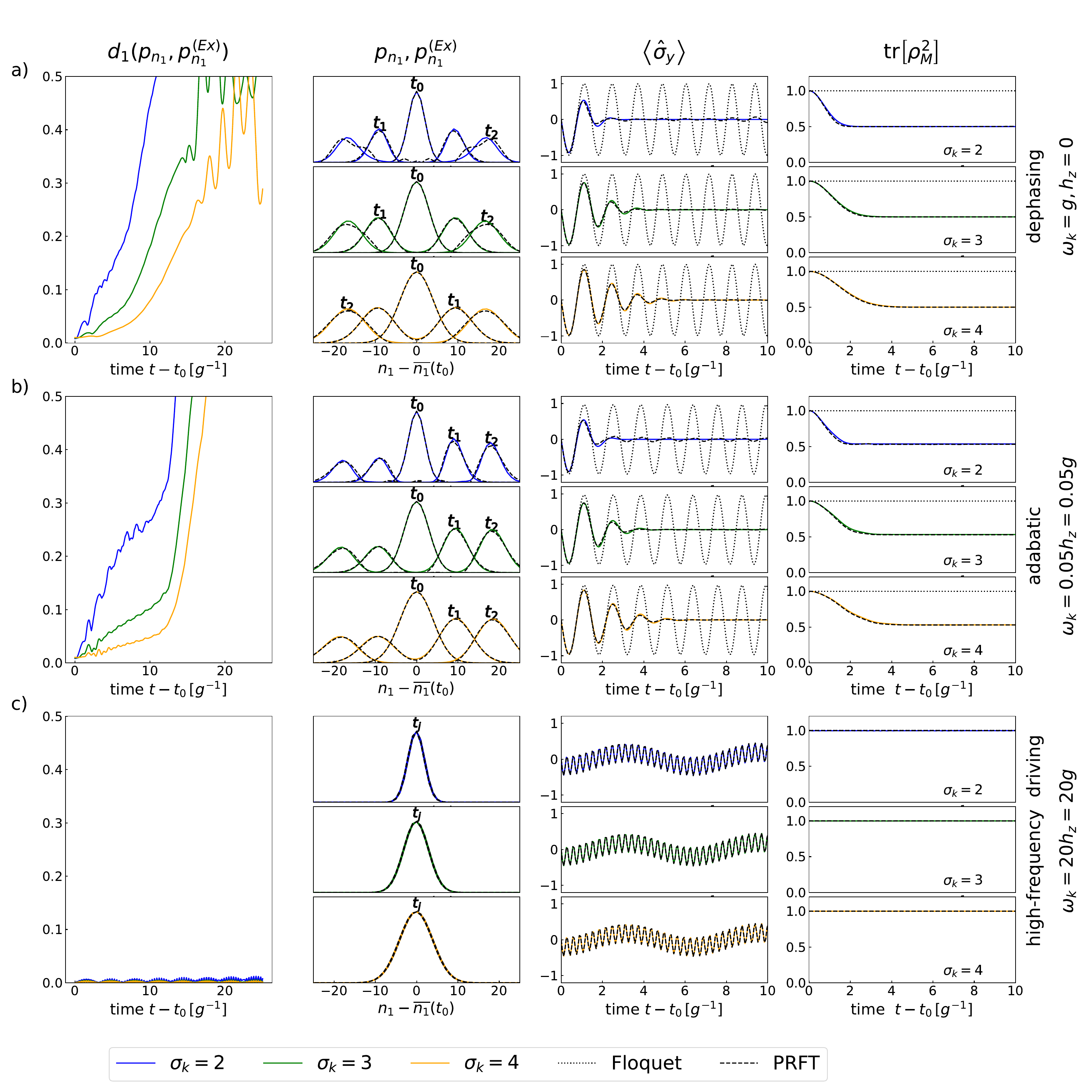}
	\caption{ Same as in Fig.~\ref{fig:benchmarkTwoModeRabiFloquetState}, but in different parameter regimes as indicated in the panels.   }
	\label{figTwoModeRabi-mixed-Sigma-SpinUp}
\end{figure*}

In Figs.~\ref{fig:RabiModelSambe}-\ref{figTwoModeRabi-mixed-Sigma-SpinUp}, we depict more benchmark calculations of the PRFT for the single-mode and  two-mode Rabi models. Here we shortly discuss the agreement of the PRFT to the exact quantum calculation, while the detailed interpretation of the physical effects exceeds the scope of this paper.

\textit{Sambe space simulation}. In Fig.~\ref{fig:RabiModelSambe}, we investigate the accuracy of the PRFT for the same parameters as in Fig.~\ref{fig:RabiModel}, but with the photonic subsystem represented in the Sambe space [introduced in Eq.~\eqref{eq:def:sambeSpace}] instead of the Fock space. In doing so, we can assess the impact of the photon number dependence of the operators $\hat a_k$ and $\hat a_k^\dagger$ on the accuracy. Here we shortly discuss the major differences between Figs.~\ref{fig:RabiModel} and ~\ref{fig:RabiModelSambe}: The variance change in Fig.~\ref{fig:RabiModelSambe}(a) predicted by PRFT agrees now perfectly to the numerical simulation. Note that the dynamics for $\sigma^2 =\overline n(t_0)$ lies exactly under the other two curves. The probability distributions in  Fig.~\ref{fig:RabiModelSambe}(b) of the PRFT and the numerical simulations agree perfectly to each other. The minor oscillations of the numerical simulation in Fig.~\ref{fig:RabiModel}(b) for $\sigma^2 =\overline n(t_0)$ have completely disappeared, as they are a consequence of the photon number dependence of $\hat a_k$ and $\hat a_k^\dagger$. The error in Fig.~\ref{fig:RabiModelSambe}(c) is significantly reduced compared to Fig.~\ref{fig:RabiModel}(c), especially for larger $\sigma$ values, demonstrating that the error is caused by the photon number dependence of $\hat a_k$ and $\hat a_k^\dagger$. Moreover, the error Fig.~\ref{fig:RabiModelSambe}(c) decreases significantly faster as function of $\overline n(t_0)$ than in Fig.~\ref{fig:RabiModel}(c). We note that  the error is bounded by $d_1> 10^{-5}$ -- $10^{-6}$ due to the computer precision in the numerical simulation.

\textit{Mean photon number}. In Fig.~\ref{figTwoModeRabi-LmCoupling-meanPhotonNmb-spinUp} we investigate the PRFT for three initial mean photon numbers $\overline n_k(t_0) = 50$, $\overline n_k(t_0) = 500$, and $\overline n_k(t_0) = 5000$ for the two photon modes $k=1,2$. The standard derivations are equally $\sigma_k =\sigma =4$. We observe that the PRFT time evolutions $\overline n_k(t_0) = 500$ and $\overline n_k(t_0) = 5000$ agree well to the exact numerical ones. We thus conclude   that the error is mainly determined by the small $\sigma$.

\textit{Mixed benchmarking}. In Fig.~\ref{figTwoModeRabi-mixed-Sigma-SpinUp} we carry out more benchmark calculations  for different photon standard deviations $\sigma_k =\sigma$ for $k=1,2$ in various parameters regimes, namely, in the dephasing, adiabatic, and high-frequency driving regime. We observe an overall precise agreement of the photonic probability distribution predicted by the PRFT to the exact numerical calculation in the full quantum model. Minor deviations  can be explained by numerical fluctuations. Interestingly, in the high-frequency driving regime $\omega_k \gg h_z,g_k$ in Fig.~\ref{figTwoModeRabi-mixed-Sigma-SpinUp}(c), we find a perfect agreement of the PRFT and exact quantum  calculations. However, in this regime there is no photon flux  between the photon modes as the two-level system cannot be resonantly excited.

\subsection{Jaynes-Cummings model} 

\label{sec:jaynesCummingsModel}

Here we  study the photonic dynamics in a solvable model of atom-photon interactions --- the Jaynes-Cummings model. This model describes a single two-level atom interacting with a near-resonant cavity mode of the photonic field~\cite{jaynes1963comparison,shore1993jaynes}:
\begin{equation}
H_{\text{JC} } =   \frac{h_z }{2}\hat \sigma_z + \omega \hat a^\dagger\hat a  + \tilde  g_1 \left( \hat \sigma_+ \hat a   + \hat \sigma_{-} \hat a^\dagger    \right),
\label{eq:ham:jcQuantum}
\end{equation}
where $\hat \sigma_{\pm} = \hat \sigma_x \pm i \hat \sigma_y$ and $\hat \sigma_z$ represents the usual Pauli matrices for spin-$1/2$ particles. Assuming the photon field to be classical, the corresponding semiclassical Hamiltonian describing the atomic subsystem is:
\begin{equation}
\mathcal H(t) =   \frac{h_z }{2}\hat \sigma_z  + g_1 \left( \hat \sigma_+ e^{-i\omega t }  + \hat \sigma_{-}e^{i\omega t }  \right), 
\label{eq:ham:jcSemCl}
\end{equation}
where $g =\tilde g\alpha$ with $\alpha$ being the amplitude of coherent photonic state. The generalized time-evolution operator is then given by Eq.~\eqref{eq:def:generalizyedTimeEvolutionOperator} where $\mathcal H_{\chi}(t)$ can be obtained from Eq.~\eqref{eq:ham:jcSemCl} by replacing $\hat \sigma_\pm  \rightarrow \hat \sigma_\pm e^{\pm i \chi}$.
In the interaction picture defined by $U_0(t)   =   \exp\left[ {-i \frac{\omega }{2}\hat \sigma_z t}\right]$, the generalized time-evolution operator becomes
\begin{eqnarray}
\mathcal U_{\chi}(t) = \cos\left( Et \right) {\mathbbm 1} +i \sin\left( Et \right) \hat \sigma_\chi ,
\end{eqnarray}
where
\begin{eqnarray}
E &=& \frac 12  \sqrt{ \left( h_z  -\omega\right)^2   + 16 g_1^2  },  \nonumber \\%
\tan \theta &=& \frac{2 g_1}  { h_z  -\omega  } \nonumber ,\\
\hat \sigma_\chi  &=&  \cos\theta \hat \sigma_z  + \sin\theta \left( \cos\chi\hat \sigma_{x }  + \sin\chi \hat \sigma_{y }   \right)  \nonumber.
\end{eqnarray}
The photon-resolved time-evolution operators can be obtained by applying Eq.~\eqref{eq:photonResolvedEvolutionOperatorSCmultiSum} and are given in Eq.~\eqref{eq:res:photonResolvedTimeEvolutionOperators}.\\

\subsection{Two-mode Jaynes-Cummings model}
\label{sec:multiModeJaynesCummings}

The two-mode generalization  of the Jaynes-Cummings model~\cite{wickenbrock2013collective,sundaresan2015beyond} allows for an analytical calculation of the counting statistics. The Hamiltonian reads as
\begin{equation}
H_{\text{TMJC} } =   \frac{h_z}{2}\hat \sigma_z + \sum_{k=1}^{2} \omega \hat a_k^\dagger\hat a_k  +\sum_{k=1}^{2}  \tilde  g_k \left( \hat \sigma_+ \hat a_k  + \hat \sigma_{-} \hat a_k^\dagger    \right)   ,
\label{eq:ham:jcQuantumMM}
\end{equation}
where both modes have the same frequency $\omega_k =\omega$, and  the photonic modes are initially in coherent states $  \left|\alpha_k e^{i\varphi_k}\right>$ with real valued $\alpha_k$ and $\varphi_k$. For a notation reason, we choose $t_0=0$ in the following calculations.\\

\textit{Moment-generating function.} Following the procedure  in Sec.~\ref{sec:overview}, we proceed to calculate the moment-generating function. The semiclassical Hamiltonian including the  counting fields is given by
\begin{eqnarray}
\mathcal H_{\boldsymbol \chi}(t) = \frac{h_z }{2}\hat \sigma_z  +  \left[\hat  \sigma_+ e^{-i\omega t  }  G(\boldsymbol \chi )+ \hat \sigma_{-}e^{i\omega t   }  G^*(\boldsymbol \chi )\right] \nonumber , 
\label{eq:generalizedHamiltonianMMJCmodel}
\end{eqnarray}
where we have defined $ G(\boldsymbol \chi ) = \sum_{k=1,2}   g_k e^{ i \chi_k   } .$
In an interaction picture defined by $U_0(t)   =   e^{-i \frac{\omega }{2}\hat \sigma_z t}$, the time-evolution operator reads as
\begin{eqnarray}
\mathcal U_{\boldsymbol \chi}(t) &=&    e^{-i H_{\boldsymbol \chi}  t} \nonumber \\
&=& \cos\left( E_{\boldsymbol \chi}  t \right) \mathbbm 1 +i \sin\left( E_{\boldsymbol \chi}  t\right) \hat \sigma_{\boldsymbol \chi} ,
\label{eq:tra:GenTimeEvOpMM}
\end{eqnarray}
where $H_{\boldsymbol \chi} = \frac{h_z }{2}\hat \sigma_z  +  \left[ \hat \sigma_+   G(\boldsymbol \chi )+\hat \sigma_{-}  G^*(\boldsymbol \chi )\right]  $ is the Floquet Hamiltonian and
\begin{eqnarray}
\hat \sigma_{\boldsymbol \chi} &=&  \cos\theta_{\boldsymbol \chi} \hat \sigma_z  + \sin\theta_{\boldsymbol \chi} \left( \cos\phi_{\boldsymbol \chi}\sigma_{x }  + \sin\phi_{\boldsymbol \chi}\hat  \sigma_{y }   \right) , \nonumber \\
E_{\boldsymbol \chi} &=& \frac{1}{2} \sqrt{ \left( h_z  -\omega\right)^2   + 16 \left| G(\boldsymbol \chi )  \right|^2  }, \nonumber \\%
\tan \theta_{\boldsymbol \chi} &=& \frac{2 \left| G(\boldsymbol \chi )  \right| }  { h_z  -\omega  }, \nonumber \\
\phi_{\boldsymbol \chi} &=& \text{ arg} \,G(\boldsymbol \chi ) .
\end{eqnarray}
The counting-field dependent Floquet states $\left| u_{\mu,\boldsymbol \chi }\right>$ are the eigenstates of $\sigma_{\boldsymbol \chi} $ with quasienergies $E_{\mu,\boldsymbol \chi} = \pm E_{\boldsymbol \chi}$. 
The dynamical moment-generating function then becomes 
\begin{widetext}
	\begin{eqnarray}
	2	M_{\text{dy} }( \boldsymbol \chi, t ) &=&  \left[\cos\left( E_{\boldsymbol \varphi }  t \right) \cos\left( E_{\boldsymbol \varphi + \boldsymbol \chi}  t \right)+  \sin\left( E_{\boldsymbol \varphi }  t \right) \sin \left( E_{\boldsymbol \varphi + \boldsymbol \chi}  t  \right) \left<\hat \sigma_{\boldsymbol \varphi } \hat \sigma_{\boldsymbol \varphi + \boldsymbol \chi} \right>_0   \right]   \nonumber  \\
	&+&  i \left[\cos\left( E_{\boldsymbol \varphi }  t \right) \sin \left( E_{\boldsymbol \varphi + \boldsymbol \chi}  t \right) \left< \sigma_{\boldsymbol \varphi + \boldsymbol \chi} \right>_0 -   \sin \left( E_{\boldsymbol \varphi}  t \right)  \cos \left( E_{\boldsymbol \varphi + \boldsymbol \chi}  t \right)\left<\hat \sigma_{\boldsymbol \varphi } \right>_0  \right] + (\text{c.c.}, \boldsymbol \chi\rightarrow -\boldsymbol\chi ) \nonumber  \\
	&\approx&  \cos\left[  \left(  E_{\boldsymbol \varphi } - E_{\boldsymbol \varphi + \boldsymbol \chi} \right) t \right]   
	+  i \sin\left[  \left(E_{\boldsymbol \varphi } - E_{\boldsymbol \varphi + \boldsymbol \chi} \right)  t \right]  \left< \hat\sigma_{\boldsymbol \varphi} \right>_0 	 + (\text{c.c.}, \boldsymbol \chi\rightarrow -\boldsymbol\chi ),
	\label{eq:res:momGenFktJaynesCummingsModel}
	\end{eqnarray}
\end{widetext}
where $\boldsymbol \varphi = \left(\varphi_1,  \varphi_2 \right)$. In the approximate expression,  we have neglected the counting fields in  $\hat \sigma_{\boldsymbol \varphi + \boldsymbol  \chi} \rightarrow \hat \sigma_{\boldsymbol \varphi }$, as it only accounts for sub-leading contributions in time.\\

\textit{Cumulants.}
For the approximated moment-generating function in Eq.~\eqref{eq:res:momGenFktJaynesCummingsModel}, the moments are:
\begin{eqnarray}
m_{2l+1}^{(1)} &=& (-1)^l\left[ \frac{2 g_1g_2 }{ E_{ \varphi} }\sin(\varphi)t \right]^{2l+1} \left< \hat \sigma_{\boldsymbol \varphi}\right>_{t_0} +\mathcal O \left(t^{2l}\right) \nonumber , \\
m_{2l}^{(1)} &=& (-1)^l\left[\frac{2 g_1g_2 }{ E_{ \varphi} } \sin(\varphi)  t\right] ^{2l} +\mathcal O \left(t^{2l-1}\right) ,
\end{eqnarray}
which  grow as $t^{2l +1}$ and $t^{2l }$ in time, respectively. In this form, we can  study the dependence of the transport processes on the relative phase $\varphi=\varphi_2 -\varphi_1$. Interestingly, only the odd moments depend on the  expectation value of $\sigma_{\boldsymbol \varphi }$.

When the initial state  is a Floquet state, the mean and the variance changes of the photon number distribution are:
\begin{eqnarray}
\left. \Delta\langle \hat   N_1 \rangle \right|_{\mu }&=&  (-1)^\mu \frac{2 g_1g_2}{ E_{ \varphi} }  \sin(\varphi)  t  +\mathcal O \left(t^0\right), \nonumber  \\
\left. \Delta \sigma_1^2 \right|_{\mu} &=&  0+\mathcal O \left(t^0\right).
\label{eq:res:meanVarDeterminisic}
\end{eqnarray}
For $\varphi\neq \left\lbrace 0,\pi \right\rbrace $, there is a net photon flux between the photon modes.  The variance vanishes up to minor  temporal fluctuations. According to Eq.~\eqref{eq:res:meanVarDeterminisic}, the direction and magnitude of the photon flux can be controlled by the relative phase  $\varphi$. In agreement with the analysis of the probability redistribution, the photon flux can be  controlled by the initial state. The variance change vanishes exactly as predicted by Eq.~\eqref{eq:vanishingVariance}.

The situation changes dramatically when the initial state is a balanced superposition of  two Floquet states $\left|\phi _{sp} \right>=( \left|u_{1,\boldsymbol \varphi} \right> + \left|u_{2,\boldsymbol \varphi} \right> )/\sqrt{2}$. In this case, the first two cumulants become
\begin{eqnarray}
\left. \Delta\langle N_1 \rangle \right|_{\left|\phi _{sp} \right> } &=&    \mathcal O \left(t^0\right)  ,\nonumber \\
\left. \Delta \sigma_1^2 \right|_{\left|\phi _{sp} \right> } &=&  \left[ \frac{2 g_1g_2 }{ E_{ \varphi} } \sin(\varphi) \right]^2 t^2+\mathcal O \left(t^1\right),
\label{eq:meanVArSchroedingerCat}
\end{eqnarray}
which shows that the mean photon flow between the two modes approximately vanishes, while the variance increases quadratically in time. The  variance change in Eq.~\eqref{eq:meanVArSchroedingerCat} is equal to the squared mean change in Eq.~\eqref{eq:res:meanVarDeterminisic}.  This time dependence is easily understood from Fig.~\ref{fig:benchmarkTwoModeRabiSpinUp}(a), as the rapid variance increase is a consequence of the linearly growing distance between the two Floquet-state-dependent peaks.\\

\begin{figure}
	\includegraphics[width=\linewidth]{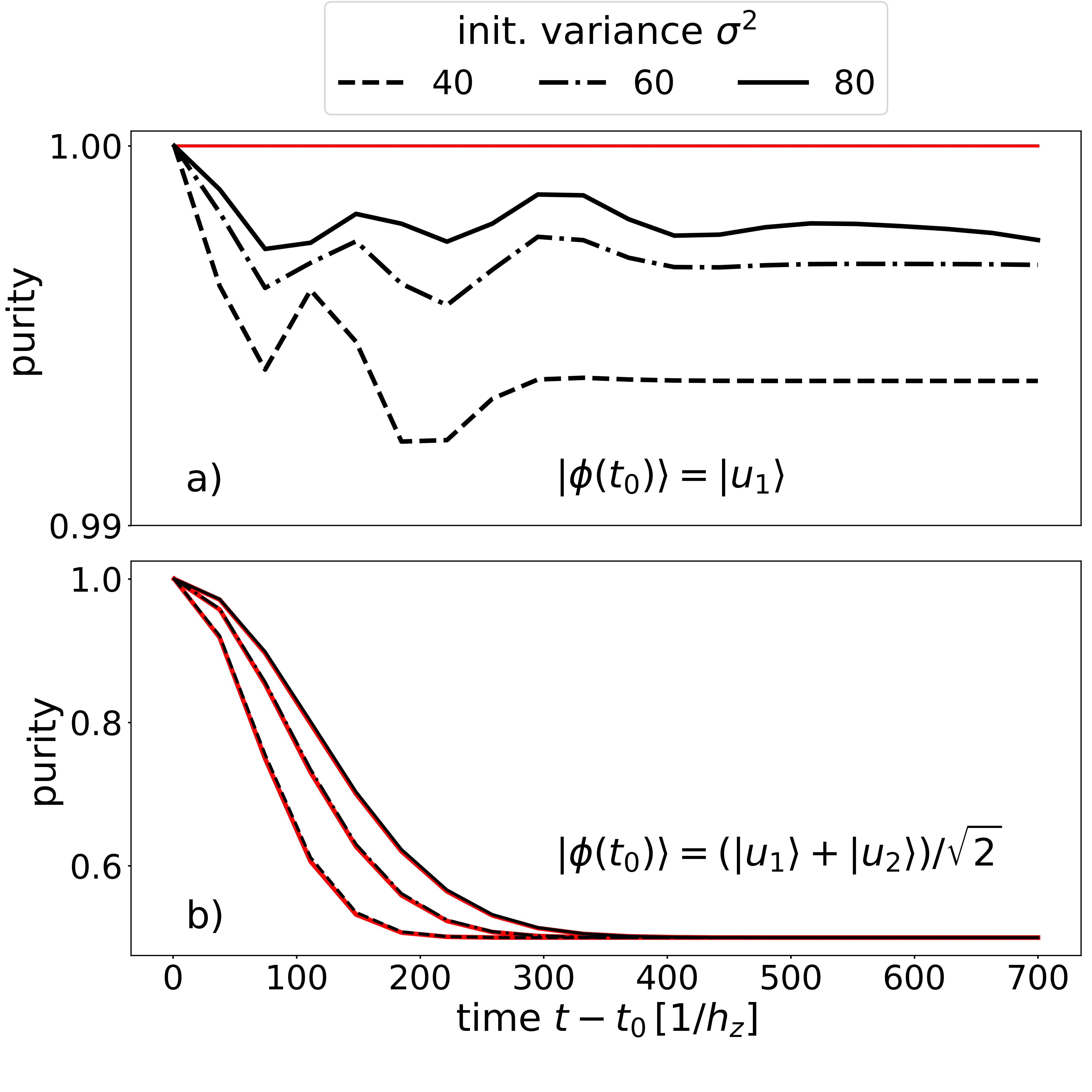}
	\caption{Purity as an entanglement measure of the light-matter state in the two-mode Jaynes-Cummings model as a function of time: (a) shows the purity for an initial Floquet state for different initial photon-number variances $\sigma^2 = \text{Var} \; \hat N_1(t_0)$, and (b) shows the purity for a balanced superposition of Floquet states as initial state. The black and red lines depict the quantum simulation and PRFT respectively. Overall parameters are $h_z =\omega = 10g$ and $\overline n_k(t_0) = 10^6$ for $k=1,2$.}
	\label{fig:PurityAnalysis}
\end{figure}

\textit{Light-matter entanglement.} Here, we apply the PRFT to calculate the purity in the Jaynes-Cummings model in order to describe the light-matter entanglement as a function of time. According to the PRFT and Eq.~\eqref{eq:lightMatterEntanglement}, the time-evolved state can be written as
\begin{equation}
\left| \Psi(t) \right> =  c_1 e^{-iE_{1,\boldsymbol \varphi} t }\left| u_{1,\boldsymbol \varphi} \right> \left| A_1   \right>  + c_2e^{-iE_{2,\boldsymbol \varphi} t } \left| u_{2,\boldsymbol \varphi}\right> \left| A_2 \right> ,
\end{equation}
with the in general non-orthogonal photonic states  $\left|A_\mu \right>$. The photon probability distribution  conditioned on the Floquet state approximately reads as $p_{ n_k\mid \mu}(t) =  e^{ (n_k-\overline n_{k,\mu}(t) )/2\sigma^2} /(\sqrt{\pi }\sigma) $, where $\sigma$ is the width, and $\overline n_{k,\mu}(t)= \overline n_{k}(0) +\kappa_{\text{dy},1\mid\mu}^{(k)}(t)$ is the time-dependent mean of mode $k=1,2$. The mean is determined by the  quasieneriges as  $\kappa_{\text{dy},1\mid\mu}^{(k)}(t) = - \partial_{\varphi_k} E_{\mu,\boldsymbol \varphi}' t \equiv   (-1)^{k}E'_{\varphi} t$, while the width stays constant.  

To calculate the purity, we evaluate the reduced density matrix of  $\rho(t) = 	\left| \Psi(t) \right> \left<  \Psi(t) \right|   $, i.e.,
\begin{eqnarray}
\rho _{\text{M}  } &=& \text{Tr}_{\text{L} } \left( \rho \right)
=
\left( 
\begin{array}{cc}
\left|c_1\right|^2 &  c_1^* c_2 \upsilon \\
c_2^* c_1 \upsilon^* & \left| c_2 \right|^2 
\end{array}
\right),
\end{eqnarray}
where $ \upsilon =\left< A_1  \mid A_2  \right>  $ is the overlap of the two photonic states. For the following calculation, we use that the phases in the Fock state expansion of both photonic states are equal, meaning that $\arg \left<  n_k \mid A_1 \right> = \arg \left<  n_k \mid A_2 \right>  $ for $k=1,2$, which is consistent with the PRFT. In this case, the overlap $\upsilon$ is completely determined by the conditional probabilities $p_{ n_k\mid\mu}(t)$. Considering the photon number as a continuous variable, we  can then evaluate
\begin{eqnarray}
\upsilon &=& \prod_{k=1,2}\left[ \frac{1}{\sigma\sqrt{\pi}}  \int dn_k \;  e^{-\frac{(n_k -\overline  n_{k,1})^2}{4 \sigma^2}} e^{-\frac{(n_k - \overline n_{k,2})^2}{4 \sigma^2}}\right] \nonumber  \\
&=&   \prod_{k=1,2}  e^{-  \frac{ \left[ \overline n_{k,1}(t) - \overline n_{k,2} (t)\right]^2 }{4 \sigma^2} }, 
\label{eq:overlapEvaluation}
\end{eqnarray}
where each photon mode $k=1,2$ has been integrated individually to obtain the correct overlap $\upsilon$.

The purity  is related to the eigenvalues $p_{r}$ of the reduced density matrix via $\mathcal P = \text{Tr}\left( \rho_{\text{M} } ^2\right) = p_1^2 + p_2^2 $, where the eigenvalues are explicitly given as
\begin{equation}
p_{1/2} = \frac{1}{2} \pm \frac{1}{2}\sqrt{ \left(\left|c_1 \right|^2-\left|c_2\right|^2 \right)^2  + 4 \left|c_1 \right|^2\left|c_2\right|^2 \left|\upsilon \right|^2 }.
\end{equation}
Using Eq.~\eqref{eq:overlapEvaluation} we thus finally obtain the time-dependent purity in the two-mode Jaynes-Cummings model
\begin{eqnarray}
\mathcal P  &=& \frac{1+ \left(\left|c_1 \right|^2-\left|c_2 \right|^2 \right)^2 }{2} + 2\left|c_1 c_2 \right|^2 e^{-  \frac{ \left(E'_{2,\varphi} -E'_{1,\varphi}   \right)^2 t^2 }{2 \sigma^2} }\nonumber . \\
\label{eq:timeDependentPurity}
\end{eqnarray}

Our results are shown in Fig.~\ref{fig:PurityAnalysis} together with the numerically calculated exact purity to confirm the accuracy of the PRFT calculation. 
We find that for an initial Floquet state (e.g., $c_1 =1$ and $c_2 =0$), the purity stays close to $\mathcal P \approx 1$ [see Fig.~\ref{fig:PurityAnalysis}(a)] in agreement with Eq.~\eqref{eq:timeDependentPurity}. On the other hand, for the balanced superposition state ($\left| c_1\right|^2 = \left| c_2\right|^2 =1/2$ and $c_2 =0$), the purity rapidly decays to   $\mathcal P \approx 1/2$ [see  Fig.~\ref{fig:PurityAnalysis}(b)], showing maximal light-matter entanglement. According to Eq.~\eqref{eq:timeDependentPurity}, the purity degrades faster for a smaller initial variance $\sigma^2$.

\bibliography{mybibliography}

\end{document}